\documentclass[prb, preprint, twocolumn, amsmath, amssymb, reprint, superscriptaddress]{revtex4-1}

\usepackage{graphicx,verbatim}
\usepackage{dcolumn}
\usepackage{bm}
\usepackage{sidecap}
\usepackage{braket}
\usepackage{color}
\usepackage{subcaption}

\usepackage{tikz}
\newcommand*\circled[1]{\tikz[baseline=(char.base)]{
            \node[shape=circle,draw,inner sep=2pt] (char) {#1};}}

\newcommand{\lsim}{\raisebox{-0.13cm}{~\shortstack{$<$ \\[-0.07cm]
      $\sim$}}~}
\newcommand{\gsim}{\raisebox{-0.13cm}{~\shortstack{$>$ \\[-0.07cm]
      $\sim$}}~}

\begin{document}

\title{
Current-induced dissociation in molecular junctions beyond the paradigm of vibrational heating: The role of anti-bonding electronic states
}

\author{A.\ Erpenbeck}
\affiliation{
Institute of Physics, Albert-Ludwig University Freiburg, Hermann-Herder-Strasse 3, 79104 Freiburg, Germany
}
\author{Y.\ Ke}
\affiliation{
Institute of Physics, Albert-Ludwig University Freiburg, Hermann-Herder-Strasse 3, 79104 Freiburg, Germany
}
\author{U.\ Peskin}
\affiliation{
Schulich Faculty of Chemistry, Technion-Israel Institute of
Technology, Haifa 32000, Israel
}
\author{M.\ Thoss}
\affiliation{
Institute of Physics, Albert-Ludwig University Freiburg, Hermann-Herder-Strasse 3, 79104 Freiburg, Germany
}
\affiliation{
EUCOR Centre for Quantum Science and Quantum Computing, Albert-Ludwig
University Freiburg, Hermann-Herder-Strasse 3, 79104 Freiburg, Germany
}

\date{\today}

\begin{abstract}
		The interaction between electronic and nuclear degrees of freedom in single-molecule junctions is an essential mechanism, which may result in the current-induced rupture of chemical bonds. As such, it is fundamental for the stability of molecular junctions and for the applicability of molecular electronic devices. In this publication, we study current-induced bond rupture in molecular junctions using a numerically exact scheme, which is based on the hierarchical quantum master equation (HQME) method in combination with a discrete variable representation for the nuclear degrees of freedom. Employing generic models for molecular junctions with dissociative nuclear potentials, we identify distinct mechanisms leading to dissociation, namely the electronic population of anti-bonding electronic states and the current-induced heating of vibrational modes. Our results reveal that the latter plays a negligible role whenever the electronic population of anti-bonding states is energetically possible. Consequently, the significance of current-induced heating as a source for dissociation in molecular junctions involving an active anti-bonding state is restricted to the non-resonant transport regime, which reframes the predominant paradigm in the field of molecular electronics. 
\end{abstract}

\maketitle

\section{Introduction}

	Transport through nanostructures is of fundamental interest for studying nonequilibrium quantum physics at the nanoscale, with the potential of a variety of future applications. 
	In particular molecular junctions, which comprise a single molecule bound to two macroscopic leads at finite bias voltage, are considered promising candidates for next-generation electronic devices.\cite{Aviram1974, Nitzan2001, Nitzan2003, Cuniberti, Galperin_Vib_Effects, Cuevas_Scheer, Zimbovskaya2011, Bergfield2013, Baldea, Su2016, Thoss2018} 
	Due to the size of molecular junctions, current-induced charge fluctuations strongly influence the nuclear (vibrational) degrees of freedom (DOFs), which in turn affects the conductivity of the molecule.\cite{Galperin_Vib_Effects,Rainer_Baldea, Galperin2005, Leijnse2008, Rainer2012, Rainer2011, Rainer2013, Wilner2014, Schinabeck2014, Erpenbeck2015, foti2018origin}
	This interplay between electronic and nuclear DOFs is the primary reason for current-induced bond rupture in molecular junctions, which is of importance for establishing reliable electronic components.
	
	A variety of experimental and theoretical studies verified the existence of current-induced vibrational excitations.\cite{Ioffe2008, Schulze2008, Schulze2008_2, deLeon2008, Huettel2009, Rainer2009, Rainer2010, Ward2010, Rainer2011, Rainer2011b, Franke2012, Schinabeck2016} 
	Typically, the level of current-induced vibrational excitation increases with bias voltage, which is in line with the fact that stable junctions are rarely observed for voltages beyond $\sim 1-2$ V.\cite{Schulze2008, Sabater2015} For larger bias voltages, the high level of current-induced vibrational excitation results in the mechanical instability of the junction, which was also observed experimentally.\cite{Venkataraman2015, Sabater2015, Venkataraman2016, Capozzi2016, Fung2019} 
	Understanding the underlying mechanisms is thus essential for the design of molecular electronic devices which remain operational at higher voltages. Moreover, it is crucial to also consider other mechanisms that can lead to the dissociation of molecular junctions beyond current-induced excitation of nuclear DOFs.

	For the theoretical assessment of the stability of molecular junctions, several different approaches have been employed. The effect of current-induced vibrational excitation is, for example, well established for models treating nuclear DOFs within the harmonic approximation.\cite{Wegewijs2005, Ryndyk2006, Benesch2008, Rainer2009, Rainer2011, Erpenbeck2016, Schinabeck2016} 
	Even though these models provide valuable indications on the stability of molecular junctions under current,\cite{Koch2006, Rainer2010, Volkovich2011, Rainer2011, Rainer2015} they fail to describe the dissociation process explicitly as this requires to account for potentials beyond the harmonic approximation. To date, this was only achieved by methodologies employing a classical description to nuclear motion \cite{Dzhioev2011, Dzhioev2013, Pozner2014, Erpenbeck_dissociation_2018} or by perturbative rate theories.\cite{Koch2006, Brisker2008}
	
	Beyond the field of molecular electronics, the effect of an electronic current on chemical bonds was also studied in the context of surface science. Experiments using a scanning tunneling microscope demonstrated that a current through a molecule can lead to desorption from the surface,\cite{Gao1995_1, Gao1995_2, Avouris1996, Gao1997, Boendgen1998, Seideman2003, Saalfrank2006, Menzel2012} 
	as well as breaking\cite{Martel1996, Stipe1997, Lauho2000, Hla2003, Roy2013} and forming\cite{Lee1999} molecular bonds.
	Along these lines, different processes responsible for these effects were identified, notably current-induced vibrational excitation\cite{Stipe1997, Seideman2003} and the population of excited electronic states.\cite{Avouris1996, Boendgen1998, Seideman2003}
	Similar processes were also considered in molecular dissociation and desorption from a surface upon laser excitation.\cite{Gao1995_1, Gao1995_2, Brandbyge1995, Saalfrank1996, Vondrak1999}
	This implies the importance of analogous processes for molecular junctions.
	From a theoretical point of view, the corresponding mechanisms were studied by models describing the nuclear DOFs in terms of truncated harmonic oscillators,\cite{Gao1995_1, Brandbyge1995, Stipe1997, Saalfrank2006}
	or Morse potentials,\cite{Brandbyge1995, Koch2006, Roy2013}
	where the dynamics was simulated by quasi-classical wave packet dynamics\cite{Avouris1996, Boendgen1998} 
	or quantum mechanical approaches using spatial grid representations.\cite{Saalfrank1996, Boendgen1998}

	In this paper, we study current-induced bond rupture in single-molecule junctions based on generic model systems employing a quantum mechanical approach to transport, which treats electrons and nuclei on a numerically exact footing. This approach, which we proposed recently [\onlinecite{Erpenbeck_2019_HQME}], is based on the numerically exact hierarchical quantum master equation (HQME) approach.\cite{Tanimura1989, Tanimura2006, Welack2006, Popescu2013, Haertle2013a, Haertle2014, Haertle2015, Wilkins2015, Wenderoth2016, Jin2008, Li2012, Zheng2013, Cheng2015, Ye2016, Schinabeck2016, Cheng2017, Hou2017, Erpenbeck_time-dep_mol-lead} Moreover, the framework uses a discrete variable representation (DVR) for the nuclear DOF, thus allowing for general potential energy surfaces (PESs) to be represented adequately.\cite{Tannor,Colbert1992}
	Here, we provide a comprehensive account of the method and apply it to study current-induced dissociation for a wide range of model parameters, ranging from the nonadiabatic regime of weak molecule-lead coupling to the adiabatic case of strong coupling.
	We note that this publication also extends our previous work in Ref.\ \onlinecite{Erpenbeck_dissociation_2018}, where we employed a mixed quantum-classical approach to current-induced bond rupture in molecular junctions thus neglecting nuclear quantum effects.
	{\color{black} In particular, the work presented in this publication accounts for current-induced excitation of the nuclear DOFs and for effects associated with tunneling of the nuclear wave packet into classical forbidden regions. These effects were not captured in the theoretical treatment employed in  Ref.\ \onlinecite{Erpenbeck_dissociation_2018}. Furthermore, the availability of a complete quantum mechanical treatment for the electrons and the nuclei allows for the validation of mixed quantum-classical approaches.} 
	
	The outline of this paper is as follows: In Sec.\ \ref{sec:theory}, we introduce the model and the transport method used in this work.
	In Sec.\ \ref{sec:results}, we present results for representative model systems. Thereby we distinguish two mechanisms leading to dissociation, namely the transient population of anti-bonding states by tunneling electrons (Sec.\ \ref{sec:DIET}) and current-induced heating of the vibrational mode (Sec.\ \ref{sec:DIMET}). Throughout this section, we comment on the validity of previous results obtained within a classical description of the nuclear DOF. Sec.\ \ref{sec:conclusion} concludes the paper with a summary.

\section{Theoretical methodology} \label{sec:theory}

	\subsection{Model}

		We study current-induced bond rupture in single-molecule junctions based on the Hamiltonian
		\begin{eqnarray}
			H	&=&	H_{\text{M}} + H_{\text{ML}} + H_{\text{MR}} + H_{\text{L}} + H_{\text{R}} ,
		\end{eqnarray}
		which consists of $H_{\text{M}}$ describing the molecule, $H_{\text{L/R}}$ modeling the left and right lead, and $H_{\text{ML/R}}$ characterizing the coupling between the molecule and the leads.

		For the sake of clarity, we apply a model of reduced dimensionality. We restrict ourselves to a single nuclear DOF describing the dissociation of a molecular bond. Further, a single electronic state of the molecule is considered, which can either be empty (in the following referred to as the neutral state of the molecule) or occupied (charged state). The corresponding molecular Hamiltonian assumes the form 
		\begin{eqnarray}
			H_{\text{M}} 	&=&	\frac{p^2}{2m} + V_0(x) (1-d^\dagger d) + V_d(x) d^\dagger d , \label{eq:H_M}
		\end{eqnarray}
		with $d^{\dagger}$/$d$ being the electronic creation/annihilation operators; $x$ and $p$ denote the position and the momentum of the nuclear mode with associated reduced mass $m$. 
		Within this model, $V_0(x)$ and $V_d(x)$ describe the PESs of the neutral and the charged state of the molecule, respectively. In Sec.\ \ref{sec:results}, we will assume that $V_0(x)$ is a bonding potential, whereas $V_d(x)$ is anti-bonding. The specific potentials used will be specified in Sec.\ \ref{sec:potentials}. We note that generalizations of this models to several vibrational modes and multiple electronic states are in principle straightforward.

		Electron transport is enabled via the coupling of the molecule to two macroscopic leads, which are modeled as reservoirs of non-interacting electrons and described by the Hamiltonian
		\begin{eqnarray}
			H_{\text{L/R}} 	&=&	\sum_{k \in {\text{L/R}}} \epsilon_k c_k^\dagger c_k . \label{eq:H_Leads}
		\end{eqnarray}
		Here, $\epsilon_k$ is the energy of lead-state $k$ and $c_k^{\dagger}$/$c_k$ the corresponding creation/annihilation operators.
		The coupling is given by
		\begin{eqnarray}
			H_{\text{ML/R}}	&=&	\sum_{k \in {\text{L/R}}} g_{\text{L/R}}(x) V_{k} c_k^\dagger d + \text{h.c.}\ . \label{eq:mol-lead_coupl}
		\end{eqnarray}
		The function $g_{\text{L/R}}(x)$ describes a position-dependent molecule-lead coupling, which allows to model scenarios, where the molecular conductance is influenced by the nuclear DOF.
		As such, the model can describe situations where the conductance of the molecule changes upon the current-induced dissociation of molecular bonds, e.g., because the conjugation across the molecule is broken or the dissociation induces geometrical changes at the contacts.
		The molecule-lead coupling of the form in Eq.\ (\ref{eq:mol-lead_coupl}) can be associated with the electronic spectral density 
		\begin{eqnarray}
			\Gamma_{\text{L/R}}(\epsilon)	&=&	2\pi\ \sum_{k \in {\text{L/R}}}  |V_{k}|^2 \delta(\epsilon_k-\epsilon)  \label{eq:def_Gamma}  ,
		\end{eqnarray}
		which is used to characterize the leads. 
		In Sec.\ \ref{sec:results}, the leads are described within the wide-band limit. Notice, that even though we used a time-independent formulation for the scope of this work, a generalization to time-dependent problems is in principle straightforward.

	\subsection{Transport theory}

		We use the method put forward in Ref.\ \onlinecite{Erpenbeck_2019_HQME} to describe current-induced dissociation in molecular junctions. This method is based on the HQME approach and describes the electronic as well as the nuclear DOFs on a numerically exact quantum level. It employs a DVR for the nuclear DOF, which facilitates the representation of generic PESs. Artificial reflections and finite-size effects are compensated for via a complex absorbing potential (CAP) and a Lindblad-like source term. 
		In order to make this publication self-contained, we present the basic concepts in the following.
		
		Generally, the HQME approach, also referred to as hierarchical equations of motion (HEOM), is a reduced density matrix scheme which describes the dynamics of a quantum system influenced by an environment. In our case, the molecule is conceived as the system whereas the leads represent the environment.
		As an extension to perturbative master equation approaches, the HQME method is capable of providing numerically exact results. 
		The approach was originally developed by Tanimura and Kubo in the context of relaxation dynamics,\cite{Tanimura1989, Tanimura2006} and was later on applied by various groups to describe transport in quantum systems.\cite{Welack2006, Popescu2013, Haertle2013a, Haertle2014, Haertle2015, Wenderoth2016, Jin2008, Li2012, Zheng2013, Cheng2015, Ye2016, Schinabeck2016, Cheng2017, Hou2017, Erpenbeck_dissociation_2018}
		For a detailed account of the HQME method, we refer to Refs.\ \onlinecite{Jin2008, Zheng2012, Haertle2013a, Tanimura2020}.

		Within the HQME framework, the influence of the leads, as described by Eqs.\ (\ref{eq:H_Leads}) and (\ref{eq:mol-lead_coupl}), is incorporated via the two-time bath correlation function,
		\begin{eqnarray}
			C_\text{L/R}^\pm(t, t',x) &=& \sum_{k \in \text{L/R}} g_{\text{L/R}}(x) |V_{k}|^2 \braket{ F^\pm_{\text{L/R}k}(t) F^\mp_{\text{L/R}k}(t') }, \nonumber\\ \label{eq:def_correl_1}
		\end{eqnarray}
		where $\braket{\dots}$ denotes the expectation value with respect to the bath DOFs.
		Thereby, the initial state of the leads is described by the density matrix $\rho_\text{L/R} = e^{-\beta(H_\text{L/R} - \mu_\text{L/R} N_\text{L/R})}/Z$, with $\beta=\frac{1}{k_{B}T}$, where $k_B$ is the Boltzmann constant, $T$ the temperature, $\mu_\text{L/R}$ the chemical potential of the respective lead, $N_\text{L/R}$ the associated particle number, and $Z$ the partition function.
		Moreover, 
		\begin{eqnarray}
			F^\pm_{\text{L/R}k}(t) &=& \exp\left(\frac{i}{\hbar} H_\text{L/R} t\right) c_k^\pm \exp\left(-\frac{i}{\hbar} H_\text{L/R} t \right) ,
		\end{eqnarray}
		with $c_k^- = c_k$ and $c_k^+ = c_k^\dagger$.
 		Notice, that we are using a slightly different definition of the correlation function compared to, for example, Refs.\ \onlinecite{Welack2006, Jin2008, Haertle2013a, Schinabeck2016}, as this simplifies the treatment of non-constant molecule-lead couplings.\cite{Erpenbeck_dissociation_2018}
		For the wide-band limit applied in the remainder of this work, 
		where $\Gamma_{\text{L/R}}(\epsilon) = \Gamma_{\text{L/R}}$,
		this two-time bath correlation function can be re-expressed as
		\begin{eqnarray}
			C_{\text{L/R}}^\pm(t,t',{x})	
			&=&	
									    \hspace*{-.05cm}
									    \int 
									    \hspace*{-.1cm}
									    d\epsilon \ e^{\pm \frac{i}{\hbar}\epsilon (t-t')} 
									    g_{\text{L/R}}(x) \Gamma_{\text{L/R}} f(\pm\epsilon, \pm\mu_\text{L/R}) . \nonumber \\ \label{eq:correlation_func_general}
		\end{eqnarray}
		Here,  $f(\epsilon, \mu) = \left( 1 + \exp(\beta(\epsilon-\mu))  \right)^{-1}$ is the Fermi distribution function.
		In order to obtain a closed set of equations within the HQME framework, it is necessary to represent Eq.\ (\ref{eq:correlation_func_general}) as a sum over exponentials,\cite{Jin2008}
		\begin{eqnarray}
			C_{\text{L/R}}^\pm(t,t',{x})	&\equiv& 	
					  \hbar \pi \ g_{\text{L/R}}(x) \Gamma_{\text{L/R}} \ \delta(t-t')	\label{eq:decomposition} \\&&
					- \sum_{p=1}^\infty \frac{2i\pi g_{\text{L/R}}(x) \Gamma_{\text{L/R}}}{\beta} \ \eta_p \ e^{-\gamma_{\text{L/R}p\pm} (t-t')} . \nonumber
		\end{eqnarray}
		Common approaches for calculating the parameters $\eta_{p}$ and $\gamma_{\text{L/R}p\pm}$ include the Matsubara,\cite{Mahan, Tanimura2006, Jin2008} the Pade,\cite{Hu2010, Hu2011} or the Chebyshev decomposition\cite{Tian2012, Popescu2015, Popescu2016} as well as more intricate schemes.\cite{Tang2015, Ye2017, Erpenbeck_RSHQME}
		The $\delta$-function in Eq.\ (\ref{eq:decomposition}) is directly associated with the wide-band approximation employed in this work. It requires an extra treatment within the HQME framework,\cite{Croy2009,Zheng2010, Zhang2013,Kwok2014, Erpenbeck_dissociation_2018} which we implement by extending the index set for the poles $p$ by zero. The details are given in the following.

		Based on the description of the bath in terms of Eqs.\ (\ref{eq:def_correl_1}) and (\ref{eq:decomposition}), the HQME method uses a set of auxiliary density operators $\rho_{j_1 \dots j_n}^{(n)}(t)$ to describe the dynamics of the system. The auxiliary density operators, which obey the equation of motion (EOM)
		\begin{eqnarray}
			\frac{\partial}{\partial t} \rho_{j_1 \dots j_n}^{(n)}(t) 	&=& 
			\left[-\frac{i}{\hbar} \left( \mathcal{L}_\text{S} + \mathcal{F}\right) -\left( \sum_{m=1}^n \gamma_{j_m} \right)  \right] \rho_{j_1 \dots j_n}^{(n)}(t)
			\nonumber \\&&
			-i \sum_{m=1}^n (-1)^{n-m} \mathcal{C}_{j_m} \rho_{j_1\dots j_{m-1} j_{m+1} \dots j_n}^{(n-1)}(t) \nonumber \\&&
			-\frac{i}{\hbar^2} \sum_{j} A^{\overline{\sigma_{j}}} \rho_{j_1 \dots j_n j}^{(n+1)}(t) \ ,
			\label{eq:EQM_nth_tier}
		\end{eqnarray}
		are operators acting on the electronic and the nuclear DOFs of the system. The multi-index is defined as $j_i = (l_i, p_i, \sigma_i)$, with $l_i \in \lbrace\text{L/R}\rbrace$, $\sigma_i = \pm 1$ and $p_i$ being the pole-index related to the decomposition in Eq.\ (\ref{eq:decomposition}). Moreover, $\overline{\sigma} = -\sigma$ and $\mathcal{L}_\text{S} O= [H_\text{S}, O]$. The operator $\mathcal{F}$, which is used to model specific aspects of the problem under investigation, is motivated and defined below (cf.\ Eq.\ (\ref{eq:7:Lindblad_explicit})).
		The auxiliary density operators with pole index $p=0$, which are related to the wide-band description of the leads, are not obtained by the EOM (\ref{eq:EQM_nth_tier}), instead they are calculated as
		\begin{eqnarray}
			\rho_{a_1 \dots a_n (l_i, 0, \sigma)}^{(n+1)} 	&=&	- \frac{i\pi\hbar}{2} 
										\cdot \left\lbrace g_{l_i}(x) \Gamma_{l_i} d^{\sigma}, \rho_{a_1 \dots a_n}^{(n)} \right\rbrace_{(-1)^{n+1}} . \nonumber\\
		\end{eqnarray}

		Within the HQME framework, the zeroth-tier auxiliary density operator $\rho^{(0)}(t)$ represents the reduced density matrix of the system. The influence of the leads on the system dynamics is encoded in the higher-tier auxiliary density operators $\rho_{j_1 \dots j_n}^{(n)}(t)$. Their respective EOMs are coupled via the operators $A^{{\sigma}}$ and $\mathcal{C}_{j}$, which couple the $n^{\text{th}}$-tier to the $(n+1)^{\text{th}}$- and $(n-1)^{\text{th}}$-tier auxiliary density operators, and which act as
		\begin{subequations}
			\begin{eqnarray}
				A^{\sigma} \rho^{(n)}	&=&	\left\lbrace g_{\text{L/R}}({x}) d^\sigma,\ \rho^{(n)}\right\rbrace_{(-1)^n} , \\
				\mathcal{C}_{j}	\rho^{(n)}	&=& - \frac{2i\pi\eta_p}{\beta}  \left\lbrace g_{\text{L/R}}(x) \Gamma_{\text{L/R}} d^{\sigma}, \rho^{(n)} \right\rbrace_{(-1)^{n+1}} , \nonumber \\
			\end{eqnarray}
		\end{subequations}
		with the shorthand notation $d^- \equiv d$ and $d^+ \equiv d^\dagger$. 
		
		The numerically exact description provided by the HQME approach relies on an infinite hierarchy of auxiliary density operators and an infinite number of poles used to represent the leads. For applications, both need to be truncated in a suitable manner. For a detailed discussion and the associated implications for applications, we refer to Refs.\ \onlinecite{Tanimura1991, Xu2005, Shi2009, Hu2010, Schinabeck2016}.

		Within the HQME formalism as presented above, all (auxiliary) density operators are also acting on the nuclear DOFs.
		In order to allow for a description of the nuclear DOFs in terms of generic PESs, we employ a DVR.\cite{Tannor, Colbert1992, Erpenbeck_2019_HQME}
		Within the DVR methodology, the nuclear wavefunction is effectively represented on a finite set of grid-points $x_i$. This methodology leads to unphysical finite size effects. To correct for these effects and in line with the method introduced in Ref.\ \onlinecite{Erpenbeck_2019_HQME}, a CAP is applied, which absorbs the parts of the wavefunction reaching the boundary of the DVR-grid.
		This, however, results in problems associated with the conservation of the particle number.\cite{Selsto2010, Kvaal2011, Prucker2018} 
		{\color{black} In particular, the action of the CAP results in a decrease of the trace of the density matrix and consequently an artificial loss of the number of electrons. A system exposed to the action of the CAP alone will tend to an improper state in which all (auxiliary) density operators are zero, whereby the predictability of the method would be lost.}
		{\color{black} To avoid these problems, we compensate for the loss of electrons due to the action of the CAP by introducing} an additional Lindblad-like source term. This source term maps the probability absorbed by the CAP to an auxiliary grid point ${x}_\infty$. Assuming that the PES and coupling strengths are constant for large values of the nuclear coordinate $x$, the additional point $x_\infty$ is chosen such that it is representative for large values of $x$,
		{\color{black} i.e.\ the value of the PES and the molecule-lead coupling strengths evaluated at $x_\infty$ are chosen in agreement with their constant values characteristic for large $x$-values.}
		This strategy is visualized in Fig.\ \ref{fig:action_Lindblad} for the parameters applied in this work.
		The CAP and the Lindblad-like source term are incorporated in the operator
		\begin{eqnarray} 
			\mathcal{F}(\rho_{j_1 \dots j_n}^{(n)}(t))	&=&	    i \lbrace W({x}), \rho_{j_1 \dots j_n}^{(n)}(t) \rbrace \label{eq:7:Lindblad_explicit} \label{eq:Lindblad_Op}\\&& 
									\hspace*{-1cm}
										  - 2 i \left( \sum_{{x}_i\in \text{grid}} 
										  \hspace{-0.2cm} W({x}_i) \braket{{x}_i| \rho_{j_1 \dots j_n}^{(n)}(t) | {x}_i} \right) \ket{{x}_\infty}\bra{{x}_\infty} , \nonumber
		\end{eqnarray}
		which enters the EOMs (\ref{eq:EQM_nth_tier}).
		Here, $W({x})$ is the CAP, the associated source-term is given by the second summand in Eq.\ (\ref{eq:Lindblad_Op}). 
		Notice, that a similar approach was, for example, employed in Ref.\ \onlinecite{Prucker2018}.

	\subsection{Observables of interest} \label{sec:observables}
	
		We employ several observables for studying current-induced bond rupture in molecular junctions.
		First of all, we investigate the current as a function of time. Within the HQME framework, the electric current between the molecule and lead $K$ is given by
		\begin{eqnarray}
			I_{K}(t)	&=&	\frac{ie}{\hbar^2} \sum_{p \in {\text{poles}}} \text{Tr}\left( g_{K}(x) \left( d \rho_{Kp+}^{(1)}(t) - d^\dagger \rho_{Kp-}^{(1)}(t) \right) \right) ,\nonumber \\ \label{eq:current} 
		\end{eqnarray}
		where Tr denotes the trace over the electronic and nuclear DOFs of the system.
		
		Another essential observable for identifying the mechanism leading to current-induced dissociation is the population of vibrational states of the bonding potential $V_0(x)$ of the neutral molecule. 
		However, in situations where the overall electronic population changes significantly with time, a direct interpretation of the population of vibrational states might be misleading. Hence, we consider the relative population of the vibrational eigenstate $\nu$ of $V_0(x)$, which is defined as
		\begin{eqnarray}
                        \eta_\nu(t) &=& \frac{
                                            \text{Tr} \Big( \rho(t)\ d d^{\dagger} \ket{\nu}\bra{\nu} \Big)
                                            }{
                                            \text{Tr} \Big( \rho(t)\ d d^{\dagger} \Big)
                                            }. \label{eq:rel_pop}
		\end{eqnarray}
		This relative vibrational population allows to probe the vibrational excitation of the bonding potential independent of the probability of being in the neutral electronic state.
		
		Finally, we consider the dissociation probability as a function of time. By design and as specified by the operator $\mathcal{F}$ in Eq.\ (\ref{eq:7:Lindblad_explicit}), the Lindblad-like source term maps the part of the wavefunction reaching the boundary of the DVR-grid to the additional grid-point $x_\infty$. 
		Therefore, the dissociation probability, which corresponds to the part of the wave packet populating the additional grid-point $x_\infty$,  is given by
		\begin{eqnarray}
			P(t)		&=&	\text{Tr} \Big( \rho(t) \ket{x_\infty}\bra{x_\infty} \Big) . \label{eq:diss_probab}
		\end{eqnarray}

\section{Results} \label{sec:results}

	In this section, we employ the methodology outlined in Sec.\ \ref{sec:theory} to study  current-induced bond rupture in molecular junctions. 
	Thereby, in Sec.\ \ref{sec:model}, we introduce the model system under investigation and provide some numerical details.
	Thereafter, we focus on specific mechanisms leading to dissociation in molecular junctions. In Sec.\ \ref{sec:DIET}, we investigate current-induced dissociation as a consequence of the transient population of anti-bonding electronic states. 
	In Sec.\ \ref{sec:DIMET}, we study current-induced excitation of the nuclear DOF and its implications for the stability of molecular junctions under transport.

	\subsection{Model system}\label{sec:model}
	
\begin{figure}[tb!]
			\includegraphics[width = 0.33\textwidth]{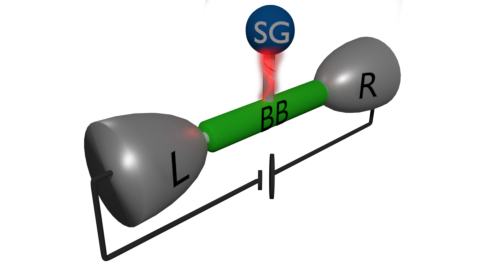}
			\captionsetup{font=small,labelfont=bf, justification=centerlast, format=plain}
			\caption{ \bf \scriptsize
				Sketch of the model under investigation. The molecular junction consists of a backbone (BB) and a side-group (SG). In case of dissociation, the side-group detaches from the backbone (highlighted in red).
				}
			\label{fig:sketch}
		\end{figure}

		The transport formalism introduced in Sec.\ \ref{sec:theory} is applicable to generic models for molecular junctions and a variety of dissociation scenarios. In the present publication and in line with our previous work in Ref.\ \onlinecite{Erpenbeck_dissociation_2018}, we focus on non-destructive current-induced bond rupture in single-molecule junctions. More specifically, we study the system sketched in Fig.\ \ref{fig:sketch}, where the molecular bridge consists of a backbone (BB) and a side-group (SG). We model the system in such a way that the current through the molecule influences the bond between the side-group and the backbone. In case of dissociation, the side-group will detach from the backbone. As such, we term this mechanism non-destructive, as the leads remain bridged by the backbone. Notice, that a similar setup was for example studied in Refs.\ \onlinecite{Brisker2008, Erpenbeck_dissociation_2018}.

		\subsubsection{Potential energy surfaces}\label{sec:potentials}

			For the model investigated in this work, we assume that the PESs for the neutral and the charged state are bonding and anti-bonding, respectively.
			The details of the potentials are in close agreement with the ones applied in Ref.\ \onlinecite{Erpenbeck_dissociation_2018}. The shape of the potentials and the corresponding parameters are inspired by data for dissociative electron attachment to CH$_3$Cl,\cite{Fabrikant1991, Fabrikant1994} H$_2$,\cite{Gertitschke1993} and CF$_3$Cl \cite{Hahndorf1994, Wilde1999}.
			However, this work applies generic model systems for studying the basic mechanisms of current-induced bond rupture and does not attempt to describe a specific molecule.

\begin{figure}[tb!]
				\vspace*{-0.3cm}
				\includegraphics[width = \linewidth]{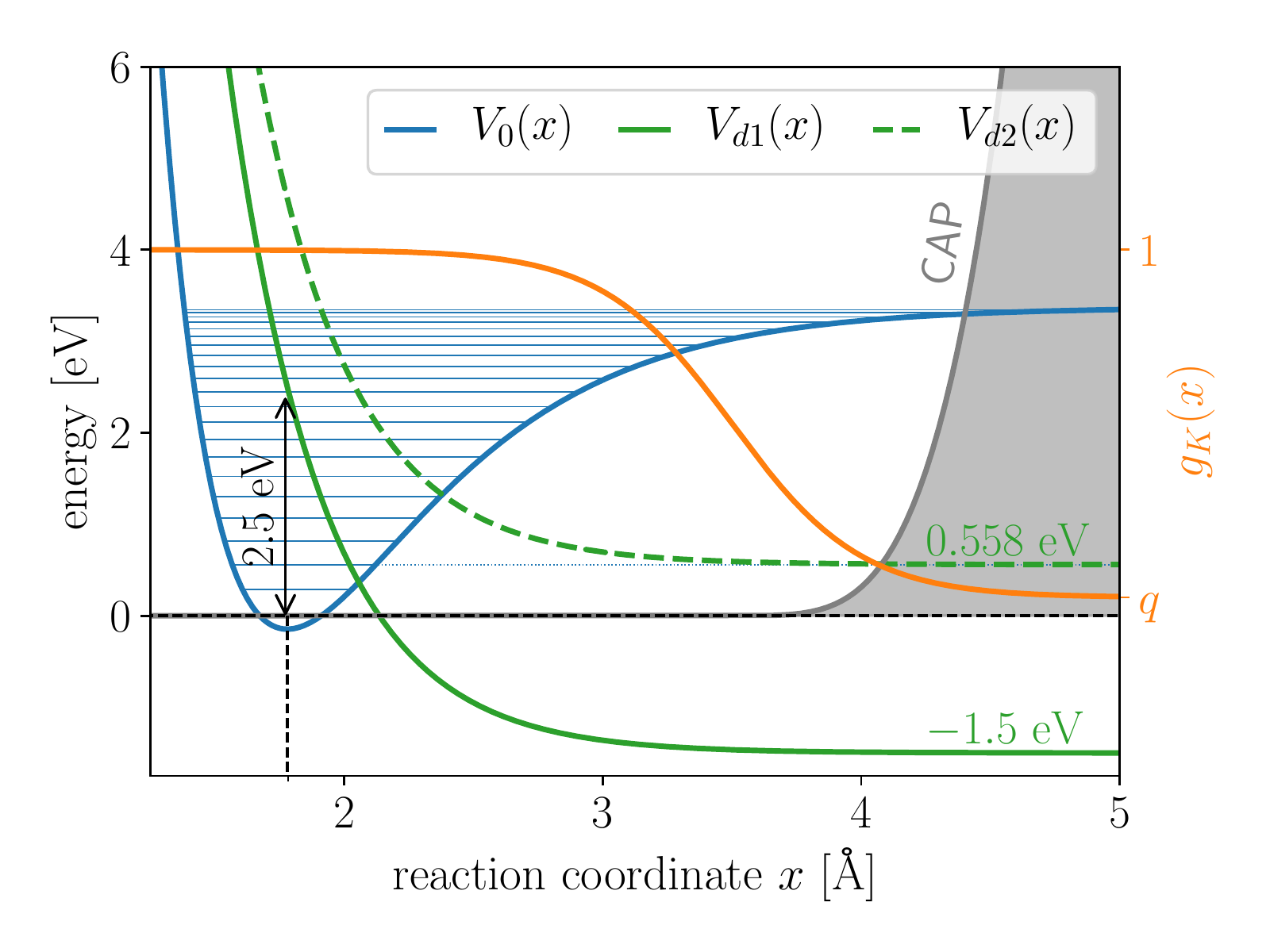}
				\captionsetup{font=small,labelfont=bf, justification=centerlast, format=plain}
				\caption{ \bf \scriptsize
					PESs used to model the bond between the backbone and the side-group of the molecular bridge in the neutral ($V_0(x)$) and the charged state ($V_d(x)$). Here, $V_{d1}(x)$ corresponds to the dissociative potential with $V_\infty=-1.5$ eV, whereas $V_{d2}(x)$ represents the dissociative potential with $V_\infty=0.558$ eV. The orange line visualizes the dependence of the molecule-lead coupling $g_{K}(x)$ on the nuclear coordinate.
					}
				\label{fig:pots_overview}
			\end{figure}

			For the neutral molecule, the bond between the backbone and the side-group is described by the binding Morse-potential, depicted in Fig.\ \ref{fig:pots_overview}, 
			\begin{eqnarray}
				V_0(x) &=& 	D \cdot\left( e^{-a(x-x_0)} -1 \right)^2 + c . \label{eq:def_V_0}
			\end{eqnarray}
			Thereby, $x_0=1.78\ \AA$ is the equilibrium bond distance, $D = 3.52$ eV the dissociation energy, $a=1.7361/\AA$ the width of the Morse potential.
			The binding potential $V_0(x)$ allows for 22 bound states.
			The corresponding energy scale for nuclear motion in the neutral state is $\hbar\omega = 297$ meV. The constant energy offset \mbox{$c= -147$ meV} is chosen such that the energy of the quantum mechanical ground state of $V_0(x)$ is zero. This implies that the Fermi energy in the leads is aligned with the vibrational ground state of the neutral molecule for $\mu_{\text{L/R}}=0$ eV. 

			The PES of the charged state is assumed to have the repulsive form
			\begin{eqnarray}
				V_d(x) &=&	D' \cdot e^{- a' (x-x_0')} + V_\infty . \label{eq:def_V_d}
			\end{eqnarray}
			Here, $D'=4.0$ eV sets the energy scale for the potential, $a'=2.758/\AA$ and $x_0'=1.78\ \AA$ determine the repulsion strength. The parameter $V_\infty$ is used to realize different dissociation mechanisms (cf.\ Secs.\ \ref{sec:DIET} and \ref{sec:DIMET}) and set to the values $-1.5$ eV and $0.558$ eV. For both these values, the corresponding potentials are depicted in Fig.\ \ref{fig:pots_overview}. For $V_\infty=-1.5$ eV, the energy of the anti-bonding potential at large distances lies well below the ground state energy of $V_0(x)$. For $V_\infty=0.558$ eV, $V_d(x\rightarrow\infty)$ is aligned with the second excited state of $V_0$.

			An important role is played by the molecule-lead coupling strength. The fact that we allow for a nuclear coordinate dependent coupling enables the description of situations where the conductance changes upon the detachment of the side-group. This can, for example, result from the destruction of a $\pi$-conjugation within the molecular backbone as a consequence of the side-group separating from the molecule. Moreover, a nuclear coordinate dependent molecule-lead coupling allows for a back-action of the nuclear motion on the electronic properties.
			For the scope of this work, we use a dependence on the nuclear coordinate of the form 
			\begin{eqnarray}
				g_{K}(x)    &=&    \left( \frac{1-q}{2} \left[ 1-\tanh\left(\frac{x-\tilde x}{\tilde a} \right) \right] + q \right) , \ \label{eq:def_mol_lead_coupling_strength}
			\end{eqnarray}
			which is also depicted in Fig.\ \ref{fig:pots_overview}.
			Here, the parameter
			$q=0.05$ determines the coupling strength in case of a dissociated molecule as $g_{K}(x\rightarrow\infty) = q$.
			The distance around which the drop in the molecule-lead coupling occurs is given by $\tilde x=3.5\ \AA$, while $\tilde a=0.5\ \AA$ sets the scale for the region of change.
			
			In the calculations reported below, the reduced mass is set to \mbox{$m=1$ amu} (atomic mass units).
			We assume that both leads are initially described by the density matrices $\rho_\text{L/R} = e^{-\beta(H_\text{L/R} - \mu_\text{L/R} N_\text{L/R})}/Z$ and have the same temperature $T=300$ K. The bias voltage $\Phi$, defined as the difference between the chemical potentials $\mu_{\text{L}}$ and $\mu_{\text{R}}$, drops symmetrically such that $\mu_{\text{L}} = -\mu_{\text{R}}=\Phi/2$. Moreover, the leads are modeled in the wide-band limit. 
			We note that under certain conditions, there can be issues related to employing the wide-band limit in cases where the coupling to the leads depends on the nuclear DOFs.\cite{Dou2016} 
			Test calculations show that these issues are not relevant for the data presented below.

		\subsubsection{Numerical details}\label{sec:numerics}	
            
\begin{figure}[tb]
				\centering
				\includegraphics[width=0.9\linewidth]{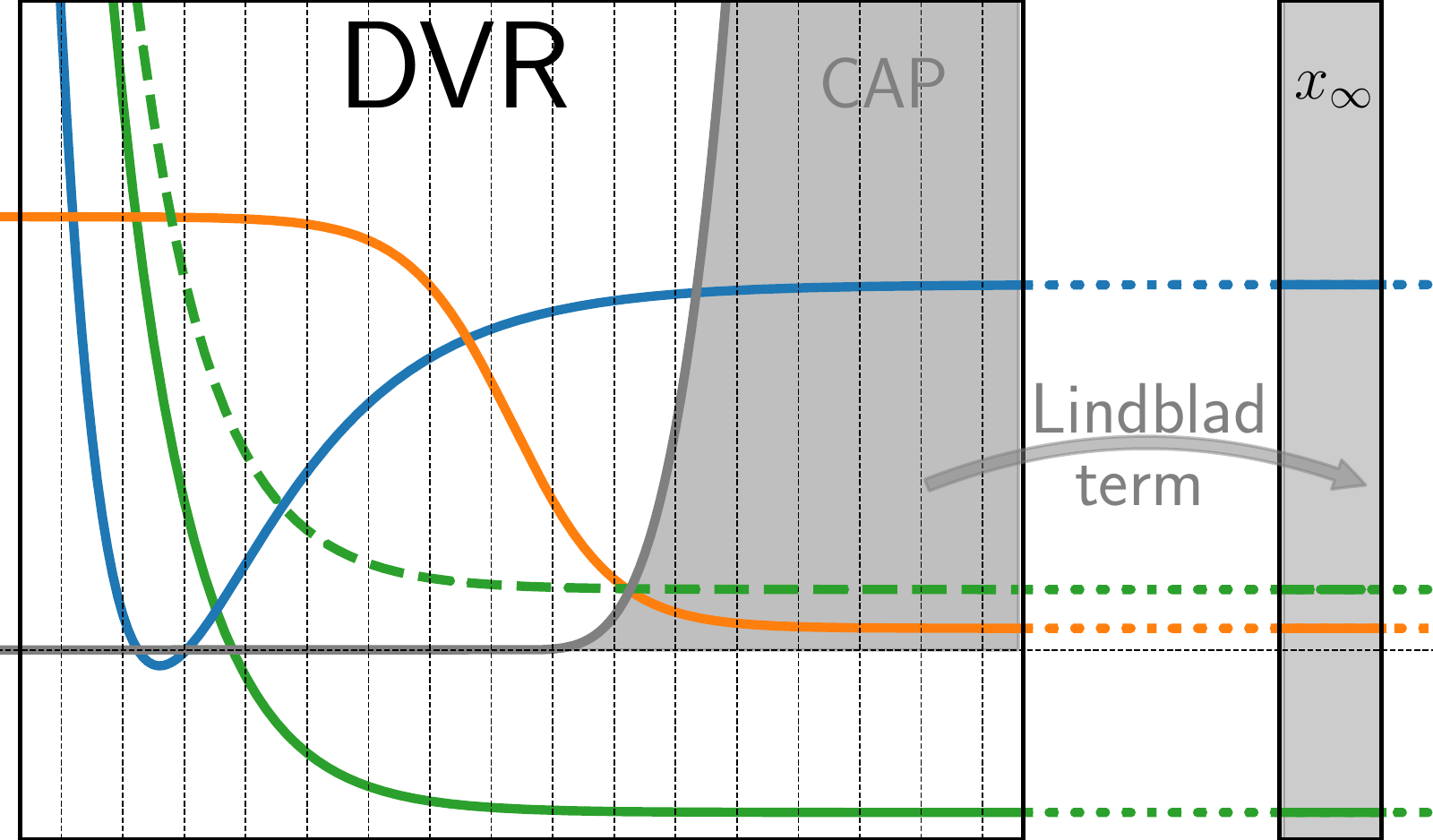}
				\captionsetup{font=small,labelfont=bf, justification=centerlast, format=plain}
				\caption{ \bf \scriptsize
					Visualization of the action of the operator $\mathcal{F}$ as defined in Eq.\ (\ref{eq:7:Lindblad_explicit}). The probability absorbed by the CAP (gray shaded area) is mapped to the representative auxiliary grid-point $x_\infty$. The vertical dotted lines outline the individual DVR-grid points. The color-coding of the potentials and coupling strengths coincides with the one used in Fig.\ \ref{fig:pots_overview}.
					}
				\label{fig:action_Lindblad}
			\end{figure}

			In order to describe the effect of dissociation in the model system detailed above, formally it is necessary to consider $x$-values from zero to infinity. To model this semi-infinite situation in terms of a finite DVR-grid, we apply a CAP for large $x$-values. The CAP prevents unphysical reflections at the boundaries of the DVR-grid. Moreover, it enters the expression for the dissociation probability (see Eqs.\ (\ref{eq:Lindblad_Op}) and (\ref{eq:diss_probab})). 
			For the scope of this work, we assume a power-law form for the CAP,
			\begin{eqnarray}
				W(x)		&=&		\alpha \left(x - x_{\text{CAP}} \right)^4  \cdot \Theta(x - x_{\text{CAP}}) ,
			\end{eqnarray}
			with the Heaviside function $\Theta$, $\alpha=5$~eV/\AA$^4$ and $x_{\text{CAP}} = 3.5$~\AA. This CAP is also depicted in Fig.\ \ref{fig:pots_overview}.
			With this functional form of the CAP, we can visualize the action of the CAP and the Lindblad-like source term in Fig.\ \ref{fig:action_Lindblad}.
			Thereby, the parameters of the power-law CAP are determined by testing the convergence of the observables of interest.\cite{Todorov2001, Kopf2004, Pshenichnyuk2013, Leitherer2017}

			The numerical data presented in the following were calculated by propagating Eqs.\ (\ref{eq:EQM_nth_tier}) using a fourth-order Runge-Kutta scheme. Thereby, it was assumed that the total density matrix factorizes at $t=0$. This implies that the contact between the molecule and the leads is established at $t=0$. 
			The leads are assumed to be described initially by the density matrices $\rho_\text{L/R} = e^{-\beta(H_\text{L/R} - \mu_\text{L/R} N_\text{L/R})}/Z$.
			The initial state of the molecule is $\ket{0}\ket{\nu=0}$, whereby $\ket{0}$ represents the unpopulated molecular electronic state and the nuclear DOF is in the ground state $\ket{\nu=0}$ of the bonding potential $V_0(x)$.
			This initial state corresponds to a stable molecule prior to the connection to the leads. A detailed discussion of the influence of these initial conditions can be found in Ref.\ \onlinecite{Erpenbeck_dissociation_2018}.
			
			For all data presented in the following, we have tested the convergence of the observables with respect to the temporal resolution used for the propagation of the equations of motion, the DVR grid, the number of poles used to represent the Fermi function in the leads, and the number of tiers considered.
			{\color{black} Specifically, we employ DVR grids consisting of varying size, depending on details of the system. The smallest DVR grids used includes $64$ grid-points, the largest $240$.}
			Consequently, we provide numerically exact results throughout this publication.

	\subsection{Dissociation upon population of anti-bonding states}\label{sec:DIET}

		In the following, we study the mechanism of current-induced dissociation as a result of the transient population of the anti-bonding state with PES $V_d(x)$ by electrons from the leads.
		This mechanism is also of fundamental interest for studying current-induced desorption of atoms and molecules from surfaces, where the mechanism is called desorption induced by electronic transitions (DIET).\cite{Dujardin1992, Martel1996, Stipe1997, Lauhon2000, Lauhon2000_2, Saalfrank2006} 
		Further, as this mechanism can also be described by a mixed quantum-classical approach under certain conditions, we thereby establish the connection and validate the results presented in our previous work in Ref.\ \onlinecite{Erpenbeck_dissociation_2018}. 
		Throughout this section, we investigate the model system with $V_\infty=-1.5$ eV (see Fig.\ \ref{fig:pots_overview}). The timescale for the electron dynamics is set by $1/\Gamma$, whereby $\Gamma$ denotes the coupling to the leads.
		We study the dissociation probability, the electronic current and the nuclear dynamics as a function of time in order to provide a profound understanding of the underlying processes and, in particular, the role of current-induced heating.

		\subsubsection{Dissociation dynamics}
		
			\begin{figure*}
				\centering
				\begin{minipage}[c]{0.32\textwidth}
					\vspace*{-0.4cm}
					\raggedright a)\\
					\centering
					\includegraphics[width=\textwidth]{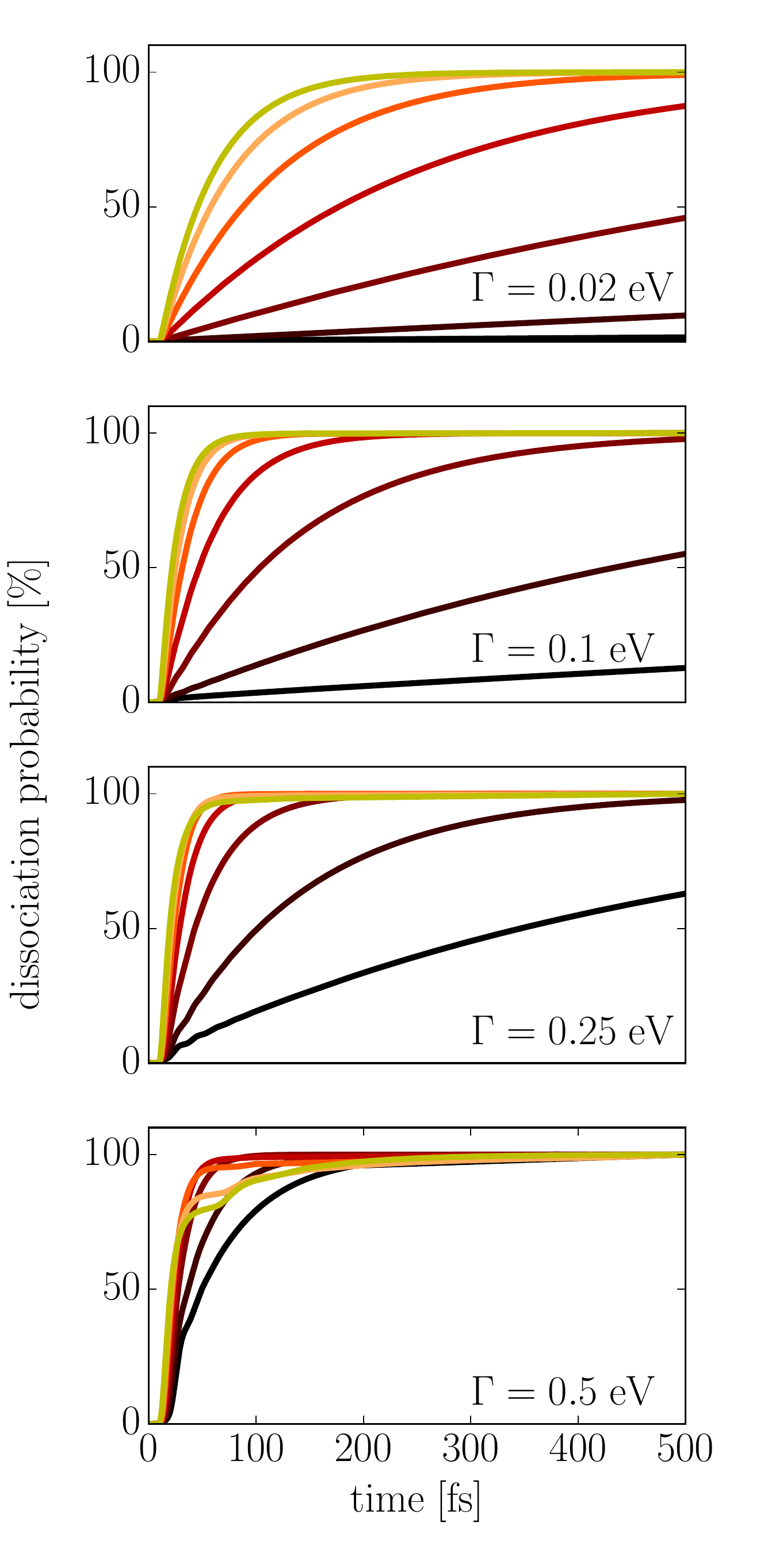}
				\end{minipage}
				\begin{minipage}[c]{0.32\textwidth}
					\vspace*{-0.4cm}
					\raggedright b)\\
					\centering
					\includegraphics[width=\textwidth]{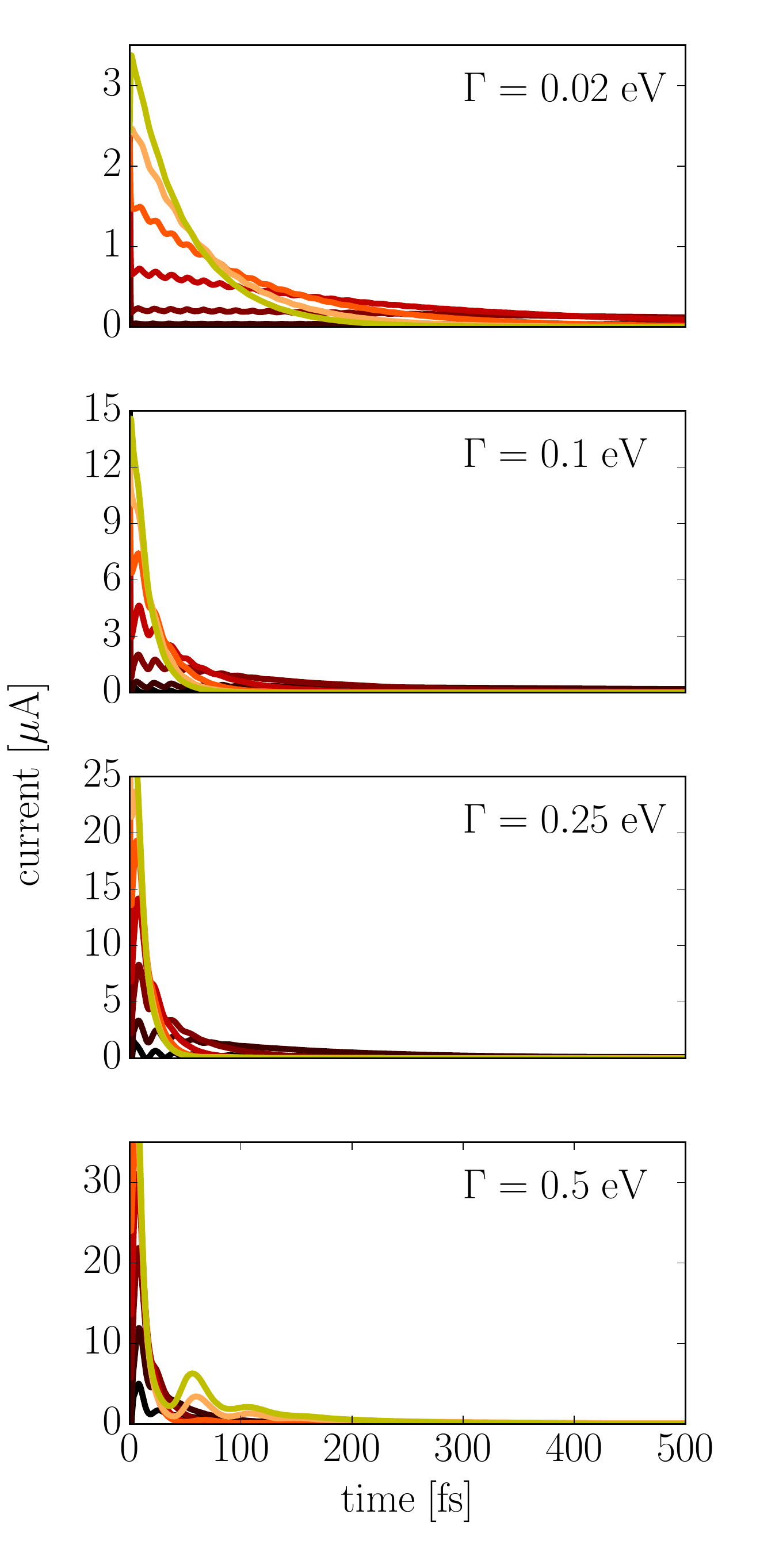}
				\end{minipage}
				\begin{minipage}[c]{0.28\textwidth}
					\raggedright c)\\
						\includegraphics[width=\textwidth]{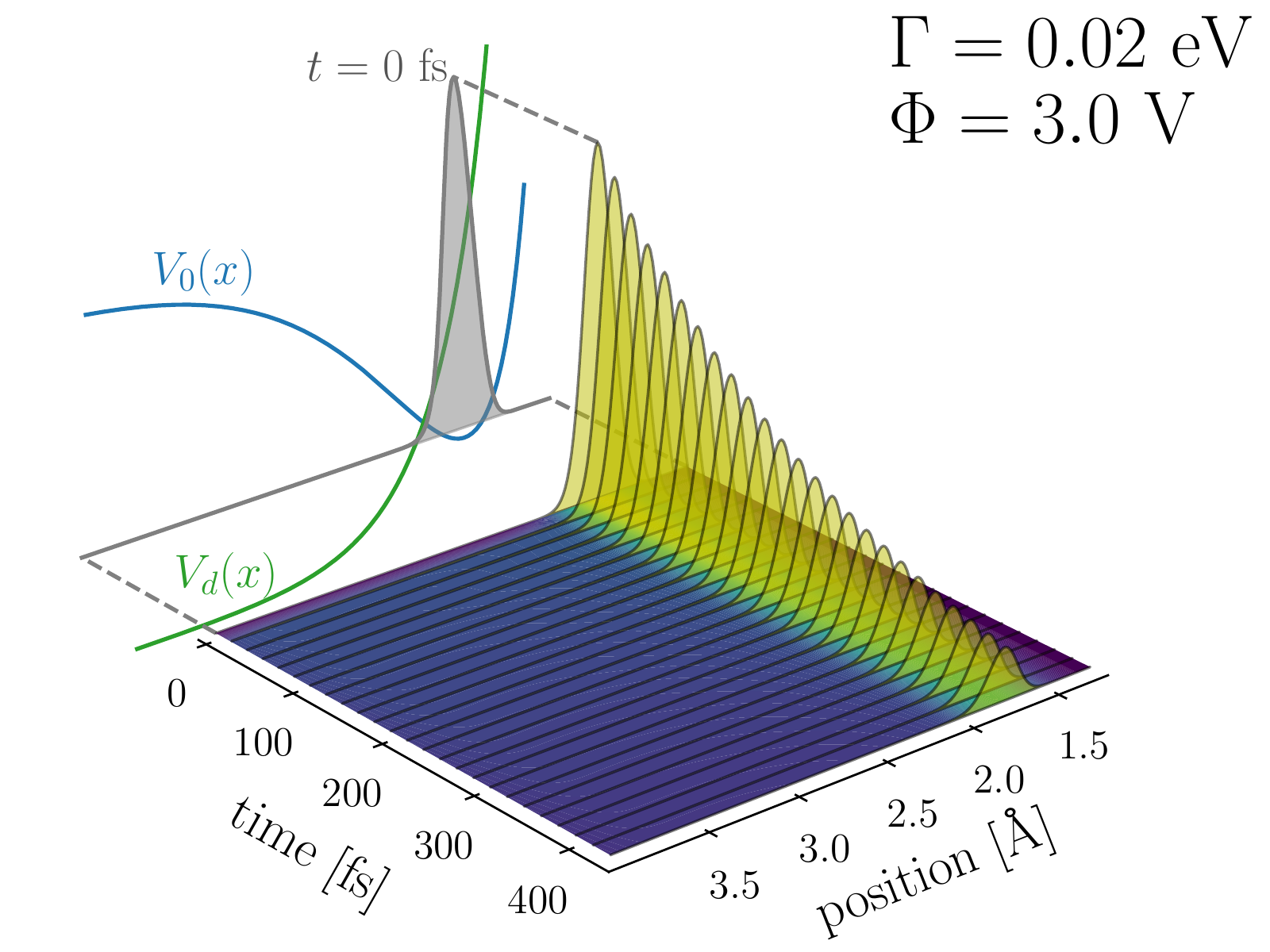}\\
					\raggedright d)\\
						\includegraphics[width=\textwidth]{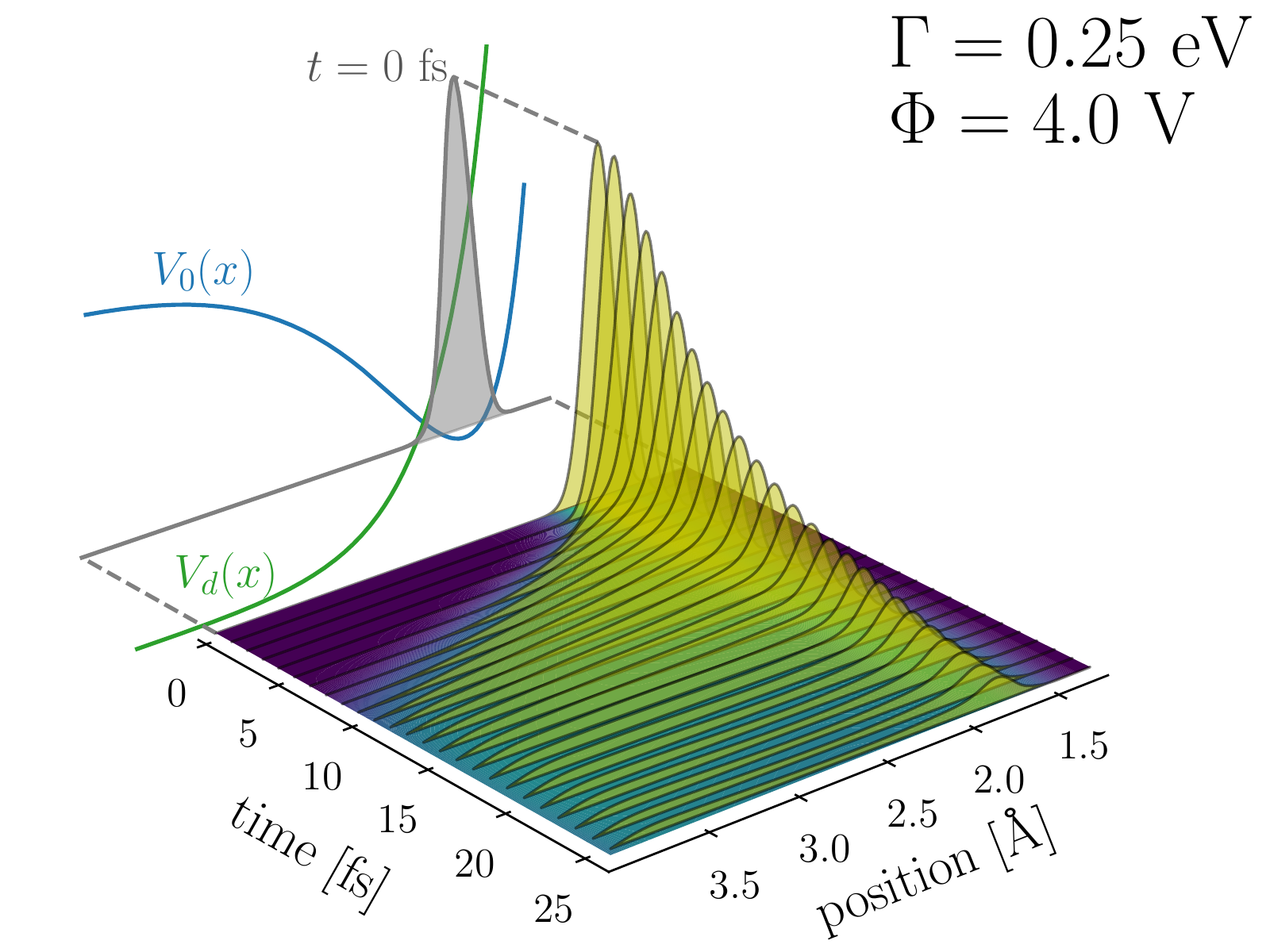}\\
					\raggedright e)\\
						\includegraphics[width=\textwidth]{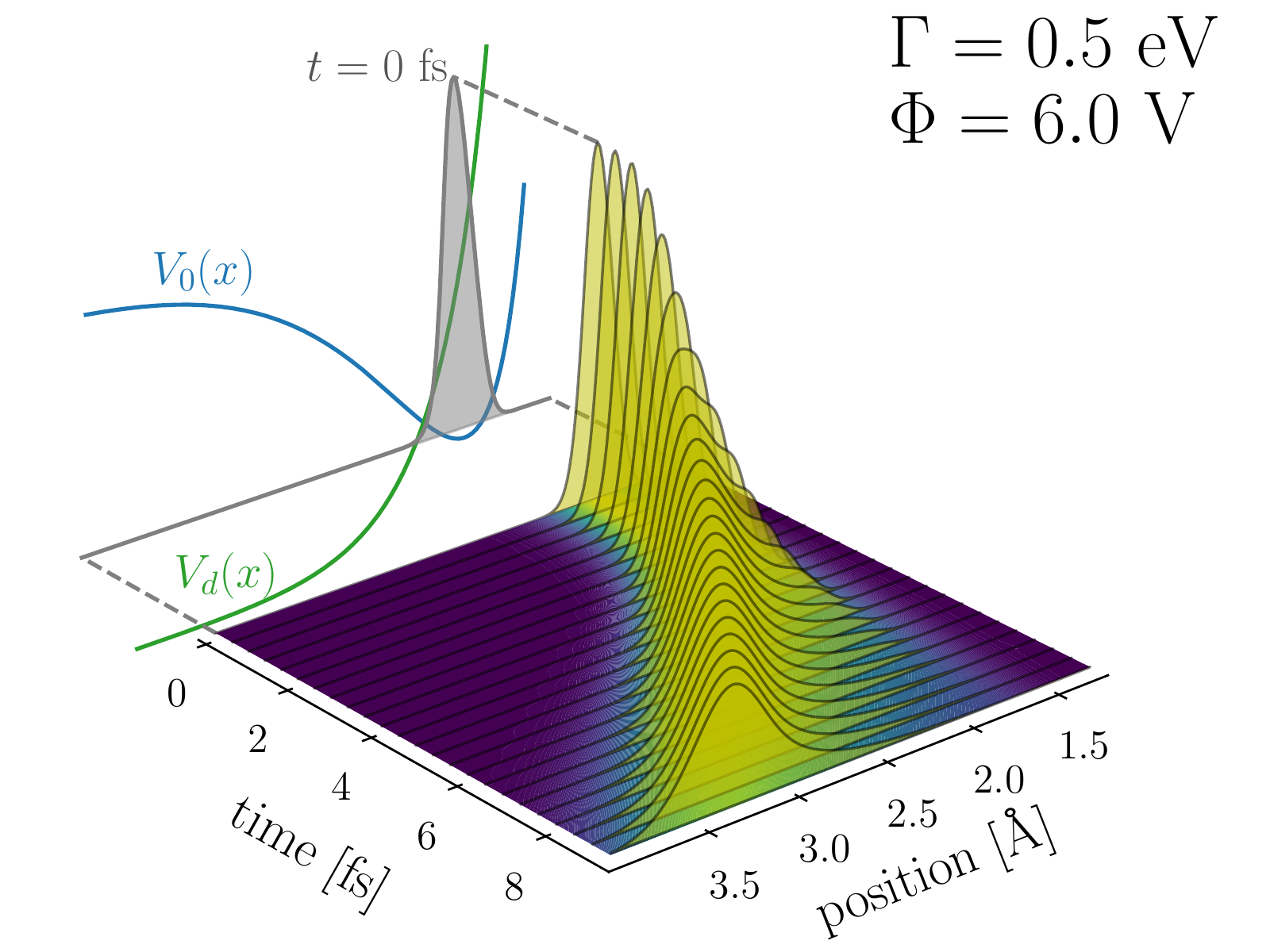}
				\end{minipage}\\
				\begin{minipage}[c]{0.62\textwidth}
					\vspace*{-0.4cm}
					\includegraphics[width=0.9\textwidth]{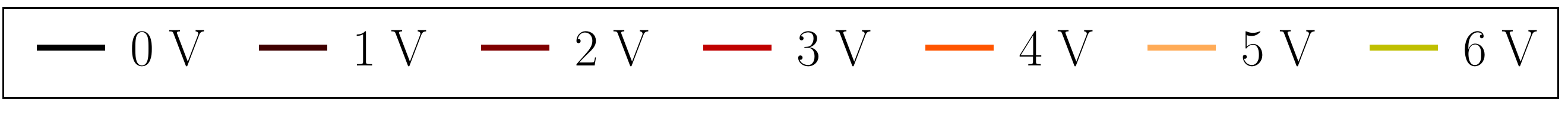}
				\end{minipage}
				\begin{minipage}[c]{0.25\textwidth}
				{\color{white}.}
				\end{minipage}
				\captionsetup{font=small,labelfont=bf, justification=centerlast, format=plain}
				\caption{ \bf \scriptsize
					Dynamics of the model system symmetrically coupled to two leads for different applied bias voltages and molecule-lead coupling strengths $\Gamma$. In all the plots, the different lines correspond to different applied bias voltages. 
					a: Dissociation probability as a function of time. 
					b: Current between the molecule and the left lead as a function of time. 
					c--e: Nuclear probability distribution as a function of time for different represent coupling strengths $\Gamma$ and bias voltages $\Phi$. The yellow shaded areas as well as the color gradient at the bottom represent the nuclear probability distribution for different times. In the background, the initial state and the potentials are depicted for reference. Notice the different timescales depicted in the different subplots c--e.
					}
				\label{fig:SYMM_data}
			\end{figure*}
			
			We start the analysis by considering a symmetric transport scenario with $\Gamma_\text{L}=\Gamma_\text{R}$. 
			Fig.\ \ref{fig:SYMM_data}a shows the dissociation probability for this system as a function of time for different molecule-lead coupling strengths $\Gamma$ and bias voltages $\Phi$.
			The dissociation probability always increases with time which is inherent to the model applied in this work. 
			This is in contrast to results obtained within a mixed quantum-classical approach, where the dissociation probability saturates at values smaller than $100\%$.\cite{Erpenbeck_dissociation_2018} Furthermore, the behavior of the dissociation as a function of time varies significantly with the molecule-lead coupling strength $\Gamma$ and the applied bias voltage. For small bias and small $\Gamma$, we observe that dissociation happens on the order of picoseconds. For large bias and large $\Gamma$, however, a dissociation probability of $100\%$ is reached within tenths of femtoseconds.
			{\color{black} A direct observation of processes on these short timescales in molecular
			junctions is challenging with current experimental techniques.
			Nevertheless, elucidating the mechanisms underlying these ultrafast
			processes is crucial to understand the stability of molecular junctions.
			Data for molecular junctions that remain stable for longer times and are indicative for experimentally resolvable timescales are discussed in Sec.\ \ref{sec:DIMET}.}

\begin{figure}
				\centering
				\vspace*{-.3cm}
				\includegraphics[width=\linewidth]{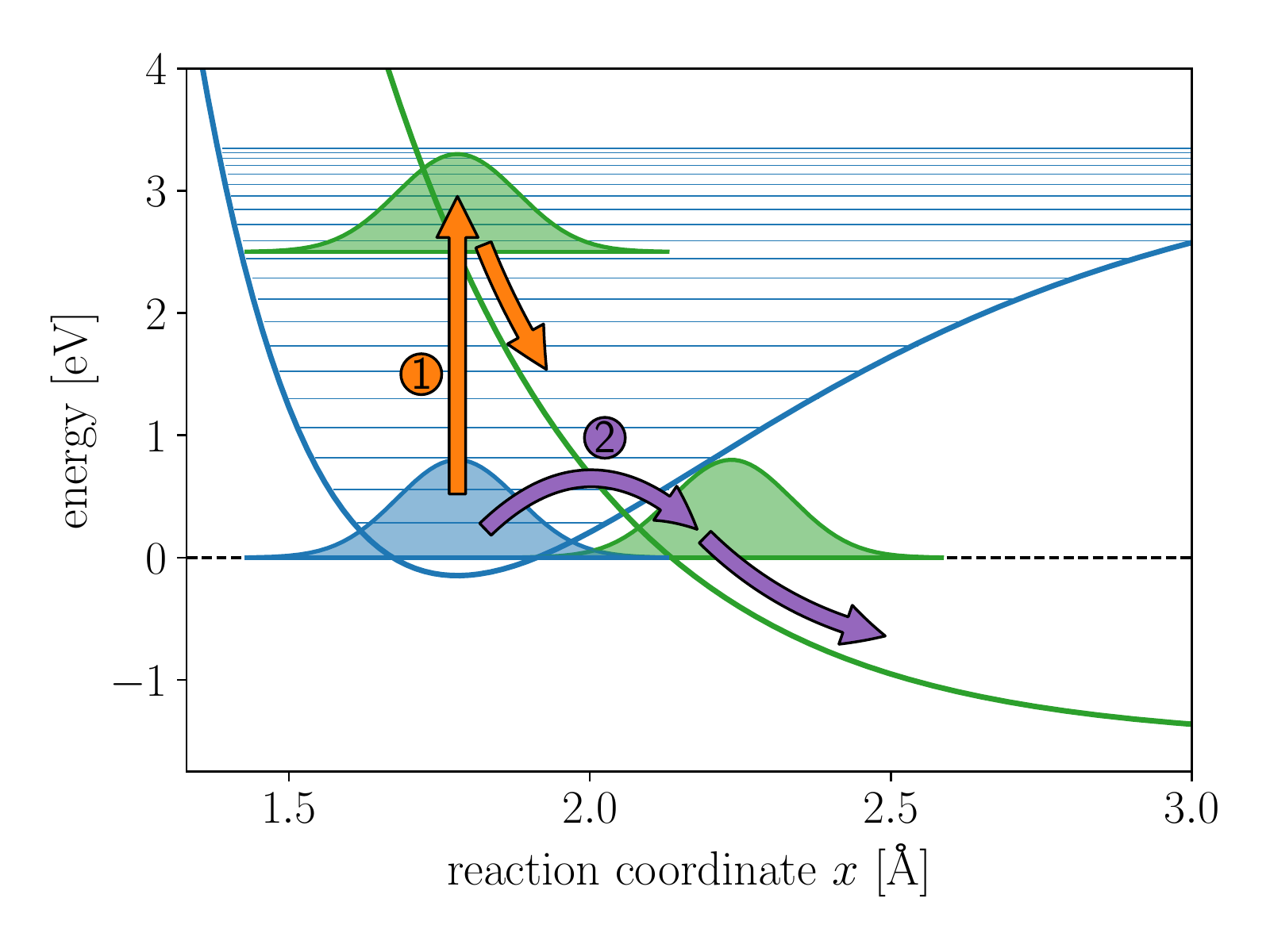}
				\captionsetup{font=small,labelfont=bf, justification=centerlast, format=plain}
				\caption{ \bf \scriptsize
					Graphical representation of two different processes leading to dissociation, which provide an explanation for the diverse timescales for dissociation. 
					(The figure sketches processes and does not contain actual data.)
					}
				\label{fig:process1}
			\end{figure}

			In order to explain the diverse timescales for dissociation, we focus on the behavior of the nuclear DOF (see Figs.\ \ref{fig:SYMM_data}c--e). In case of slow dissociation on the order of picoseconds, which is predominantly found for small $\Phi$ and small $\Gamma$ and which is exemplified in Fig.\ \ref{fig:SYMM_data}c, the overall shape of the wave packet is approximately constant in time. However, its amplitude decreases with time. In this case, the exponentially suppressed tail of the nuclear DOF reaching out into the classical forbidden regime is responsible for dissociation. This is sketched as process $\circled{2}$ in Fig.\ \ref{fig:process1}. 
			Notice that this process suggests a nonadiabatic picture, {\color{black} where the timescale of the electronic dynamics related to the charging/decharging of the molecule, given by $1/\Gamma$, is long compared to the typical timescale of nuclear motion}, which is applicable for the present system for $\Gamma\ll0.3$ eV.
			Moreover, this process can, in principle, be described by a dissociation rate, however, it is beyond approaches that treat the nuclear DOF classically.
			
			In case of fast dissociation on the order of tenth of femtoseconds, which is predominantly found for large bias and large $\Gamma$, we find that the entire wave packet propagates in positive $x$-direction, as is exemplified in Fig.\ \ref{fig:SYMM_data}e. This behavior of the nuclear probability distribution suggests that the nuclear motion is governed by the anti-bonding potential $V_d(x)$, which means that the molecule gets populated by an electron and dissociates directly. This is sketched as process $\circled{1}$ in Fig.\ \ref{fig:process1}. 
			Within an adiabatic picture, {\color{black} where the electronic dynamics is fast compared to the nuclear motion, which is the case for $\Gamma\gg0.3$ eV}, this process can also be interpreted in terms of an average electronic background.
			The corresponding average PES, only displays a small potential barrier, which decreases with applied bias voltage (data not shown).
 			Notice that this process cannot be accounted for by a dissociation rate, however, it can be described by approaches employing a classical description for the nuclear DOF.\cite{Erpenbeck_dissociation_2018}
			
			In the intermediate regime between fast and slow dissociation, we find that the wave packet adopts a mixed behavior. 
			A part of the wave packet propagates in positive $x$-direction, whereas the remaining part approximately preserves its shape. 
			The corresponding dynamics of the nuclear probability distribution is exemplified in Fig.\ \ref{fig:SYMM_data}d.

			Apart from the different timescales for dissociation, we note that the dissociation probability shows a step-like structure in the large $\Gamma$-regime. This behavior originates from the dependence of the molecule-lead coupling strength on the nuclear DOF and relates to the wave packet being reflected at regions exhibiting pronounced changes in $\Gamma$. This effect will be investigated in detail in future work. The general implications of a variable molecule-lead coupling strength in a transport context were also addressed in the recent publication Ref.\ \onlinecite{Preston2020}.

\begin{figure}[t!]
				\raggedright a) \hspace*{3.85cm} b)\\
						\vspace*{-0.075cm}
						\hspace*{-0.25cm}
						\begin{minipage}[c]{0.24\textwidth}
							\centering
							\includegraphics[width=1\textwidth]{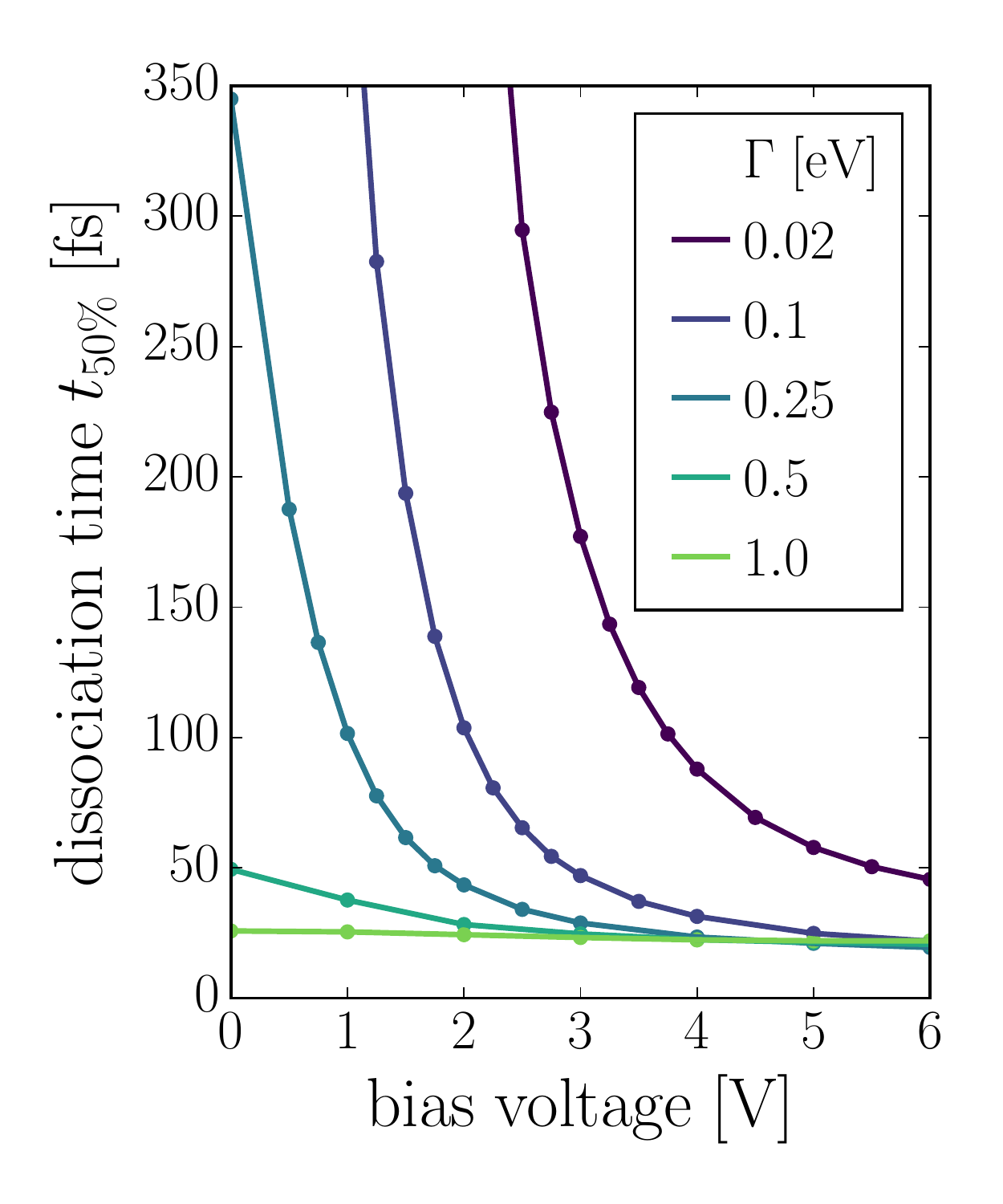}
						\end{minipage}
						\begin{minipage}[c]{0.24\textwidth}
							\centering
							\includegraphics[width=1\textwidth]{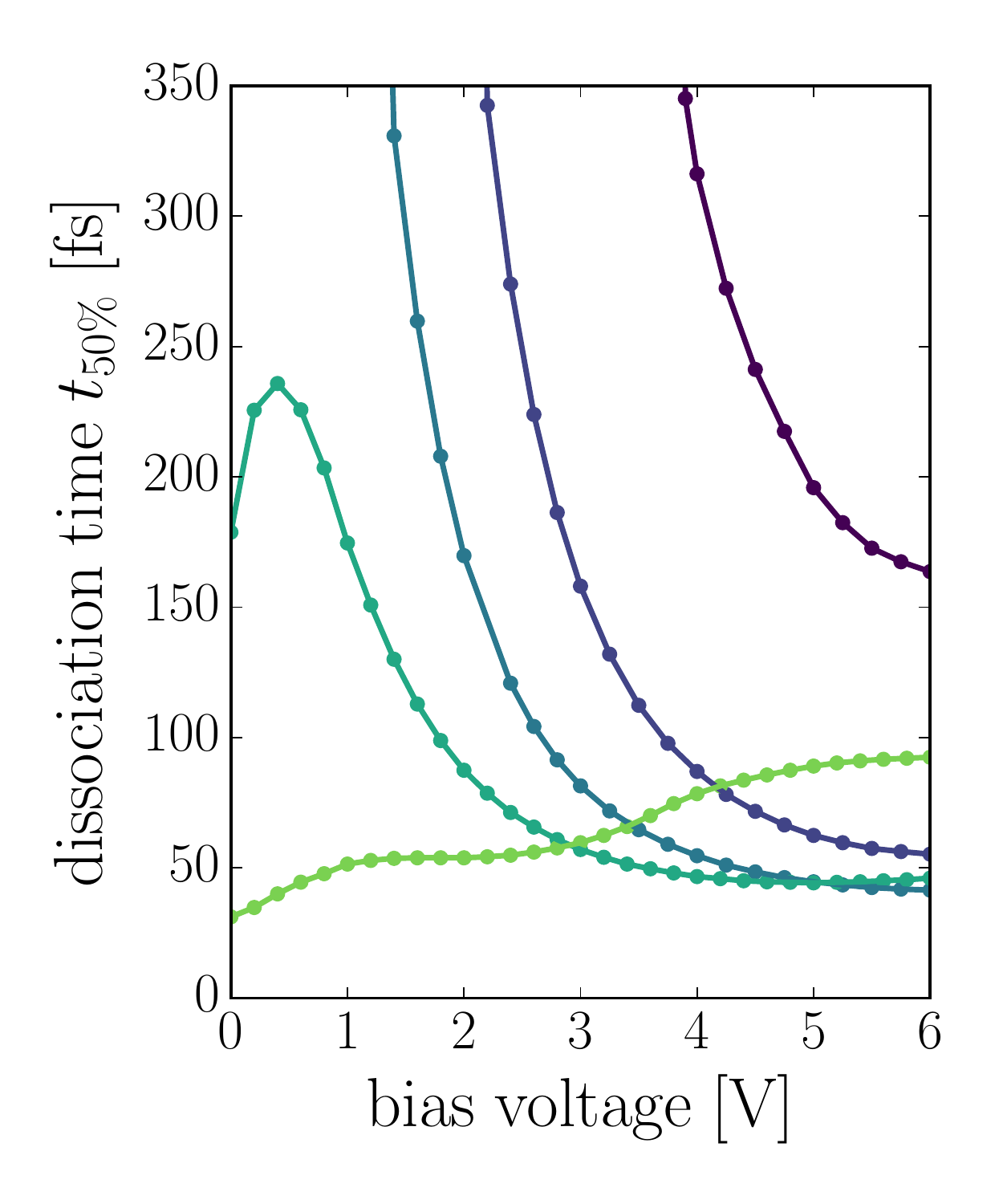}
						\end{minipage}
						\\
				\raggedright c) \hspace*{3.85cm} d)\\
						\vspace*{-0.075cm}
						\hspace*{-0.25cm}
						\begin{minipage}[c]{0.24\textwidth}
							\centering
							\includegraphics[width=1\textwidth]{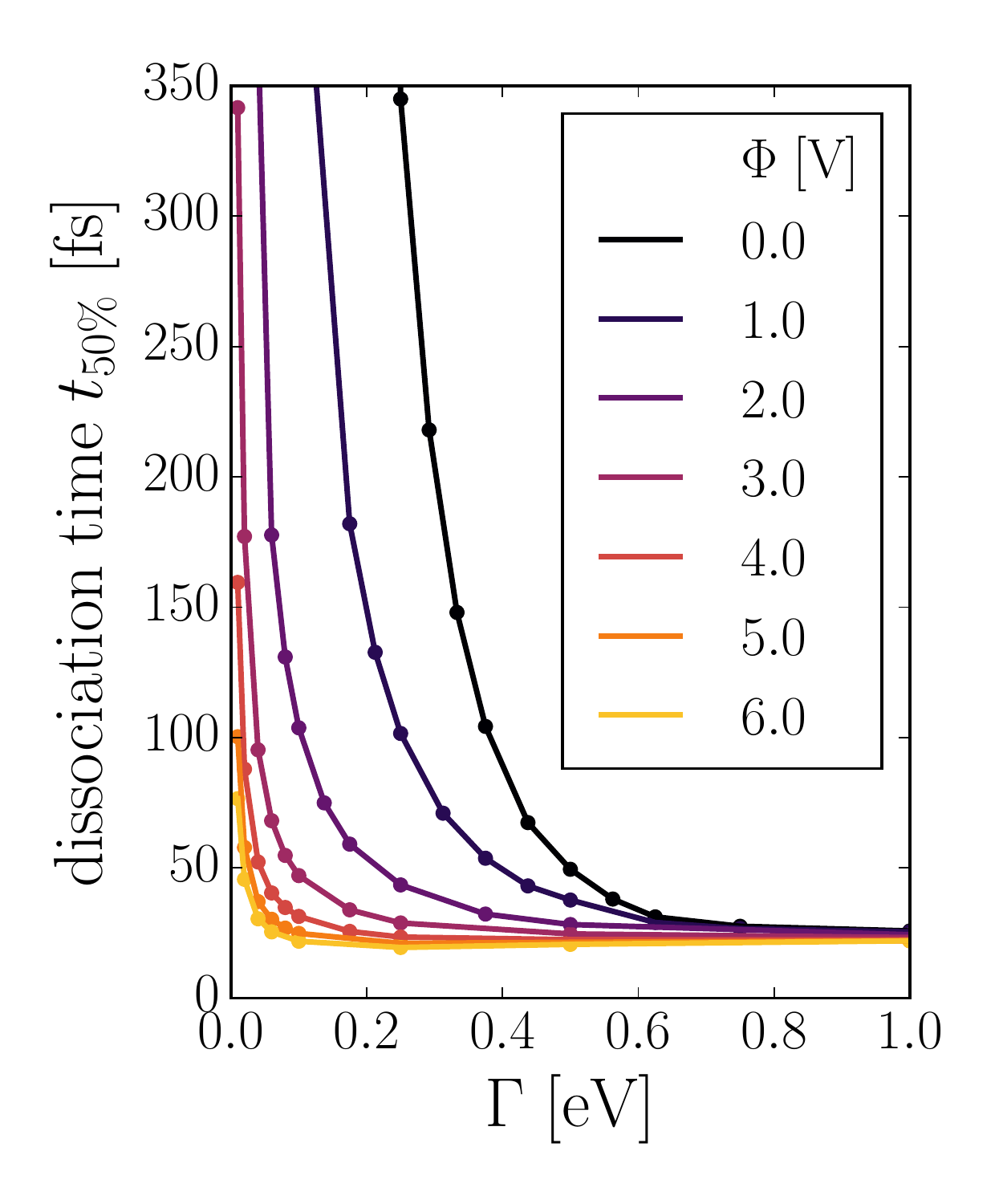}
						\end{minipage}
						\begin{minipage}[c]{0.24\textwidth}
							\centering
							\includegraphics[width=1\textwidth]{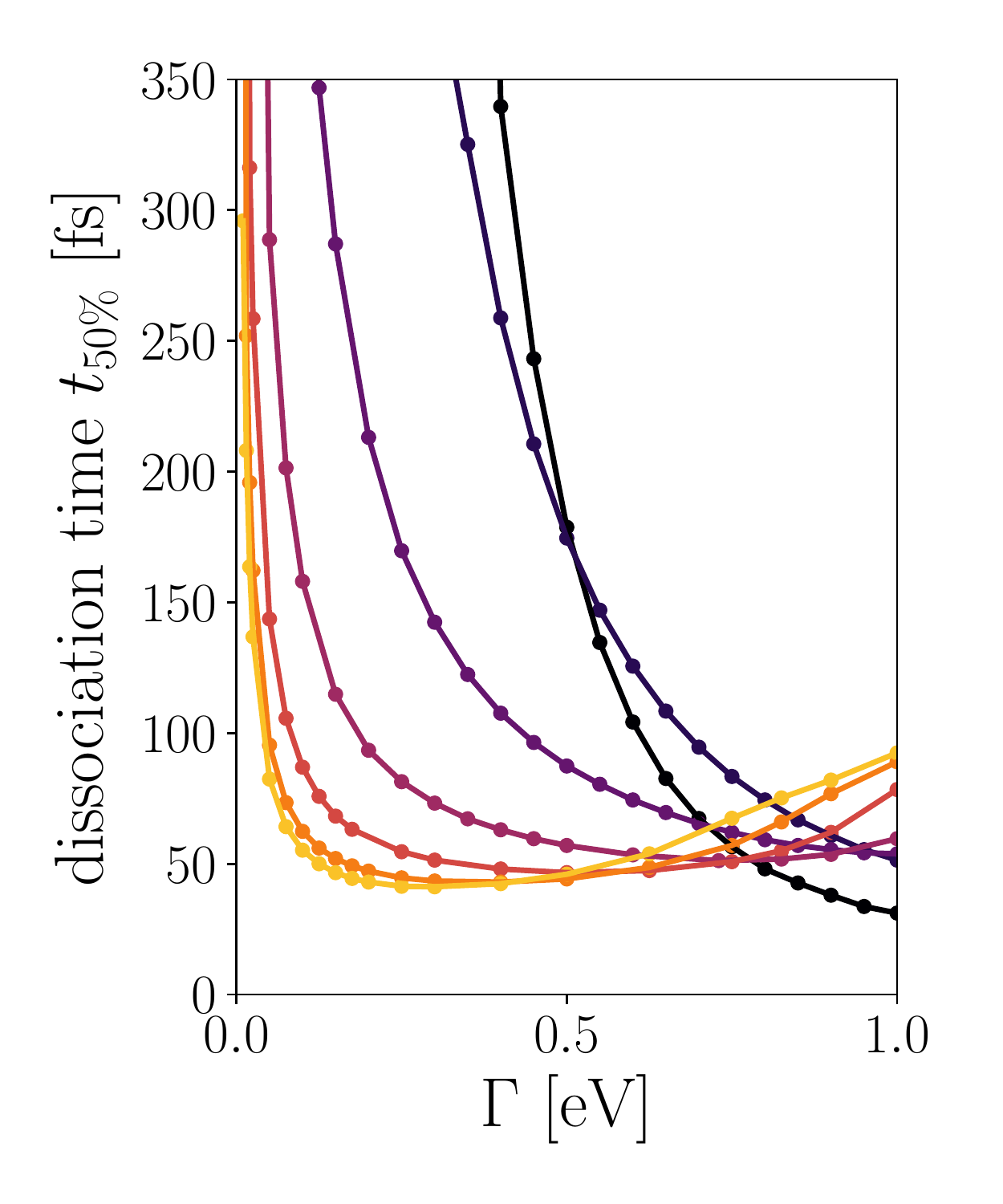}
						\end{minipage}
				\captionsetup{font=small,labelfont=bf, justification=centerlast, format=plain}
				\caption{ \bf \scriptsize
					Dissociation time $t_{50\%}$ as a function of bias voltage $\Phi$ and molecule-lead coupling $\Gamma$ for the symmetric model (left) and the asymmetric model (right). Points mark the actual data, the lines serve as a guide for the eye.
					a: Dissociation time $t_{50\%}$ of the symmetric model ($\Gamma_\text{L} = \Gamma_\text{R}$) as a function of bias voltage for different values of $\Gamma$. 
					b: Dissociation time $t_{50\%}$ of the asymmetric model ($\Gamma_\text{L} = 0.25\cdot\Gamma_\text{R}$) as a function of bias voltage for different values of $\Gamma$. 
					c: Dissociation time $t_{50\%}$ of the symmetric model as a function of $\Gamma$ for different bias voltages. 
					d: Dissociation time $t_{50\%}$ of the asymmetric model as a function of $\Gamma$ for different bias voltages. 
					}
				\label{fig:comparison_DIET_1}
			\end{figure}

			The physics of the dissociation process is encoded in the timescale at which dissociation occurs. In the following, we study the time $t_{50\%}$ at which a dissociation probability of 50\% is reached. This rather simple observable does not rely on any assumption for the dissociation process and is therefore a suitable quantity to distinguish different mechanisms.
			Figs.\ \ref{fig:comparison_DIET_1} displays the dissociation time $t_{50\%}$ as a function of bias $\Phi$ and molecule-lead coupling $\Gamma$. In addition to the symmetric coupling scenario discussed so far (Figs.\ \ref{fig:comparison_DIET_1}a and c), we also examine the asymmetric coupling case with $\Gamma_\text{L} = 0.25\cdot\Gamma_\text{R} = 0.25\Gamma$ (Figs.\ \ref{fig:comparison_DIET_1}b and d). We note that the overall dynamics of the symmetric and the asymmetric model is quite similar (data not shown).
			Considering $t_{50\%}$ as a function of applied bias voltage (Figs.\ \ref{fig:comparison_DIET_1}a and b), we find a sharp decrease from long dissociation times at lower bias voltages to short dissociation times at higher bias voltages for low to intermediate coupling strengths $\Gamma$. For the strong coupling regime, the dissociation time of the symmetrically coupled system depends only weakly on $\Phi$. Still, the dissociation time decreases with an increase in bias. In contrast to that, the dissociation time for the asymmetrically coupled system changes its behavior profoundly from the weak to the strong coupling regime. For large $\Gamma$, an increase in bias can even stabilize the junction.

			First, we discuss the sharp decrease from long and short dissociation times, which is the dominant feature, especially for $\Gamma\lsim0.3$ eV.
			For small bias voltages, the dissociation is mediated by the ''slow'' process relying on the exponentially suppressed part of the nuclear wave packet leaking into the classically forbidden regime (process $\circled{2}$ in Fig.\ \ref{fig:process1}). 
			For large bias voltages beyond the onset of resonant transport, dissociation is dominated by the resonant population of the anti-bonding electronic state (process $\circled{1}$ in Fig.\ \ref{fig:process1}). Once this process becomes energetically possible by the applied bias voltage, the dissociation time $t_{50\%}$ is almost independent of the applied bias voltage. The bias values, at which the transition between the two processes occurs, depend on all parameters of the system and is related to the onset of resonant transport beyond a certain voltage. Moreover, in the regime of small to intermediate $\Gamma$, the dissociation time $t_{50\%}$ decreases with increasing coupling strength. This behavior is caused by the broadening and the enhanced availability of electrons from the leads with increasing $\Gamma$.

			Next, we analyze the dissociation behavior of the asymmetrically coupled system. 
			Generally, the asymmetric setup exhibits longer dissociation times, which is explained by its molecule-lead coupling scenario which favors the neutral molecule.
			Most notably, however, is the behavior of this coupling scenario for large $\Gamma$ within the adiabatic regime, {\color{black} where the electronic dynamics if fast compared to nuclear motion}. 
			In particular for $\Gamma=1.0$ eV, the data reveal an increase in dissociation time with bias voltage.  For large $\Gamma\gsim0.3$ eV, the notion of an averaged electronic background and an adiabatic PES, which governs the nuclear dynamics, becomes applicable.
			Given that the asymmetric model favors an empty electronic state under transport, the adiabatic PES of the asymmetric model exhibits a potential barrier, which increases with applied bias voltage, resulting in a stabilization of the junction with applied bias voltage.
			Notice that this effect was also observed in our previous work.\cite{Erpenbeck_dissociation_2018} This validates the findings obtained by mixed quantum-classical approaches.
			For $\Gamma=0.5$ eV, the system shows a transition behavior between stabilization for low bias voltages and the sharp decrease of the dissociation time towards larger bias voltages.

			Figs.\ \ref{fig:comparison_DIET_1}c and d display the dissociation time as a function of the molecule-lead coupling strength $\Gamma$ for different bias voltages.
			Generally, the data exhibits a nonlinear relationship between dissociation time and the molecule-lead coupling strength. Moreover, the dissociation time exhibits a sharp decrease around a certain $\Gamma$-range, which depends on the applied bias voltage. 
			For the symmetric coupling scenario (Fig.\ \ref{fig:comparison_DIET_1}c), the dissociation time always decreases with increasing molecule-lead coupling strength. Further, with increasing bias, the dissociation time becomes constant over a wide range of molecule-lead coupling strengths, which can be explained by the fact that the leads can always provide electrons in the high-bias regime. 
			When considering the dissociation time for the asymmetric setup (Fig.\ \ref{fig:comparison_DIET_1}d), $t_{50\%}$ depends in a non-monotonous way on $\Gamma$  and displays a distinct crossover between the weak coupling behavior and the strong coupling region around $\Gamma\sim0.7$ eV.
			This can be interpreted in terms of the crossover between the nonadiabatic regime, where dissociation is mediated by the exponentially suppressed part of the wavefunction, and the adiabatic regime, where the notion of an adiabatic PES becomes meaningful and the voltage-stabilization effect becomes active.

		\subsubsection{Electronic current}

			Fig.\ \ref{fig:SYMM_data}b shows the electronic current between the molecule and the left lead of the symmetric system as a function of time for different coupling strengths $\Gamma$ and applied bias voltages $\Phi$.
			For short times, the current exhibits a peak which is associated with establishing the contact between the molecule and the leads. This initial spike in the current decreases to a small but nonzero value with time. 
			Partial oscillations in the current are reminiscent of the nuclear excitation, which stems from closing the contact between molecule and leads at time $t=0$. This can be validated upon monitoring the motion of the nuclear wave packet as a function of time (data not shown). Moreover, notice that the associated oscillation period of $\sim 20$ fs is in line with the frequency of the harmonic approximation to $V_0(x)$.
			The long-time current value is determined by the conductance properties of the dissociated molecule. For intermediate times, however, the current can be smaller for a higher applied bias voltage. This behavior is explained by considering the current in correlation with the dissociation probability. As the dissociation probability increases with time, also the average conductance properties of the molecule change. This leads to a steady decrease in current until the dissociation probability is close to 100\% and the current assumes the value corresponding to the dissociated system. As higher bias voltages yield larger dissociation probabilities at intermediate times, they also display smaller currents. 
			
			Finally, we remark that neither the current nor the dynamics of the wave packet (Figs.\ \ref{fig:SYMM_data}c--e) reveal signatures that suggest that current-induced nuclear excitation would play a pronounced role for the dissociation dynamics. This is in line with the observation of a dissociation process dominated by the direct charging of the molecule, ie.\ the population of the anti-bonding electronic state.

		\subsubsection{Comparison with mixed quantum-classical approaches}\label{sec:validity_Ehrenfest_DIET}

			A common approach to study current-induced dissociation in molecular junctions is to invoke a mixed quantum-classical description, where the nuclear DOFs are described classically.\cite{Horsfield2004_3, Verdozzi2006, Dundas2009, Dzhioev2013, Todorov2014, Cunningham2015, Erpenbeck_dissociation_2018, Leitherer2019}
			This is due to the fact that these approaches are usually numerically less demanding and allow for an intuitive interpretation in terms of forces acting on the nuclei.
			Generally, mixed quantum-classical frameworks are considered to provide reliable results for situations where the electronic motion is fast compared to the nuclear dynamics. This justifies the description of the nuclear motion governed by an averaged electronic background which is inherent to these schemes. 
			Nuclear quantum effects are not captured in mixed quantum-classical approaches and essential effects such as Joule heating are often not included adequately.\cite{Horsfield2004_3, Horsfield2004_2}
			
			In a previous publication,\cite{Erpenbeck_dissociation_2018} we have employed a mixed quantum-classical framework to a similar model in order to study the dissociation dynamics in molecular junctions. The validity of the results in Ref.\ \onlinecite{Erpenbeck_dissociation_2018} can be assessed upon comparing its findings to the outcome of this section.
			One of the main results of Ref.\ \onlinecite{Erpenbeck_dissociation_2018} is a threshold-like onset of dissociation beyond a certain applied bias voltage. This finding is also recovered in the full quantum data as a steep decrease in dissociation time beyond a given bias voltage (see Figs.\ \ref{fig:comparison_DIET_1}a and b). Despite this similarity, there are also deviations between the outcomes of the two frameworks, in particular for low bias voltages and small molecule-lead coupling strengths within the non-adiabatic regime. Here, the mixed quantum-classical framework predicts a (at least partially) stable junction, whereas the quantum results reveal that the junction dissociates due to the exponentially suppressed part of the nuclear wavefunction leaking in the classically forbidden regime.
			Notice that the lack of Joule heating in mixed quantum-classical method used in Ref.\ \onlinecite{Erpenbeck_dissociation_2018} does not constitute a problem as current-induced nuclear excitation plays a negligible role for the model considered in this section. 
			
			Another integral finding of the present work and Ref.\ \onlinecite{Erpenbeck_dissociation_2018} is the current-stabilization effect found in asymmetrically coupled systems favoring a sparsely populated electronic state under transport. 
			This stabilization observed in the classical calculations translates into a time-delay in the dissociation in the fully quantum mechanical case. 
			However, the mixed quantum-classical methodology predicts this effect for any molecule-lead coupling strength, whereas the quantum mechanical results suggest that this effect is limited to the adiabatic regime of large $\Gamma$. This enforces the notion that the validity of the mixed quantum-classical framework is restricted to the adiabatic regime.

	\subsection{Dissociation upon current-induced heating}\label{sec:DIMET}
	
		In the model system considered in Sec.\ \ref{sec:DIET}, current-induced excitation of the nuclear DOF on the ground state PES by tunneling electrons plays a negligible role. In this section, we aim at identifying model systems and parameter regimes where current-induced excitation of the nuclear DOF is the predominant reason for dissociation. This allows us to embed our findings into the commonly used paradigm, which is that current-induced dissociation in molecular junctions occurs as a consequence of current-induced heating of vibrational modes.\cite{Koch2006, Rainer2011, Dzhioev2013}
		The same mechanism was also considered for current-induced desorption of atoms and molecules from surfaces using a scanning tunneling microscope. In this context, the mechanism is called desorption induced by multiple electronic transitions (DIMET) and is most pronounced in the non-resonant transport regime.\cite{Saalfrank2006}
		The mechanism is characterized by a nonlinear relationship between the electronic current and the desorption rate and typically leads to much smaller desorption rates than the dissociation induced by the transient population of anti-bonding electronic states.\cite{Saalfrank2006, Shen1995}
		
\begin{figure}[tb!]
			\centering
			\vspace*{-0.3cm}
			\includegraphics[width=\linewidth]{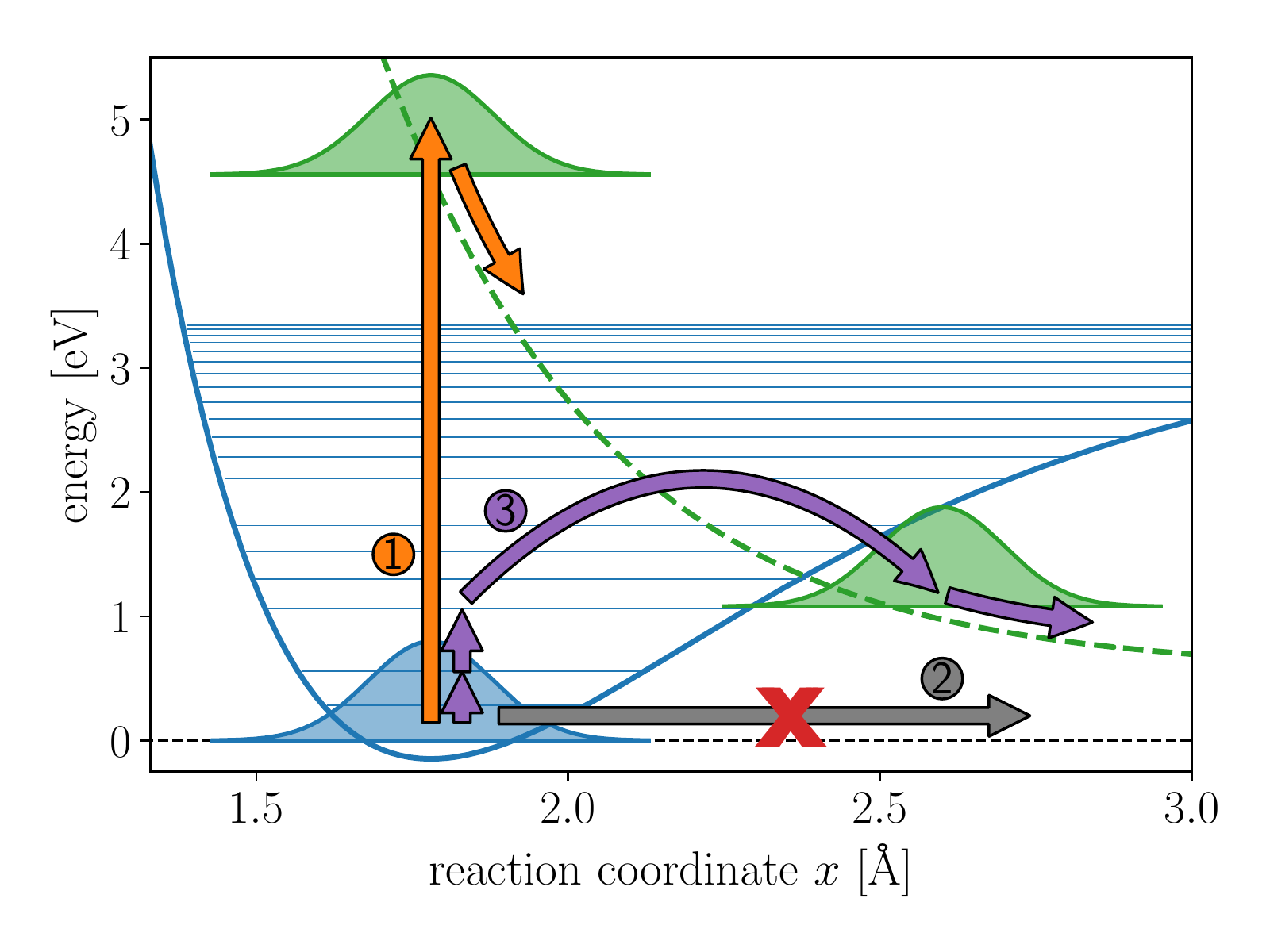}
			\captionsetup{font=small,labelfont=bf, justification=centerlast, format=plain}
			\caption{ \bf \scriptsize
				Graphical representation of three processes leading to dissociation.
				For the case of $V_\infty=0.558$ eV considered here, a direct dissociation 
				(process $2$) is energetically not possible as highlighted by the red cross. As such, current-induced dissociation is required to enable this effect (process $3$).
				(The figure sketches processes and does not contain actual data.)
				}
			\label{fig:process2}
		\end{figure}

		In the following, we consider the model system introduced in Sec.\ \ref{sec:model} with $V_\infty = 0.558$ eV as depicted in Fig.\ \ref{fig:pots_overview}. This shift in $V_\infty$ has two effects on the system dynamics as compared to the analysis provided in Sec.\ \ref{sec:DIET}. 
		First, the energy difference between $V_d(x)$ and $V_0(x)$ is increased such that the resonant population of the electronic state is suppressed and its onset is shifted to larger bias voltages (see process $\circled{1}$ in Fig.\ \ref{fig:process2}). 
		Second, the dissociation of the junction upon the exponentially suppressed part of the vibrational ground state  wavefunction leaking into the classically forbidden regime is energetically excluded. This is sketched as process $\circled{2}$ in Fig.\ \ref{fig:process2}. Consequently, for low bias voltages, dissociation is induced by heating of the nuclear DOF by inelastic co-tunneling processes. This is depicted as process $\circled{3}$ in Fig.\ \ref{fig:process2}. 
		
		In the following, we study the dissociation probability, the electronic population, the electronic current, and the relative population of the eigenstates of the bonding potential $V_0(x)$ to analyze the dissociation dynamics. 
		To avoid obscuring the influence of current-induced heating by broadening effects, we resort to the small $\Gamma$ regime. Moreover, we exclusively study the symmetrically coupled system to provide evidence for dissociation upon current-induced heating of the nuclear DOF  and refrain from comparing the different coupling scenarios. A transfer of the findings to asymmetric systems should be straightforward.

		\subsubsection{Dissociation dynamics}

\begin{figure*}
				\begin{minipage}[c]{0.58\textwidth}
					\raggedright \quad a) \hspace{4.5cm} b) \\
					\centering
					\includegraphics[width=\linewidth]{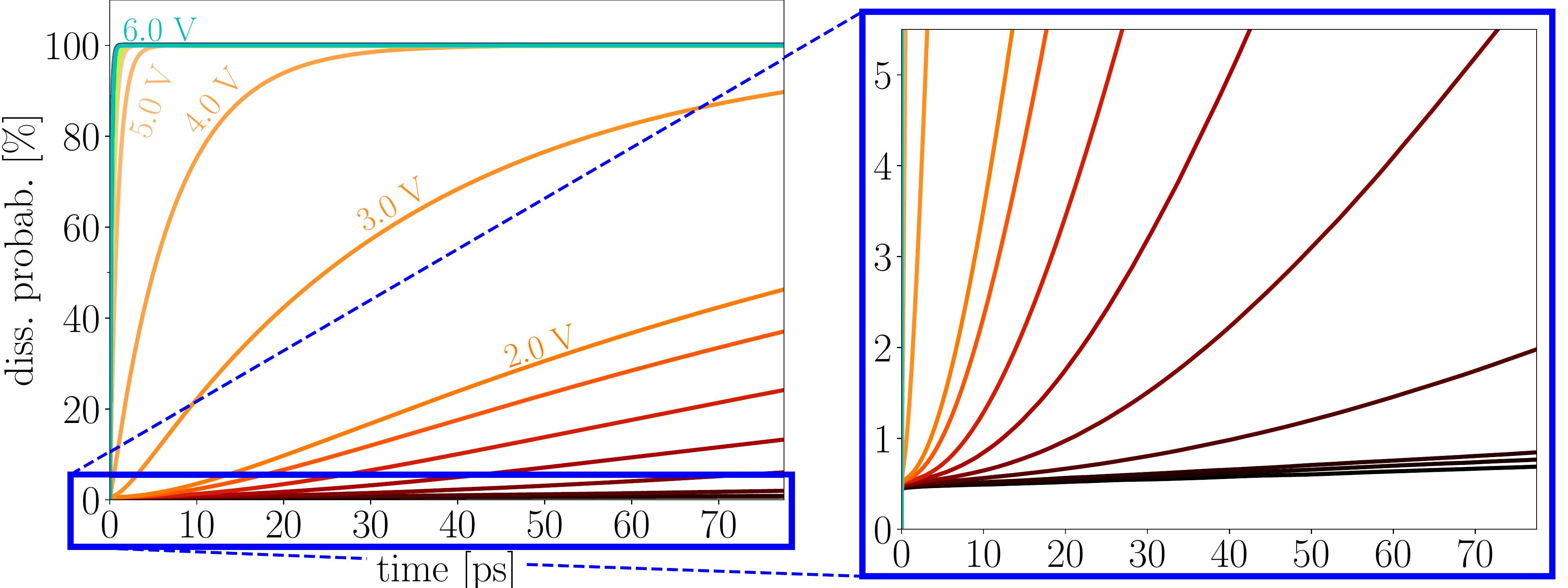}
				\end{minipage}
				\begin{minipage}[c]{0.29\textwidth}
					\raggedright c)\\
					\centering
					\includegraphics[width=1.025\linewidth]{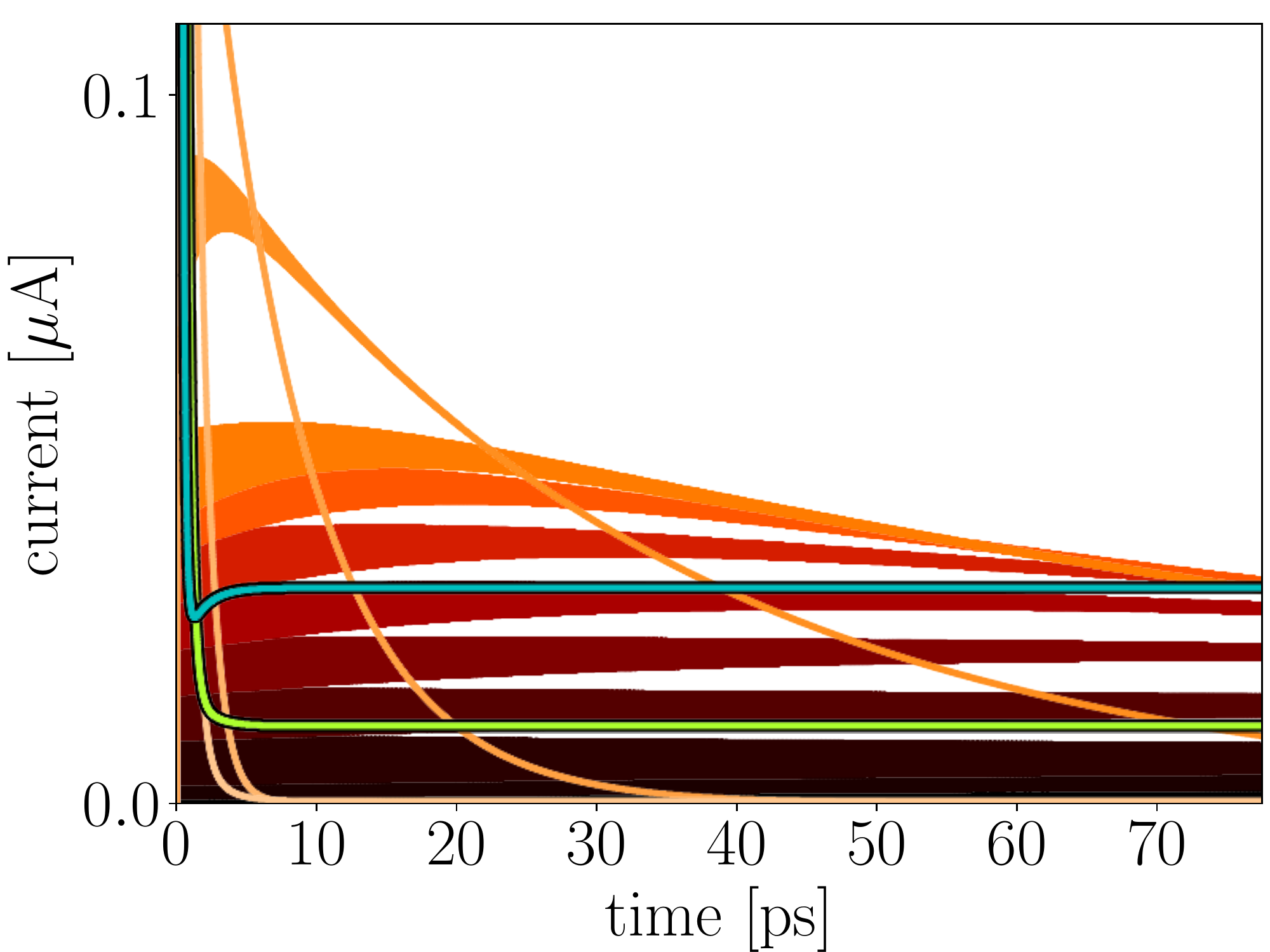}
				\end{minipage}
				\begin{minipage}[r]{0.08\textwidth}
					\hspace*{0.3cm}
					\vspace*{+0.18cm}
					\includegraphics[width=0.85\linewidth]{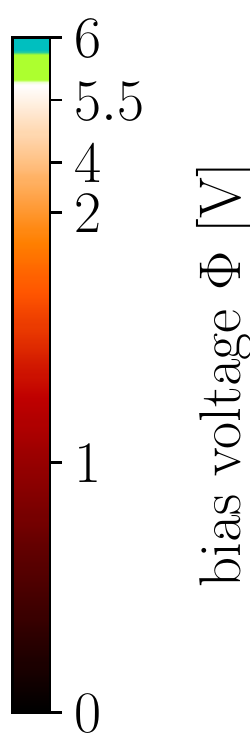}
				\end{minipage}
				\captionsetup{font=small,labelfont=bf, justification=centerlast, format=plain}
				\caption{ \bf \scriptsize
					Dynamics of the model system with $V_\infty=0.558$ eV symmetrically coupled to two leads for different applied bias voltages and $\Gamma=0.1$ eV. In all the plots, the different lines correspond to different applied bias voltages. 
					a: Dissociation probability as a function of time for different applied bias voltages. 
					b: Closeup of Fig.\ \ref{fig:DIMET_data_diss}a. 
					c: Current between the molecule and the left lead as a function of time for different applied bias voltages.
					}
				\label{fig:DIMET_data_diss}
			\end{figure*}

			Fig.\ \ref{fig:DIMET_data_diss}a depicts the dissociation probability for different applied bias voltages $\Phi$ for the symmetrically coupled system with $\Gamma=0.1$ eV.
			As rationalized before, the dissociation probability always increases with time which is inherent to the model. 
			For larger bias voltages, we find the familiar behavior discussed in Sec.\ \ref{sec:DIET}. Once the population of the anti-bonding state becomes energetically possible by the applied bias voltage, dissociation is dominated by this effect and occurs on the timescale of tenths of femtoseconds to nanoseconds.
			For low bias voltages, however, we find molecular junctions which are stable on the order of several nanoseconds and beyond. These will be the junctions most relevant for experimental observations. Fig.\ \ref{fig:DIMET_data_diss}b provides a closeup on the dissociation probability as a function of time emphasizing the behavior in the low bias regime. 
			
			For low bias voltages, the dissociation probability almost instantly assumes a small but non-vanishing value. This is reminiscent of the molecule-lead connection being established at time $t=0$. Thereafter, the dissociation probability increases very slowly and in an almost linear way with time. Moreover, the dissociation is enhanced on the picosecond timescale by an increased bias voltage. This particular behavior in the low bias regime is substantially different from what was discussed in Sec.\ \ref{sec:DIET} and is therefore beyond the effect of dissociation induced by the population of the anti-bonding state.
			In the following, we focus on this regime and argue that in this case, dissociation is mediated by the excitation of the nuclear DOF by inelastic co-tunneling in the non-resonant transport regime.

		\subsubsection{Electronic current and different transport regimes}
		
			The electronic current as a function of time for different applied bias voltages $\Phi$ for the symmetric model with $\Gamma=0.1$ eV is depicted in Fig.\ \ref{fig:DIMET_data_diss}c.
			For large bias voltages, we recover the behavior already discussed in Sec.\ \ref{sec:DIET}, which comprises an initial peak in the current associated with connecting the molecule and leads at time $t=0$, followed by a steady decrease in current which is related to an increase in dissociation probability. 
			Due to the almost instant dissociation resulting from the population of the dissociative PES, the current in the long-time limit is given by the transport characteristics of the dissociated molecule, which is represented by an electronic single-level system with a charging energy of $\lim_{x\rightarrow\infty} V_d(x)-V_0(x) \approx -2.82$ eV (see Fig.\ \ref{fig:pots_overview}).
			Thus, for the voltages explicitly considered here, a distinct role is attributed to the bias voltage of $6$ V, which allows for resonant transport through the dissociated molecule and therefore exhibits a pronounced non-zero steady-state current. 
			{\color{black} Moreover, we consider the bias voltages $\Phi=5.4$ V and $\Phi=5.6$ V, which are at the onset of the resonant transport regime.}
			Notice that this is only possible as we are studying a non-destructive dissociative model system, where in case of dissociation, the molecular backbone still links the two leads (see Fig.\ \ref{fig:sketch}).
			
			For the low bias regime, the current also displays an onset behavior, however for long times, it approaches a value which is approximately constant. Thereby, the lines for the current for low bias values in Fig.\ \ref{fig:DIMET_data_diss}c appear broadened as the current exhibits oscillations that are fast compared to the timescales under investigation.
			The oscillations in the current are reminiscent of the nuclear excitation originating from closing the contact between molecule and leads at time $t=0$. This excitation results in an oscillation of the wave packet about the minimum of $V_0(x)$. Notice that an adequate description of this dynamics requires a coherent description of the nuclear DOF.\cite{Schmidt2008, Jorn2009, Albrecht2013}
			Considering the nonzero average current for long times, the data reveals a dependency on the applied bias voltage which is monotonically increasing. These values for the average current are in line with what is to be expected for the non-resonant transport regime. This implies that the current is related to co-tunneling processes, thus avoiding the population of the anti-bonding electronic state.

			\begin{figure}
				\centering
				\includegraphics[width=0.45\textwidth]{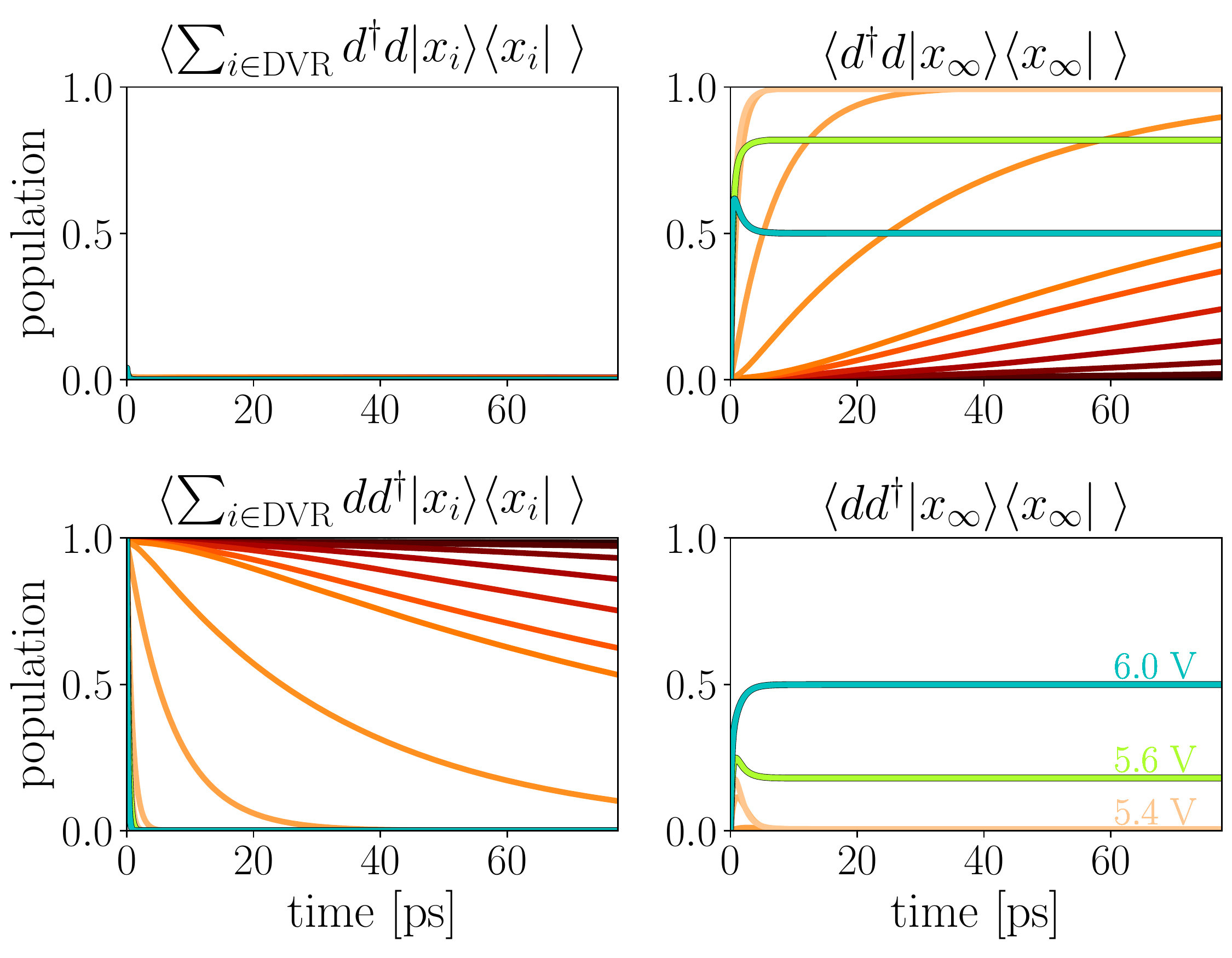}
				\\
				\vspace*{-6.66cm}
				\raggedright \quad a) \hspace{3.5cm} b) \\
				\vspace*{2.66cm}
				\raggedright \quad c) \hspace{3.5cm} d) \\
				\vspace*{3.33cm}
				\captionsetup{font=small,labelfont=bf, justification=centerlast, format=plain}
				\caption{ \bf \scriptsize
						    Population of the electronic states as a function of time for different bias voltages.
						    The different lines correspond to the different bias voltages, the color-scale coincides with the one used in Fig.\ \ref{fig:DIMET_data_diss}.
						    a: Population of the charged state of the {\color{black}non-dissociated} molecule.
						    b: Population of the charged state of the dissociated molecule.
						    c: Population of the neutral state of the {\color{black}non-dissociated} molecule.
						    d: Population of the neutral state of the dissociated molecule.
					}
				\label{fig:el_pop_VS_time}
			\end{figure}
				    
				\begin{figure*}
					    \begin{minipage}[c]{0.32\textwidth}
						    \raggedright a)\\
						    \centering
						    \includegraphics[width=\textwidth]{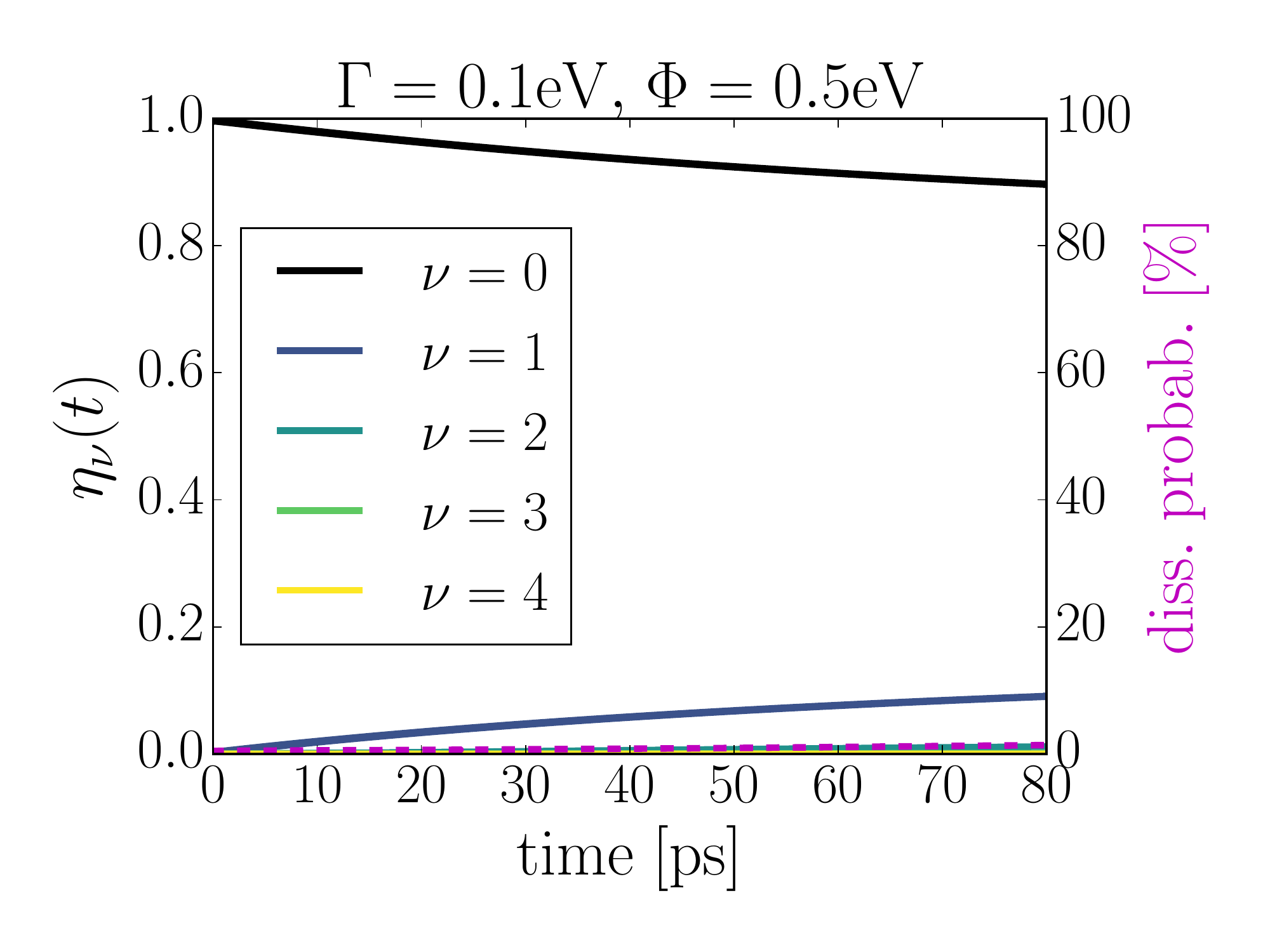}\\
					    \end{minipage}
					    \begin{minipage}[c]{0.32\textwidth}
						    \raggedright b)\\
						    \centering
						    \includegraphics[width=\textwidth]{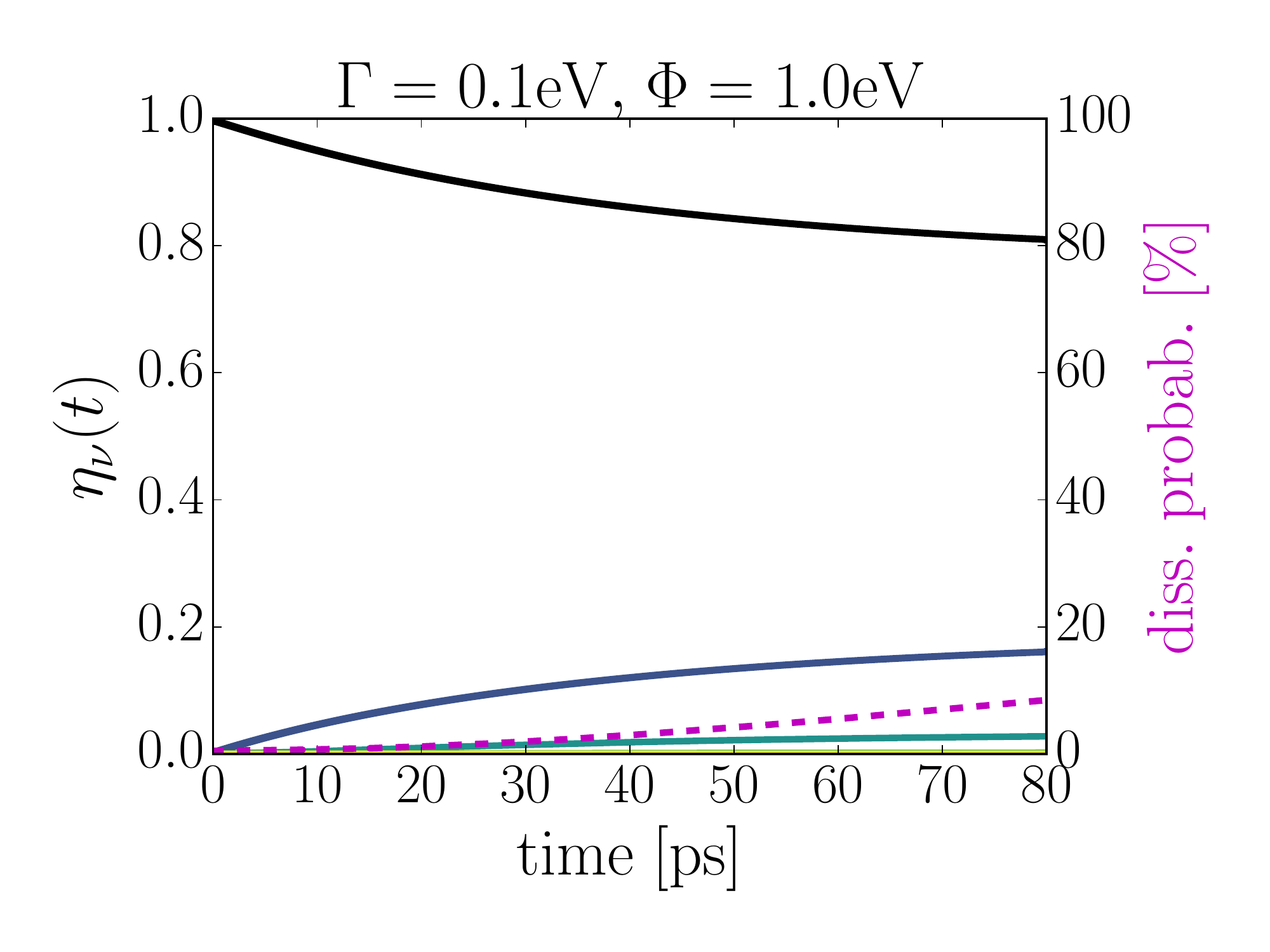}\\
					    \end{minipage}
					    \begin{minipage}[c]{0.32\textwidth}
						    \raggedright c)\\
						    \centering
						    \includegraphics[width=\textwidth]{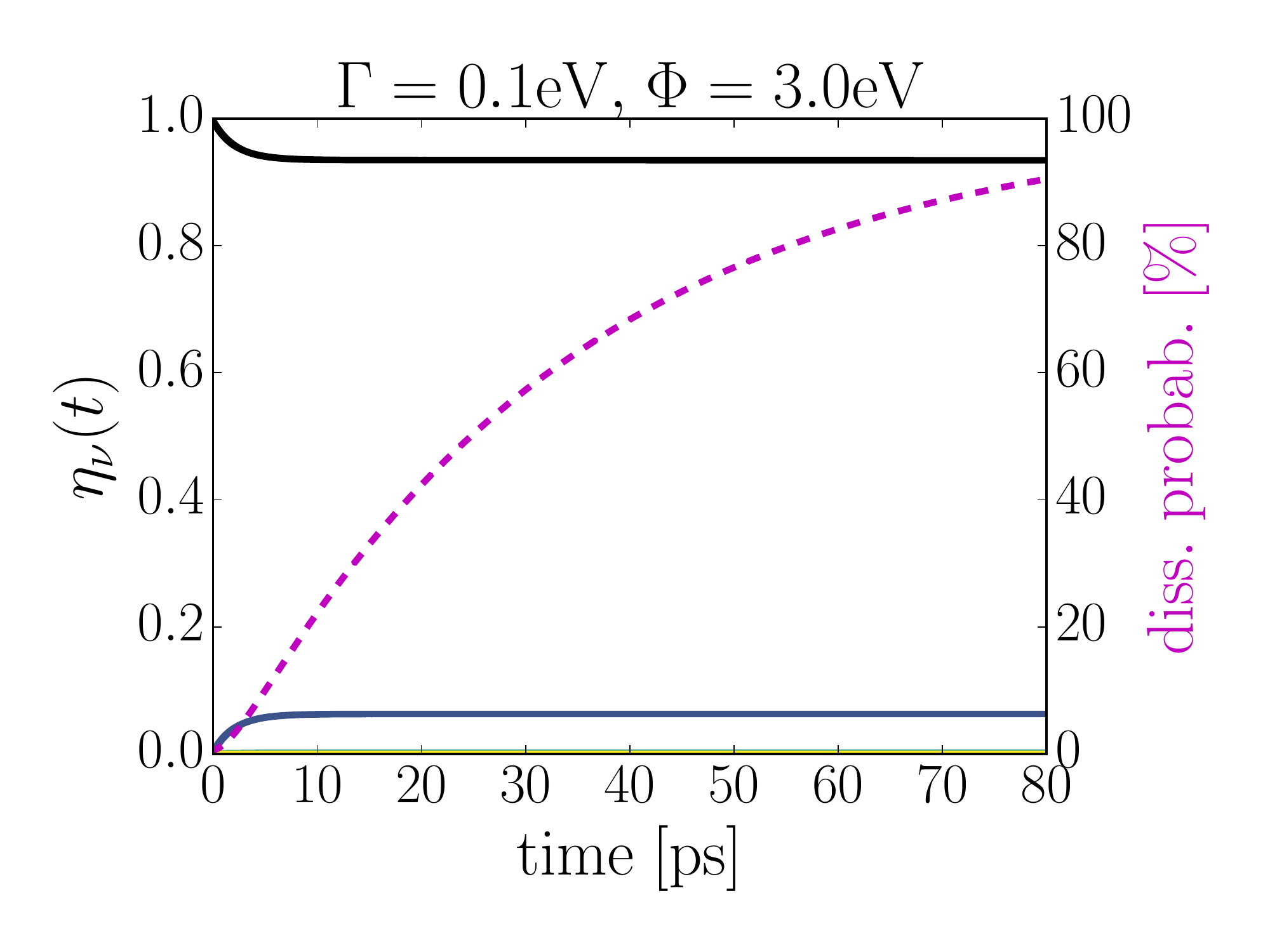}\\
					    \end{minipage}
					    \captionsetup{font=small,labelfont=bf, justification=centerlast, format=plain}
					    \caption{\bf \scriptsize
						    Relative nuclear population $\eta_\nu(t)$ as a function of time of the five lowest eigenstates $\nu$ of $V_0(x)$ for representative bias voltages and $\Gamma=0.1$ eV. For timescale referencing, the dissociation probability as a function of time is depicted as a purple dashed line.}
					    \label{fig:system2_SYMM_nuc_excitation}
				\end{figure*}

			To support the notion of the realization of different transport regimes, we consider the electronic population in Fig.\ \ref{fig:el_pop_VS_time}. Thereby, we distinguish between the charge states of the {\color{black}non-dissociated} molecule (Figs.\ \ref{fig:el_pop_VS_time}a and c) and of the dissociated molecule (Figs.\ \ref{fig:el_pop_VS_time}b and d).
			We find that, initial effects aside, the population starting in the neutral state of the {\color{black}non-dissociated} molecule is steadily transferred to the charged state of the dissociated molecule. 
			For the low bias regime, the data suggests that the molecule is predominantly in the neutral state of the {\color{black}non-dissociated} molecule and partially in the charged state of the dissociated molecule, whereby the probabilities of being in the charged state of the dissociated molecule grows with time.
			As the population of the charged state of the {\color{black}non-dissociated} molecule is avoided in this process, the current found for this system is associated with non-resonant co-tunneling processes.
			For larger bias voltages, this transfer from the neutral state of the {\color{black}non-dissociated} molecule to the charged state of the dissociated molecule happens much faster. For bias voltages  {\color{black} $\lsim 5.4$ V, the dissociated molecule is in the non-resonant transport regime, which explains the low current associated with these systems and the negligible population of the neutral state of the dissociated molecule.
			For $\Phi=5.6$ V and beyond, the dissociated molecule enters the resonant transport regime.}
			Consequently, the population of the neutral and the charge state of the dissociated molecule {\color{black} increases with bias and} assumes a value of $0.5$ {\color{black} for $6$ V, which is representative for the system being in the resonant transport regime.}

		\subsubsection{Current-induced excitations of the nuclear DOF}
		
			In the following, we establish the role of current-induced excitation of the nuclear DOF and its impact on the stability of molecular junctions. To this end, we study the relative nuclear population $\eta_\nu(t)$ of the bound states of the binding potential $V_0(x)$ of the neutral molecule
			for three representative bias voltages.
			For low bias voltages in the deep non-resonant transport regime (Fig.\ \ref{fig:system2_SYMM_nuc_excitation}a), we find a built-up of the relative population of the first excited state with time, reflecting current-induced nuclear excitation in this regime. Moreover, the dissociation probability increases also very slightly.  However, the full impact of this process on the stability of the molecular junction is to be expected at a timescale far beyond tenths of picoseconds, when the nuclear excitation levels out.
			For intermediate bias voltages, still in the non-resonant transport regime (Fig.\ \ref{fig:system2_SYMM_nuc_excitation}b), we find a pronounced population of the first, but also of the second excited state. The dissociation probability increases noticeably as the nuclear excitation increases. This suggests that current-induced excitation is responsible for the dissociation process (see process $\circled{3}$ in Fig.\ \ref{fig:process2}).
			For even higher bias voltages
			(Fig.\ \ref{fig:system2_SYMM_nuc_excitation}c), the relative population of the first excited state assumes a constant value after a short initial time, which is smaller than for the lower bias voltages considered before. Moreover, the relative population of the higher excited vibrational states is close to zero. 
			This is remarkable as the effect of current-induced excitation usually increases with bias voltage. 
			{\color{black} Similar influences on the vibrational distribution or the heat transport characteristics have been found in the context of co-tunneling assisted resonant transport.\cite{Huettel2009, Gergs2015, Gaudenzi2017}}
			Moreover, as the second excited state is aligned with the limiting value $V_d(x\rightarrow\infty)$ (see Fig.\ \ref{fig:pots_overview}), which renders a direct dissociation from the first excited state energetically impossible, the data suggests that current-induced excitation of the nuclear DOF alone is not able to explain the dissociation dynamics.
			Hence, the effect of dissociation upon the population of the anti-bonding electronic state needs to play a non-negligible role. 
			The dissociation dynamics is therefore influenced by the competition between current-induced excitation and the population of the anti-bonding state.
			However, as the process of populating the anti-bonding states happens on shorter timescales, it effectively depletes the population of the excited vibrational states, leading to a reduced relative population of the excited vibrational states with increasing bias voltage.
			Consequently, the influence of current-induced nuclear excitation on the dissociation dynamics is reduced to the point where it plays a subordinate role and the dissociation dynamics is dominated by the resonant population of the anti-bonding electronic state.

			After the discussion of the behavior of the nuclear excitation as a function of time, we systematically analyze its impact on the dissociation mechanism and its dependence on the applied bias voltage $\Phi$.
			To this end, we consider the relative population of the vibrational eigenstates $\nu$ of the binding potential as defined in Eq.\ (\ref{eq:rel_pop}).
			\begin{figure}
				\centering
				\begin{minipage}[c]{0.425\textwidth}
					\centering
					\vspace*{-.3cm}
					\includegraphics[width=\textwidth]{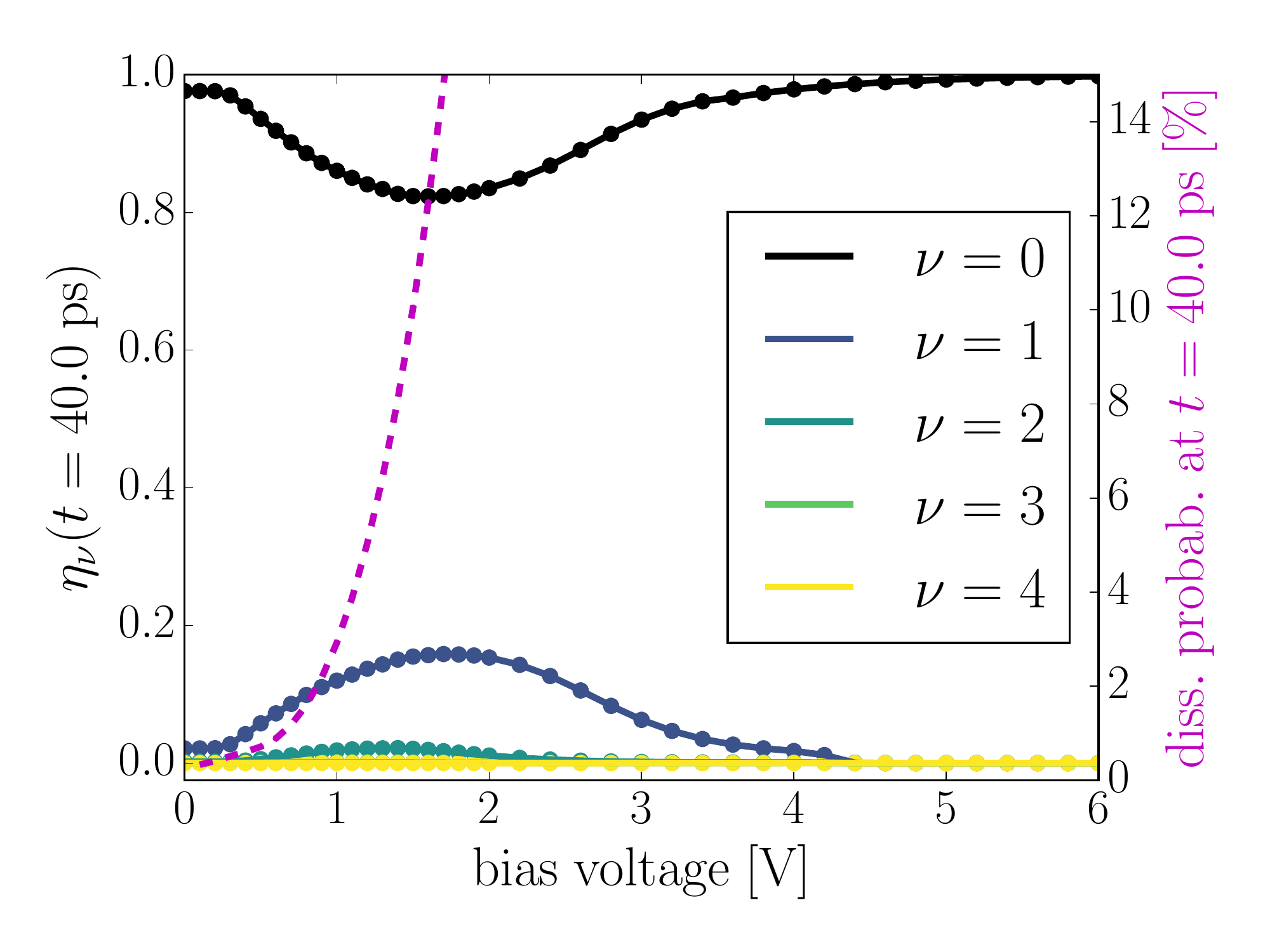}
				\end{minipage}
				\captionsetup{font=small,labelfont=bf, justification=centerlast, format=plain}
				\caption{ \bf \scriptsize
					Relative nuclear population $\eta_\nu(t)$ of the five lowest eigenstates of $V_0(x)$ at time $t=40$ ps for $\Gamma=0.1$ eV.
					The different lines correspond to the different states.
					Points mark the actual data, the lines serve as a guide for the eye.
					The purple dashed line, which represents the dissociation probability at time $t=40$ ps for the low bias regime, serves as a reference.
					}
				\label{fig:nuc_pop_VS_bias}
			\end{figure}
			Notice that the onset of dissociation as a consequence of the population of the anti-bonding state drastically diminishes the probabilities of being in the neutral electronic state. Consequently, the population of the state $\nu$ tends to zero for large bias voltages, whereas $\eta_\nu$ stays finite allowing for an analysis of the nuclear excitation independent of the dissociation mechanism.
			Fig.\ \ref{fig:nuc_pop_VS_bias} shows $\eta_\nu$ at time $t=40$ ps for the five lowest eigenstates of $V_0(x)$ for different bias voltages. Moreover, the dissociation probability at this very time is also depicted in Fig.\ \ref{fig:nuc_pop_VS_bias}. Thereby, the qualitative appearance of the plot is independent of the time chosen, as long as the time was selected such that the influence of the current-induced excitation is active.

			Considering Fig.\ \ref{fig:nuc_pop_VS_bias}, we find that even for zero or very low bias voltages, there is vibrational excitation present in the system. As discussed before, this is related to the initial conditions and the finite temperature.
			At about $\Phi\sim0.3$ V, the data exhibits an increase in the population of the excited vibrational states and a decrease in the occupation of the nuclear ground state. This redistribution of the relative population of the vibrational states grows with increasing bias voltage and correlates with a profound rise in dissociation probability.
			This behavior is explained by the onset of inelastic transport, which is expected at a bias voltage of $\Phi=0.2855$ V for the given system.
			For voltages beyond $\Phi\sim1.8$ V, the trend changes such that the relative population of the excited states decreases with increasing bias voltage, whereas the relative population of the ground state approaches towards one. This observation, together with the behavior of the current and the dissociation probability, indicates the takeover of the population of the anti-bonding state as the dominant mechanism for dissociation. As the population of the anti-bonding state happens on short timescales as compared to current-induced vibrational excitation, the onset of the population of the anti-bonding state depletes the vibrational excited states, leading to a decrease of the relative vibrational populations of the excited states with bias voltage. 
			As such, the notion of current-induced excitation of the nuclear DOFs as the source of dissociation in molecular junctions is restricted to the non-resonant transport regime. Beyond the non-resonant transport regime, the population of the anti-bonding state gives rise to dissociation on much faster timescales, rendering molecular junctions in this regime unstable.

\section{Conclusion}\label{sec:conclusion}

	We have investigated current-induced bond rupture in single-molecule junctions involving a dissociative PES. 
	To this end, we applied a numerically exact framework to generic models for molecular junctions comprising a bonding and an anti-bonding electronic state. We considered a wide range of physical parameters ranging from the nonadiabatic regime of weak molecule-lead coupling to the adiabatic case of strong coupling, as well as different molecule-lead coupling scenarios. 
	
	Overall, we identified different mechanisms leading to dissociation on distinct timescales, namely the population of the anti-bonding PES and current-induced excitation of the nuclear DOF.
	Whenever the population of the anti-bonding PES is energetically possible, dissociation is mediated by this process and occurs on the timescale of tenths of femtoseconds to picoseconds. 
	This is the case if the applied bias voltage exceeds a certain limit, which depends on the details of the PES. 
	Moreover, in situations where the model system under investigation does not exhibit a pronounced non-resonant transport regime, the systems showed to be unstable beyond the nanosecond timescale due to this effect. This realization can be used as a guide to preselect certain molecules for experiments.
	For model systems exhibiting a pronounced non-resonant transport regime, current-induced heating is the dominant process leading to dissociation in the non-resonant transport regime on the timescale of tenth of nanoseconds and beyond. However, once these systems enter the resonant transport regime, dissociation is again mediated by the population of the anti-bonding PES and is not necessarily related to high levels of nuclear excitation on the ground state PES.
	
	Given the numerically exact approach employed in this work, we were also able to validate and extend previous results using mixed quantum-classical methods.\cite{Erpenbeck_dissociation_2018}
	Generally, the applicability of mixed quantum-classical frameworks in the adiabatic regime of strong molecule-lead coupling could be confirmed. In particular, we were able to confirm the existence of current-stabilization in molecular junctions favoring the neutral state under transport, where in certain parameter ranges, an increase in the current across a molecular junction can increase its stability. However, for low bias voltages and weak molecule-lead coupling strengths, mixed quantum-classical frameworks fail as they miss the influence of nuclear quantum effects on the stability of molecular junctions. In these situations, it is necessary to apply a framework which treats the electrons and the nuclei on a quantum level.
	{\color{black} In general, the HQME approach employed in this work can provide valuable benchmark results, which may also help to improve other, more approximate transport methods and thus contribute to the development of methods that allow an accurate description of current-induced nuclear dynamics in molecular junctions even beyond basic models.}

\section*{Acknowledgement}

	We thank C.\ Schinabeck, C.\ Kaspar and J.\ B\"atge for helpful discussions.
	This work was supported by the German Research Foundation (DFG) and the Israel Science Foundation (ISF).
	Y.K. gratefully acknowledges a Research Fellowship of the Alexander von Humboldt Foundation.

\section*{Conflicts of interest}
	There are no conflicts to declare.

\bibliography{Bib}

\begin{thebibliography}{134}%
\makeatletter
\providecommand \@ifxundefined [1]{%
 \@ifx{#1\undefined}
}%
\providecommand \@ifnum [1]{%
 \ifnum #1\expandafter \@firstoftwo
 \else \expandafter \@secondoftwo
 \fi
}%
\providecommand \@ifx [1]{%
 \ifx #1\expandafter \@firstoftwo
 \else \expandafter \@secondoftwo
 \fi
}%
\providecommand \natexlab [1]{#1}%
\providecommand \enquote  [1]{``#1''}%
\providecommand \bibnamefont  [1]{#1}%
\providecommand \bibfnamefont [1]{#1}%
\providecommand \citenamefont [1]{#1}%
\providecommand \href@noop [0]{\@secondoftwo}%
\providecommand \href [0]{\begingroup \@sanitize@url \@href}%
\providecommand \@href[1]{\@@startlink{#1}\@@href}%
\providecommand \@@href[1]{\endgroup#1\@@endlink}%
\providecommand \@sanitize@url [0]{\catcode `\\12\catcode `\$12\catcode
  `\&12\catcode `\#12\catcode `\^12\catcode `\_12\catcode `\%12\relax}%
\providecommand \@@startlink[1]{}%
\providecommand \@@endlink[0]{}%
\providecommand \url  [0]{\begingroup\@sanitize@url \@url }%
\providecommand \@url [1]{\endgroup\@href {#1}{\urlprefix }}%
\providecommand \urlprefix  [0]{URL }%
\providecommand \Eprint [0]{\href }%
\providecommand \doibase [0]{http://dx.doi.org/}%
\providecommand \selectlanguage [0]{\@gobble}%
\providecommand \bibinfo  [0]{\@secondoftwo}%
\providecommand \bibfield  [0]{\@secondoftwo}%
\providecommand \translation [1]{[#1]}%
\providecommand \BibitemOpen [0]{}%
\providecommand \bibitemStop [0]{}%
\providecommand \bibitemNoStop [0]{.\EOS\space}%
\providecommand \EOS [0]{\spacefactor3000\relax}%
\providecommand \BibitemShut  [1]{\csname bibitem#1\endcsname}%
\let\auto@bib@innerbib\@empty
\bibitem [{\citenamefont {Aviram}\ and\ \citenamefont
  {Ratner}(1974)}]{Aviram1974}%
  \BibitemOpen
  \bibfield  {author} {\bibinfo {author} {\bibfnamefont {A.}~\bibnamefont
  {Aviram}}\ and\ \bibinfo {author} {\bibfnamefont {M.~A.}\ \bibnamefont
  {Ratner}},\ }\href@noop {} {\bibfield  {journal} {\bibinfo  {journal} {Chem.
  Phys. Lett.}\ }\textbf {\bibinfo {volume} {29}},\ \bibinfo {pages} {277 }
  (\bibinfo {year} {1974})}\BibitemShut {NoStop}%
\bibitem [{\citenamefont {Nitzan}(2001)}]{Nitzan2001}%
  \BibitemOpen
  \bibfield  {author} {\bibinfo {author} {\bibfnamefont {A.}~\bibnamefont
  {Nitzan}},\ }\href@noop {} {\bibfield  {journal} {\bibinfo  {journal} {Annu.
  Rev. Phys. Chem.}\ }\textbf {\bibinfo {volume} {52}},\ \bibinfo {pages} {681}
  (\bibinfo {year} {2001})}\BibitemShut {NoStop}%
\bibitem [{\citenamefont {Nitzan}\ and\ \citenamefont
  {Ratner}(2003)}]{Nitzan2003}%
  \BibitemOpen
  \bibfield  {author} {\bibinfo {author} {\bibfnamefont {A.}~\bibnamefont
  {Nitzan}}\ and\ \bibinfo {author} {\bibfnamefont {M.~A.}\ \bibnamefont
  {Ratner}},\ }\href@noop {} {\bibfield  {journal} {\bibinfo  {journal}
  {Science}\ }\textbf {\bibinfo {volume} {300}},\ \bibinfo {pages} {1384}
  (\bibinfo {year} {2003})}\BibitemShut {NoStop}%
\bibitem [{\citenamefont {Cuniberti}\ \emph {et~al.}(2005)\citenamefont
  {Cuniberti}, \citenamefont {Fagas},\ and\ \citenamefont
  {Richter}}]{Cuniberti}%
  \BibitemOpen
  \bibfield  {author} {\bibinfo {author} {\bibfnamefont {G.}~\bibnamefont
  {Cuniberti}}, \bibinfo {author} {\bibfnamefont {G.}~\bibnamefont {Fagas}}, \
  and\ \bibinfo {author} {\bibfnamefont {K.}~\bibnamefont {Richter}},\
  }\href@noop {} {\emph {\bibinfo {title} {Introducing Molecular
  Electronics}}}\ (\bibinfo  {publisher} {Springer, Heidelberg},\ \bibinfo
  {year} {2005})\BibitemShut {NoStop}%
\bibitem [{\citenamefont {Galperin}\ \emph {et~al.}(2007)\citenamefont
  {Galperin}, \citenamefont {Ratner},\ and\ \citenamefont
  {Nitzan}}]{Galperin_Vib_Effects}%
  \BibitemOpen
  \bibfield  {author} {\bibinfo {author} {\bibfnamefont {M.}~\bibnamefont
  {Galperin}}, \bibinfo {author} {\bibfnamefont {M.~A.}\ \bibnamefont
  {Ratner}}, \ and\ \bibinfo {author} {\bibfnamefont {A.}~\bibnamefont
  {Nitzan}},\ }\href@noop {} {\bibfield  {journal} {\bibinfo  {journal} {J.
  Phys. Condens. Matter}\ }\textbf {\bibinfo {volume} {19}},\ \bibinfo {pages}
  {103201} (\bibinfo {year} {2007})}\BibitemShut {NoStop}%
\bibitem [{\citenamefont {Cuevas}\ and\ \citenamefont
  {Scheer}(2010)}]{Cuevas_Scheer}%
  \BibitemOpen
  \bibfield  {author} {\bibinfo {author} {\bibfnamefont {J.~C.}\ \bibnamefont
  {Cuevas}}\ and\ \bibinfo {author} {\bibfnamefont {E.}~\bibnamefont
  {Scheer}},\ }\href@noop {} {\emph {\bibinfo {title} {Molecular Electronics -
  An Introduction to Theory and Experiment}}}\ (\bibinfo  {publisher} {World
  Scientific},\ \bibinfo {year} {2010})\BibitemShut {NoStop}%
\bibitem [{\citenamefont {Zimbovskaya}\ and\ \citenamefont
  {Pederson}(2011)}]{Zimbovskaya2011}%
  \BibitemOpen
  \bibfield  {author} {\bibinfo {author} {\bibfnamefont {N.~A.}\ \bibnamefont
  {Zimbovskaya}}\ and\ \bibinfo {author} {\bibfnamefont {M.~R.}\ \bibnamefont
  {Pederson}},\ }\href@noop {} {\bibfield  {journal} {\bibinfo  {journal}
  {Phys. Reports}\ }\textbf {\bibinfo {volume} {509}},\ \bibinfo {pages} {1 }
  (\bibinfo {year} {2011})}\BibitemShut {NoStop}%
\bibitem [{\citenamefont {Bergfield}\ and\ \citenamefont
  {Ratner}(2013)}]{Bergfield2013}%
  \BibitemOpen
  \bibfield  {author} {\bibinfo {author} {\bibfnamefont {J.~P.}\ \bibnamefont
  {Bergfield}}\ and\ \bibinfo {author} {\bibfnamefont {M.~A.}\ \bibnamefont
  {Ratner}},\ }\href@noop {} {\bibfield  {journal} {\bibinfo  {journal} {Phys.
  Status Solidi B}\ }\textbf {\bibinfo {volume} {250}},\ \bibinfo {pages}
  {2249} (\bibinfo {year} {2013})}\BibitemShut {NoStop}%
\bibitem [{\citenamefont {Baldea}(2015)}]{Baldea}%
  \BibitemOpen
  \bibinfo {editor} {\bibfnamefont {I.}~\bibnamefont {Baldea}},\ ed.,\
  \href@noop {} {\emph {\bibinfo {title} {Molecular Electronics: A Theoretical
  and Experimental Approach}}}\ (\bibinfo  {publisher} {Pan Stanford,
  Singapore},\ \bibinfo {year} {2015})\BibitemShut {NoStop}%
\bibitem [{\citenamefont {Su}\ \emph {et~al.}(2016)\citenamefont {Su},
  \citenamefont {Neupane}, \citenamefont {Steigerwald}, \citenamefont
  {Venkataraman},\ and\ \citenamefont {Nuckolls}}]{Su2016}%
  \BibitemOpen
  \bibfield  {author} {\bibinfo {author} {\bibfnamefont {T.~A.}\ \bibnamefont
  {Su}}, \bibinfo {author} {\bibfnamefont {M.}~\bibnamefont {Neupane}},
  \bibinfo {author} {\bibfnamefont {M.~L.}\ \bibnamefont {Steigerwald}},
  \bibinfo {author} {\bibfnamefont {L.}~\bibnamefont {Venkataraman}}, \ and\
  \bibinfo {author} {\bibfnamefont {C.}~\bibnamefont {Nuckolls}},\ }\href@noop
  {} {\bibfield  {journal} {\bibinfo  {journal} {Nat. Rev. Mat.}\ }\textbf
  {\bibinfo {volume} {1}},\ \bibinfo {pages} {16002} (\bibinfo {year}
  {2016})}\BibitemShut {NoStop}%
\bibitem [{\citenamefont {Thoss}\ and\ \citenamefont
  {Evers}(2018)}]{Thoss2018}%
  \BibitemOpen
  \bibfield  {author} {\bibinfo {author} {\bibfnamefont {M.}~\bibnamefont
  {Thoss}}\ and\ \bibinfo {author} {\bibfnamefont {F.}~\bibnamefont {Evers}},\
  }\href@noop {} {\bibfield  {journal} {\bibinfo  {journal} {J. Chem. Phys.}\
  }\textbf {\bibinfo {volume} {148}},\ \bibinfo {pages} {030901} (\bibinfo
  {year} {2018})}\BibitemShut {NoStop}%
\bibitem [{\citenamefont {H\"artle}\ and\ \citenamefont
  {Thoss}(2015)}]{Rainer_Baldea}%
  \BibitemOpen
  \bibfield  {author} {\bibinfo {author} {\bibfnamefont {R.}~\bibnamefont
  {H\"artle}}\ and\ \bibinfo {author} {\bibfnamefont {M.}~\bibnamefont
  {Thoss}},\ }in\ \href@noop {} {\emph {\bibinfo {booktitle} {Molecular
  Electronics: A Theoretical and Experimental Approach}}},\ \bibinfo {editor}
  {edited by\ \bibinfo {editor} {\bibfnamefont {I.}~\bibnamefont {Baldea}}}\
  (\bibinfo  {publisher} {Pan Stanford, Singapore},\ \bibinfo {year}
  {2015})\BibitemShut {NoStop}%
\bibitem [{\citenamefont {Galperin}\ \emph {et~al.}(2005)\citenamefont
  {Galperin}, \citenamefont {Ratner},\ and\ \citenamefont
  {Nitzan}}]{Galperin2005}%
  \BibitemOpen
  \bibfield  {author} {\bibinfo {author} {\bibfnamefont {M.}~\bibnamefont
  {Galperin}}, \bibinfo {author} {\bibfnamefont {M.~A.}\ \bibnamefont
  {Ratner}}, \ and\ \bibinfo {author} {\bibfnamefont {A.}~\bibnamefont
  {Nitzan}},\ }\href@noop {} {\bibfield  {journal} {\bibinfo  {journal} {Nano
  Lett.}\ }\textbf {\bibinfo {volume} {5}},\ \bibinfo {pages} {125} (\bibinfo
  {year} {2005})}\BibitemShut {NoStop}%
\bibitem [{\citenamefont {Leijnse}\ and\ \citenamefont
  {Wegewijs}(2008)}]{Leijnse2008}%
  \BibitemOpen
  \bibfield  {author} {\bibinfo {author} {\bibfnamefont {M.}~\bibnamefont
  {Leijnse}}\ and\ \bibinfo {author} {\bibfnamefont {M.~R.}\ \bibnamefont
  {Wegewijs}},\ }\href@noop {} {\bibfield  {journal} {\bibinfo  {journal}
  {Phys. Rev. B}\ }\textbf {\bibinfo {volume} {78}},\ \bibinfo {pages} {235424}
  (\bibinfo {year} {2008})}\BibitemShut {NoStop}%
\bibitem [{\citenamefont {Ballmann}\ \emph {et~al.}(2012)\citenamefont
  {Ballmann}, \citenamefont {H\"artle}, \citenamefont {Coto}, \citenamefont
  {Elbing}, \citenamefont {Mayor}, \citenamefont {Bryce}, \citenamefont
  {Thoss},\ and\ \citenamefont {Weber}}]{Rainer2012}%
  \BibitemOpen
  \bibfield  {author} {\bibinfo {author} {\bibfnamefont {S.}~\bibnamefont
  {Ballmann}}, \bibinfo {author} {\bibfnamefont {R.}~\bibnamefont {H\"artle}},
  \bibinfo {author} {\bibfnamefont {P.~B.}\ \bibnamefont {Coto}}, \bibinfo
  {author} {\bibfnamefont {M.}~\bibnamefont {Elbing}}, \bibinfo {author}
  {\bibfnamefont {M.}~\bibnamefont {Mayor}}, \bibinfo {author} {\bibfnamefont
  {M.~R.}\ \bibnamefont {Bryce}}, \bibinfo {author} {\bibfnamefont
  {M.}~\bibnamefont {Thoss}}, \ and\ \bibinfo {author} {\bibfnamefont {H.~B.}\
  \bibnamefont {Weber}},\ }\href@noop {} {\bibfield  {journal} {\bibinfo
  {journal} {Phys. Rev. Lett.}\ }\textbf {\bibinfo {volume} {109}},\ \bibinfo
  {pages} {056801} (\bibinfo {year} {2012})}\BibitemShut {NoStop}%
\bibitem [{\citenamefont {H\"artle}\ and\ \citenamefont
  {Thoss}(2011{\natexlab{a}})}]{Rainer2011}%
  \BibitemOpen
  \bibfield  {author} {\bibinfo {author} {\bibfnamefont {R.}~\bibnamefont
  {H\"artle}}\ and\ \bibinfo {author} {\bibfnamefont {M.}~\bibnamefont
  {Thoss}},\ }\href@noop {} {\bibfield  {journal} {\bibinfo  {journal} {Phys.
  Rev. B}\ }\textbf {\bibinfo {volume} {83}},\ \bibinfo {pages} {115414}
  (\bibinfo {year} {2011}{\natexlab{a}})}\BibitemShut {NoStop}%
\bibitem [{\citenamefont {H\"artle}\ \emph
  {et~al.}(2013{\natexlab{a}})\citenamefont {H\"artle}, \citenamefont
  {Butzin},\ and\ \citenamefont {Thoss}}]{Rainer2013}%
  \BibitemOpen
  \bibfield  {author} {\bibinfo {author} {\bibfnamefont {R.}~\bibnamefont
  {H\"artle}}, \bibinfo {author} {\bibfnamefont {M.}~\bibnamefont {Butzin}}, \
  and\ \bibinfo {author} {\bibfnamefont {M.}~\bibnamefont {Thoss}},\
  }\href@noop {} {\bibfield  {journal} {\bibinfo  {journal} {Phys. Rev. B}\
  }\textbf {\bibinfo {volume} {87}},\ \bibinfo {pages} {085422} (\bibinfo
  {year} {2013}{\natexlab{a}})}\BibitemShut {NoStop}%
\bibitem [{\citenamefont {Wilner}\ \emph {et~al.}(2014)\citenamefont {Wilner},
  \citenamefont {Wang}, \citenamefont {Thoss},\ and\ \citenamefont
  {Rabani}}]{Wilner2014}%
  \BibitemOpen
  \bibfield  {author} {\bibinfo {author} {\bibfnamefont {E.~Y.}\ \bibnamefont
  {Wilner}}, \bibinfo {author} {\bibfnamefont {H.}~\bibnamefont {Wang}},
  \bibinfo {author} {\bibfnamefont {M.}~\bibnamefont {Thoss}}, \ and\ \bibinfo
  {author} {\bibfnamefont {E.}~\bibnamefont {Rabani}},\ }\href@noop {}
  {\bibfield  {journal} {\bibinfo  {journal} {Phys. Rev. B}\ }\textbf {\bibinfo
  {volume} {89}},\ \bibinfo {pages} {205129} (\bibinfo {year}
  {2014})}\BibitemShut {NoStop}%
\bibitem [{\citenamefont {Schinabeck}\ \emph {et~al.}(2014)\citenamefont
  {Schinabeck}, \citenamefont {H\"artle}, \citenamefont {Weber},\ and\
  \citenamefont {Thoss}}]{Schinabeck2014}%
  \BibitemOpen
  \bibfield  {author} {\bibinfo {author} {\bibfnamefont {C.}~\bibnamefont
  {Schinabeck}}, \bibinfo {author} {\bibfnamefont {R.}~\bibnamefont
  {H\"artle}}, \bibinfo {author} {\bibfnamefont {H.~B.}\ \bibnamefont {Weber}},
  \ and\ \bibinfo {author} {\bibfnamefont {M.}~\bibnamefont {Thoss}},\
  }\href@noop {} {\bibfield  {journal} {\bibinfo  {journal} {Phys. Rev. B}\
  }\textbf {\bibinfo {volume} {90}},\ \bibinfo {pages} {075409} (\bibinfo
  {year} {2014})}\BibitemShut {NoStop}%
\bibitem [{\citenamefont {Erpenbeck}\ \emph {et~al.}(2015)\citenamefont
  {Erpenbeck}, \citenamefont {H\"artle},\ and\ \citenamefont
  {Thoss}}]{Erpenbeck2015}%
  \BibitemOpen
  \bibfield  {author} {\bibinfo {author} {\bibfnamefont {A.}~\bibnamefont
  {Erpenbeck}}, \bibinfo {author} {\bibfnamefont {R.}~\bibnamefont {H\"artle}},
  \ and\ \bibinfo {author} {\bibfnamefont {M.}~\bibnamefont {Thoss}},\
  }\href@noop {} {\bibfield  {journal} {\bibinfo  {journal} {Phys. Rev. B}\
  }\textbf {\bibinfo {volume} {91}},\ \bibinfo {pages} {195418} (\bibinfo
  {year} {2015})}\BibitemShut {NoStop}%
\bibitem [{\citenamefont {Foti}\ and\ \citenamefont
  {Vazquez}(2018)}]{foti2018origin}%
  \BibitemOpen
  \bibfield  {author} {\bibinfo {author} {\bibfnamefont {G.}~\bibnamefont
  {Foti}}\ and\ \bibinfo {author} {\bibfnamefont {H.}~\bibnamefont {Vazquez}},\
  }\href@noop {} {\bibfield  {journal} {\bibinfo  {journal} {J. Phys. Chem.
  Lett.}\ }\textbf {\bibinfo {volume} {9}},\ \bibinfo {pages} {2791} (\bibinfo
  {year} {2018})}\BibitemShut {NoStop}%
\bibitem [{\citenamefont {Ioffe}\ \emph {et~al.}(2008)\citenamefont {Ioffe},
  \citenamefont {Shamai}, \citenamefont {Ophir}, \citenamefont {Noy},
  \citenamefont {Yutsis}, \citenamefont {Kfir}, \citenamefont {Cheshnovsky},\
  and\ \citenamefont {Selzer}}]{Ioffe2008}%
  \BibitemOpen
  \bibfield  {author} {\bibinfo {author} {\bibfnamefont {Z.}~\bibnamefont
  {Ioffe}}, \bibinfo {author} {\bibfnamefont {T.}~\bibnamefont {Shamai}},
  \bibinfo {author} {\bibfnamefont {A.}~\bibnamefont {Ophir}}, \bibinfo
  {author} {\bibfnamefont {G.}~\bibnamefont {Noy}}, \bibinfo {author}
  {\bibfnamefont {I.}~\bibnamefont {Yutsis}}, \bibinfo {author} {\bibfnamefont
  {K.}~\bibnamefont {Kfir}}, \bibinfo {author} {\bibfnamefont {O.}~\bibnamefont
  {Cheshnovsky}}, \ and\ \bibinfo {author} {\bibfnamefont {Y.}~\bibnamefont
  {Selzer}},\ }\href@noop {} {\bibfield  {journal} {\bibinfo  {journal} {Nat.\
  Nano}\ }\textbf {\bibinfo {volume} {3}},\ \bibinfo {pages} {727} (\bibinfo
  {year} {2008})}\BibitemShut {NoStop}%
\bibitem [{\citenamefont {Schulze}\ \emph
  {et~al.}(2008{\natexlab{a}})\citenamefont {Schulze}, \citenamefont {Franke},
  \citenamefont {Gagliardi}, \citenamefont {Romano}, \citenamefont {Lin},
  \citenamefont {Rosa}, \citenamefont {Niehaus}, \citenamefont {Frauenheim},
  \citenamefont {Di~Carlo}, \citenamefont {Pecchia},\ and\ \citenamefont
  {Pascual}}]{Schulze2008}%
  \BibitemOpen
  \bibfield  {author} {\bibinfo {author} {\bibfnamefont {G.}~\bibnamefont
  {Schulze}}, \bibinfo {author} {\bibfnamefont {K.~J.}\ \bibnamefont {Franke}},
  \bibinfo {author} {\bibfnamefont {A.}~\bibnamefont {Gagliardi}}, \bibinfo
  {author} {\bibfnamefont {G.}~\bibnamefont {Romano}}, \bibinfo {author}
  {\bibfnamefont {C.~S.}\ \bibnamefont {Lin}}, \bibinfo {author} {\bibfnamefont
  {A.~L.}\ \bibnamefont {Rosa}}, \bibinfo {author} {\bibfnamefont {T.~A.}\
  \bibnamefont {Niehaus}}, \bibinfo {author} {\bibfnamefont {T.}~\bibnamefont
  {Frauenheim}}, \bibinfo {author} {\bibfnamefont {A.}~\bibnamefont
  {Di~Carlo}}, \bibinfo {author} {\bibfnamefont {A.}~\bibnamefont {Pecchia}}, \
  and\ \bibinfo {author} {\bibfnamefont {J.~I.}\ \bibnamefont {Pascual}},\
  }\href@noop {} {\bibfield  {journal} {\bibinfo  {journal} {Phys. Rev. Lett.}\
  }\textbf {\bibinfo {volume} {100}},\ \bibinfo {pages} {136801} (\bibinfo
  {year} {2008}{\natexlab{a}})}\BibitemShut {NoStop}%
\bibitem [{\citenamefont {Schulze}\ \emph
  {et~al.}(2008{\natexlab{b}})\citenamefont {Schulze}, \citenamefont {Franke},\
  and\ \citenamefont {Pascual}}]{Schulze2008_2}%
  \BibitemOpen
  \bibfield  {author} {\bibinfo {author} {\bibfnamefont {G.}~\bibnamefont
  {Schulze}}, \bibinfo {author} {\bibfnamefont {K.~J.}\ \bibnamefont {Franke}},
  \ and\ \bibinfo {author} {\bibfnamefont {J.~I.}\ \bibnamefont {Pascual}},\
  }\href@noop {} {\bibfield  {journal} {\bibinfo  {journal} {New J. Phys.}\
  }\textbf {\bibinfo {volume} {10}},\ \bibinfo {pages} {065005} (\bibinfo
  {year} {2008}{\natexlab{b}})}\BibitemShut {NoStop}%
\bibitem [{\citenamefont {de~Leon}\ \emph {et~al.}(2008)\citenamefont
  {de~Leon}, \citenamefont {Liang}, \citenamefont {Gu},\ and\ \citenamefont
  {Park}}]{deLeon2008}%
  \BibitemOpen
  \bibfield  {author} {\bibinfo {author} {\bibfnamefont {N.~P.}\ \bibnamefont
  {de~Leon}}, \bibinfo {author} {\bibfnamefont {W.}~\bibnamefont {Liang}},
  \bibinfo {author} {\bibfnamefont {Q.}~\bibnamefont {Gu}}, \ and\ \bibinfo
  {author} {\bibfnamefont {H.}~\bibnamefont {Park}},\ }\href@noop {} {\bibfield
   {journal} {\bibinfo  {journal} {Nano Lett.}\ }\textbf {\bibinfo {volume}
  {8}},\ \bibinfo {pages} {2963} (\bibinfo {year} {2008})}\BibitemShut
  {NoStop}%
\bibitem [{\citenamefont {H\"uttel}\ \emph {et~al.}(2009)\citenamefont
  {H\"uttel}, \citenamefont {Witkamp}, \citenamefont {Leijnse}, \citenamefont
  {Wegewijs},\ and\ \citenamefont {van~der Zant}}]{Huettel2009}%
  \BibitemOpen
  \bibfield  {author} {\bibinfo {author} {\bibfnamefont {A.~K.}\ \bibnamefont
  {H\"uttel}}, \bibinfo {author} {\bibfnamefont {B.}~\bibnamefont {Witkamp}},
  \bibinfo {author} {\bibfnamefont {M.}~\bibnamefont {Leijnse}}, \bibinfo
  {author} {\bibfnamefont {M.~R.}\ \bibnamefont {Wegewijs}}, \ and\ \bibinfo
  {author} {\bibfnamefont {H.~S.~J.}\ \bibnamefont {van~der Zant}},\
  }\href@noop {} {\bibfield  {journal} {\bibinfo  {journal} {Phys. Rev. Lett.}\
  }\textbf {\bibinfo {volume} {102}},\ \bibinfo {pages} {225501} (\bibinfo
  {year} {2009})}\BibitemShut {NoStop}%
\bibitem [{\citenamefont {H\"artle}\ \emph {et~al.}(2009)\citenamefont
  {H\"artle}, \citenamefont {Benesch},\ and\ \citenamefont
  {Thoss}}]{Rainer2009}%
  \BibitemOpen
  \bibfield  {author} {\bibinfo {author} {\bibfnamefont {R.}~\bibnamefont
  {H\"artle}}, \bibinfo {author} {\bibfnamefont {C.}~\bibnamefont {Benesch}}, \
  and\ \bibinfo {author} {\bibfnamefont {M.}~\bibnamefont {Thoss}},\
  }\href@noop {} {\bibfield  {journal} {\bibinfo  {journal} {Phys. Rev. Lett.}\
  }\textbf {\bibinfo {volume} {102}},\ \bibinfo {pages} {146801} (\bibinfo
  {year} {2009})}\BibitemShut {NoStop}%
\bibitem [{\citenamefont {H\"artle}\ \emph {et~al.}(2010)\citenamefont
  {H\"artle}, \citenamefont {Volkovich}, \citenamefont {Thoss},\ and\
  \citenamefont {Peskin}}]{Rainer2010}%
  \BibitemOpen
  \bibfield  {author} {\bibinfo {author} {\bibfnamefont {R.}~\bibnamefont
  {H\"artle}}, \bibinfo {author} {\bibfnamefont {R.}~\bibnamefont {Volkovich}},
  \bibinfo {author} {\bibfnamefont {M.}~\bibnamefont {Thoss}}, \ and\ \bibinfo
  {author} {\bibfnamefont {U.}~\bibnamefont {Peskin}},\ }\href@noop {}
  {\bibfield  {journal} {\bibinfo  {journal} {J. Chem. Phys.}\ }\textbf
  {\bibinfo {volume} {133}},\ \bibinfo {pages} {081102} (\bibinfo {year}
  {2010})}\BibitemShut {NoStop}%
\bibitem [{\citenamefont {Ward}\ \emph {et~al.}(2010)\citenamefont {Ward},
  \citenamefont {Corley}, \citenamefont {Tour},\ and\ \citenamefont
  {Natelson}}]{Ward2010}%
  \BibitemOpen
  \bibfield  {author} {\bibinfo {author} {\bibfnamefont {D.~R.}\ \bibnamefont
  {Ward}}, \bibinfo {author} {\bibfnamefont {D.~A.}\ \bibnamefont {Corley}},
  \bibinfo {author} {\bibfnamefont {J.~M.}\ \bibnamefont {Tour}}, \ and\
  \bibinfo {author} {\bibfnamefont {D.}~\bibnamefont {Natelson}},\ }\href@noop
  {} {\bibfield  {journal} {\bibinfo  {journal} {Nat.\ Nano}\ }\textbf
  {\bibinfo {volume} {6}},\ \bibinfo {pages} {33} (\bibinfo {year}
  {2010})}\BibitemShut {NoStop}%
\bibitem [{\citenamefont {H\"artle}\ and\ \citenamefont
  {Thoss}(2011{\natexlab{b}})}]{Rainer2011b}%
  \BibitemOpen
  \bibfield  {author} {\bibinfo {author} {\bibfnamefont {R.}~\bibnamefont
  {H\"artle}}\ and\ \bibinfo {author} {\bibfnamefont {M.}~\bibnamefont
  {Thoss}},\ }\href@noop {} {\bibfield  {journal} {\bibinfo  {journal} {Phys.
  Rev. B}\ }\textbf {\bibinfo {volume} {83}},\ \bibinfo {pages} {125419}
  (\bibinfo {year} {2011}{\natexlab{b}})}\BibitemShut {NoStop}%
\bibitem [{\citenamefont {Franke}\ and\ \citenamefont
  {Pascual}(2012)}]{Franke2012}%
  \BibitemOpen
  \bibfield  {author} {\bibinfo {author} {\bibfnamefont {K.~J.}\ \bibnamefont
  {Franke}}\ and\ \bibinfo {author} {\bibfnamefont {J.~I.}\ \bibnamefont
  {Pascual}},\ }\href@noop {} {\bibfield  {journal} {\bibinfo  {journal} {J.
  Phys. Condens. Matter}\ }\textbf {\bibinfo {volume} {24}},\ \bibinfo {pages}
  {394002} (\bibinfo {year} {2012})}\BibitemShut {NoStop}%
\bibitem [{\citenamefont {Schinabeck}\ \emph {et~al.}(2016)\citenamefont
  {Schinabeck}, \citenamefont {Erpenbeck}, \citenamefont {H\"artle},\ and\
  \citenamefont {Thoss}}]{Schinabeck2016}%
  \BibitemOpen
  \bibfield  {author} {\bibinfo {author} {\bibfnamefont {C.}~\bibnamefont
  {Schinabeck}}, \bibinfo {author} {\bibfnamefont {A.}~\bibnamefont
  {Erpenbeck}}, \bibinfo {author} {\bibfnamefont {R.}~\bibnamefont {H\"artle}},
  \ and\ \bibinfo {author} {\bibfnamefont {M.}~\bibnamefont {Thoss}},\
  }\href@noop {} {\bibfield  {journal} {\bibinfo  {journal} {Phys. Rev. B}\
  }\textbf {\bibinfo {volume} {94}},\ \bibinfo {pages} {201407} (\bibinfo
  {year} {2016})}\BibitemShut {NoStop}%
\bibitem [{\citenamefont {Sabater}\ \emph {et~al.}(2015)\citenamefont
  {Sabater}, \citenamefont {Untiedt},\ and\ \citenamefont {van
  Ruitenbeek}}]{Sabater2015}%
  \BibitemOpen
  \bibfield  {author} {\bibinfo {author} {\bibfnamefont {C.}~\bibnamefont
  {Sabater}}, \bibinfo {author} {\bibfnamefont {C.}~\bibnamefont {Untiedt}}, \
  and\ \bibinfo {author} {\bibfnamefont {J.~M.}\ \bibnamefont {van
  Ruitenbeek}},\ }\href@noop {} {\bibfield  {journal} {\bibinfo  {journal}
  {Beilstein J. Nanotechnol.}\ }\textbf {\bibinfo {volume} {6}},\ \bibinfo
  {pages} {2338} (\bibinfo {year} {2015})}\BibitemShut {NoStop}%
\bibitem [{\citenamefont {Li}\ \emph {et~al.}(2015)\citenamefont {Li},
  \citenamefont {Su}, \citenamefont {Zhang}, \citenamefont {Steigerwald},
  \citenamefont {Nuckolls},\ and\ \citenamefont
  {Venkataraman}}]{Venkataraman2015}%
  \BibitemOpen
  \bibfield  {author} {\bibinfo {author} {\bibfnamefont {H.}~\bibnamefont
  {Li}}, \bibinfo {author} {\bibfnamefont {T.~A.}\ \bibnamefont {Su}}, \bibinfo
  {author} {\bibfnamefont {V.}~\bibnamefont {Zhang}}, \bibinfo {author}
  {\bibfnamefont {M.~L.}\ \bibnamefont {Steigerwald}}, \bibinfo {author}
  {\bibfnamefont {C.}~\bibnamefont {Nuckolls}}, \ and\ \bibinfo {author}
  {\bibfnamefont {L.}~\bibnamefont {Venkataraman}},\ }\href@noop {} {\bibfield
  {journal} {\bibinfo  {journal} {J. Am. Chem. Soc.}\ }\textbf {\bibinfo
  {volume} {137}},\ \bibinfo {pages} {5028} (\bibinfo {year}
  {2015})}\BibitemShut {NoStop}%
\bibitem [{\citenamefont {Li}\ \emph {et~al.}(2016)\citenamefont {Li},
  \citenamefont {Kim}, \citenamefont {Su}, \citenamefont {Steigerwald},
  \citenamefont {Nuckolls}, \citenamefont {Darancet}, \citenamefont
  {Leighton},\ and\ \citenamefont {Venkataraman}}]{Venkataraman2016}%
  \BibitemOpen
  \bibfield  {author} {\bibinfo {author} {\bibfnamefont {H.}~\bibnamefont
  {Li}}, \bibinfo {author} {\bibfnamefont {N.~T.}\ \bibnamefont {Kim}},
  \bibinfo {author} {\bibfnamefont {T.~A.}\ \bibnamefont {Su}}, \bibinfo
  {author} {\bibfnamefont {M.~L.}\ \bibnamefont {Steigerwald}}, \bibinfo
  {author} {\bibfnamefont {C.}~\bibnamefont {Nuckolls}}, \bibinfo {author}
  {\bibfnamefont {P.}~\bibnamefont {Darancet}}, \bibinfo {author}
  {\bibfnamefont {J.~L.}\ \bibnamefont {Leighton}}, \ and\ \bibinfo {author}
  {\bibfnamefont {L.}~\bibnamefont {Venkataraman}},\ }\href@noop {} {\bibfield
  {journal} {\bibinfo  {journal} {J. Am. Chem. Soc.}\ }\textbf {\bibinfo
  {volume} {138}},\ \bibinfo {pages} {16159} (\bibinfo {year}
  {2016})}\BibitemShut {NoStop}%
\bibitem [{\citenamefont {Capozzi}\ \emph {et~al.}(2016)\citenamefont
  {Capozzi}, \citenamefont {Low}, \citenamefont {Xia}, \citenamefont {Liu},
  \citenamefont {Neaton}, \citenamefont {Campos},\ and\ \citenamefont
  {Venkataraman}}]{Capozzi2016}%
  \BibitemOpen
  \bibfield  {author} {\bibinfo {author} {\bibfnamefont {B.}~\bibnamefont
  {Capozzi}}, \bibinfo {author} {\bibfnamefont {J.~Z.}\ \bibnamefont {Low}},
  \bibinfo {author} {\bibfnamefont {J.}~\bibnamefont {Xia}}, \bibinfo {author}
  {\bibfnamefont {Z.-F.}\ \bibnamefont {Liu}}, \bibinfo {author} {\bibfnamefont
  {J.~B.}\ \bibnamefont {Neaton}}, \bibinfo {author} {\bibfnamefont {L.~M.}\
  \bibnamefont {Campos}}, \ and\ \bibinfo {author} {\bibfnamefont
  {L.}~\bibnamefont {Venkataraman}},\ }\href@noop {} {\bibfield  {journal}
  {\bibinfo  {journal} {Nano Lett.}\ }\textbf {\bibinfo {volume} {16}},\
  \bibinfo {pages} {3949} (\bibinfo {year} {2016})}\BibitemShut {NoStop}%
\bibitem [{\citenamefont {Fung}\ \emph {et~al.}(2019)\citenamefont {Fung},
  \citenamefont {Gelbwaser}, \citenamefont {Taylor}, \citenamefont {Low},
  \citenamefont {Xia}, \citenamefont {Davydenko}, \citenamefont {Campos},
  \citenamefont {Marder}, \citenamefont {Peskin},\ and\ \citenamefont
  {Venkataraman}}]{Fung2019}%
  \BibitemOpen
  \bibfield  {author} {\bibinfo {author} {\bibfnamefont {E.-D.}\ \bibnamefont
  {Fung}}, \bibinfo {author} {\bibfnamefont {D.}~\bibnamefont {Gelbwaser}},
  \bibinfo {author} {\bibfnamefont {J.}~\bibnamefont {Taylor}}, \bibinfo
  {author} {\bibfnamefont {J.}~\bibnamefont {Low}}, \bibinfo {author}
  {\bibfnamefont {J.}~\bibnamefont {Xia}}, \bibinfo {author} {\bibfnamefont
  {I.}~\bibnamefont {Davydenko}}, \bibinfo {author} {\bibfnamefont {L.~M.}\
  \bibnamefont {Campos}}, \bibinfo {author} {\bibfnamefont {S.}~\bibnamefont
  {Marder}}, \bibinfo {author} {\bibfnamefont {U.}~\bibnamefont {Peskin}}, \
  and\ \bibinfo {author} {\bibfnamefont {L.}~\bibnamefont {Venkataraman}},\
  }\href@noop {} {\bibfield  {journal} {\bibinfo  {journal} {Nano Lett.}\
  }\textbf {\bibinfo {volume} {19}},\ \bibinfo {pages} {2555} (\bibinfo {year}
  {2019})}\BibitemShut {NoStop}%
\bibitem [{\citenamefont {Wegewijs}\ and\ \citenamefont
  {Nowack}(2005)}]{Wegewijs2005}%
  \BibitemOpen
  \bibfield  {author} {\bibinfo {author} {\bibfnamefont {M.~R.}\ \bibnamefont
  {Wegewijs}}\ and\ \bibinfo {author} {\bibfnamefont {K.~C.}\ \bibnamefont
  {Nowack}},\ }\href@noop {} {\bibfield  {journal} {\bibinfo  {journal} {New J.
  Phys.}\ }\textbf {\bibinfo {volume} {7}},\ \bibinfo {pages} {239} (\bibinfo
  {year} {2005})}\BibitemShut {NoStop}%
\bibitem [{\citenamefont {Ryndyk}\ \emph {et~al.}(2006)\citenamefont {Ryndyk},
  \citenamefont {Hartung},\ and\ \citenamefont {Cuniberti}}]{Ryndyk2006}%
  \BibitemOpen
  \bibfield  {author} {\bibinfo {author} {\bibfnamefont {D.~A.}\ \bibnamefont
  {Ryndyk}}, \bibinfo {author} {\bibfnamefont {M.}~\bibnamefont {Hartung}}, \
  and\ \bibinfo {author} {\bibfnamefont {G.}~\bibnamefont {Cuniberti}},\
  }\href@noop {} {\bibfield  {journal} {\bibinfo  {journal} {Phys. Rev. B}\
  }\textbf {\bibinfo {volume} {73}},\ \bibinfo {pages} {045420} (\bibinfo
  {year} {2006})}\BibitemShut {NoStop}%
\bibitem [{\citenamefont {Benesch}\ \emph {et~al.}(2008)\citenamefont
  {Benesch}, \citenamefont {Čížek}, \citenamefont {Klimeš}, \citenamefont
  {Kondov}, \citenamefont {Thoss},\ and\ \citenamefont {Domcke}}]{Benesch2008}%
  \BibitemOpen
  \bibfield  {author} {\bibinfo {author} {\bibfnamefont {C.}~\bibnamefont
  {Benesch}}, \bibinfo {author} {\bibfnamefont {M.}~\bibnamefont {Čížek}},
  \bibinfo {author} {\bibfnamefont {J.}~\bibnamefont {Klimeš}}, \bibinfo
  {author} {\bibfnamefont {I.}~\bibnamefont {Kondov}}, \bibinfo {author}
  {\bibfnamefont {M.}~\bibnamefont {Thoss}}, \ and\ \bibinfo {author}
  {\bibfnamefont {W.}~\bibnamefont {Domcke}},\ }\href@noop {} {\bibfield
  {journal} {\bibinfo  {journal} {J. Phys. Chem. C}\ }\textbf {\bibinfo
  {volume} {112}},\ \bibinfo {pages} {9880} (\bibinfo {year}
  {2008})}\BibitemShut {NoStop}%
\bibitem [{\citenamefont {Erpenbeck}\ \emph {et~al.}(2016)\citenamefont
  {Erpenbeck}, \citenamefont {H\"artle}, \citenamefont {Bockstedte},\ and\
  \citenamefont {Thoss}}]{Erpenbeck2016}%
  \BibitemOpen
  \bibfield  {author} {\bibinfo {author} {\bibfnamefont {A.}~\bibnamefont
  {Erpenbeck}}, \bibinfo {author} {\bibfnamefont {R.}~\bibnamefont {H\"artle}},
  \bibinfo {author} {\bibfnamefont {M.}~\bibnamefont {Bockstedte}}, \ and\
  \bibinfo {author} {\bibfnamefont {M.}~\bibnamefont {Thoss}},\ }\href@noop {}
  {\bibfield  {journal} {\bibinfo  {journal} {Phys. Rev. B}\ }\textbf {\bibinfo
  {volume} {93}},\ \bibinfo {pages} {115421} (\bibinfo {year}
  {2016})}\BibitemShut {NoStop}%
\bibitem [{\citenamefont {Koch}\ \emph {et~al.}(2006)\citenamefont {Koch},
  \citenamefont {Semmelhack}, \citenamefont {von Oppen},\ and\ \citenamefont
  {Nitzan}}]{Koch2006}%
  \BibitemOpen
  \bibfield  {author} {\bibinfo {author} {\bibfnamefont {J.}~\bibnamefont
  {Koch}}, \bibinfo {author} {\bibfnamefont {M.}~\bibnamefont {Semmelhack}},
  \bibinfo {author} {\bibfnamefont {F.}~\bibnamefont {von Oppen}}, \ and\
  \bibinfo {author} {\bibfnamefont {A.}~\bibnamefont {Nitzan}},\ }\href@noop {}
  {\bibfield  {journal} {\bibinfo  {journal} {Phys. Rev. B}\ }\textbf {\bibinfo
  {volume} {73}},\ \bibinfo {pages} {155306} (\bibinfo {year}
  {2006})}\BibitemShut {NoStop}%
\bibitem [{\citenamefont {Volkovich}\ \emph {et~al.}(2011)\citenamefont
  {Volkovich}, \citenamefont {Hartle}, \citenamefont {Thoss},\ and\
  \citenamefont {Peskin}}]{Volkovich2011}%
  \BibitemOpen
  \bibfield  {author} {\bibinfo {author} {\bibfnamefont {R.}~\bibnamefont
  {Volkovich}}, \bibinfo {author} {\bibfnamefont {R.}~\bibnamefont {Hartle}},
  \bibinfo {author} {\bibfnamefont {M.}~\bibnamefont {Thoss}}, \ and\ \bibinfo
  {author} {\bibfnamefont {U.}~\bibnamefont {Peskin}},\ }\href@noop {}
  {\bibfield  {journal} {\bibinfo  {journal} {Phys. Chem. Chem. Phys.}\
  }\textbf {\bibinfo {volume} {13}},\ \bibinfo {pages} {14333} (\bibinfo {year}
  {2011})}\BibitemShut {NoStop}%
\bibitem [{\citenamefont {H\"artle}\ and\ \citenamefont
  {Kulkarni}(2015)}]{Rainer2015}%
  \BibitemOpen
  \bibfield  {author} {\bibinfo {author} {\bibfnamefont {R.}~\bibnamefont
  {H\"artle}}\ and\ \bibinfo {author} {\bibfnamefont {M.}~\bibnamefont
  {Kulkarni}},\ }\href@noop {} {\bibfield  {journal} {\bibinfo  {journal}
  {Phys. Rev. B}\ }\textbf {\bibinfo {volume} {91}},\ \bibinfo {pages} {245429}
  (\bibinfo {year} {2015})}\BibitemShut {NoStop}%
\bibitem [{\citenamefont {Dzhioev}\ and\ \citenamefont
  {Kosov}(2011)}]{Dzhioev2011}%
  \BibitemOpen
  \bibfield  {author} {\bibinfo {author} {\bibfnamefont {A.~A.}\ \bibnamefont
  {Dzhioev}}\ and\ \bibinfo {author} {\bibfnamefont {D.~S.}\ \bibnamefont
  {Kosov}},\ }\href@noop {} {\bibfield  {journal} {\bibinfo  {journal} {J.
  Chem. Phys.}\ }\textbf {\bibinfo {volume} {135}},\ \bibinfo {pages} {074701}
  (\bibinfo {year} {2011})}\BibitemShut {NoStop}%
\bibitem [{\citenamefont {Dzhioev}\ \emph {et~al.}(2013)\citenamefont
  {Dzhioev}, \citenamefont {Kosov},\ and\ \citenamefont {von
  Oppen}}]{Dzhioev2013}%
  \BibitemOpen
  \bibfield  {author} {\bibinfo {author} {\bibfnamefont {A.~A.}\ \bibnamefont
  {Dzhioev}}, \bibinfo {author} {\bibfnamefont {D.~S.}\ \bibnamefont {Kosov}},
  \ and\ \bibinfo {author} {\bibfnamefont {F.}~\bibnamefont {von Oppen}},\
  }\href@noop {} {\bibfield  {journal} {\bibinfo  {journal} {J. Chem. Phys.}\
  }\textbf {\bibinfo {volume} {138}},\ \bibinfo {pages} {134103} (\bibinfo
  {year} {2013})}\BibitemShut {NoStop}%
\bibitem [{\citenamefont {Pozner}\ \emph {et~al.}(2014)\citenamefont {Pozner},
  \citenamefont {Lifshitz},\ and\ \citenamefont {Peskin}}]{Pozner2014}%
  \BibitemOpen
  \bibfield  {author} {\bibinfo {author} {\bibfnamefont {R.}~\bibnamefont
  {Pozner}}, \bibinfo {author} {\bibfnamefont {E.}~\bibnamefont {Lifshitz}}, \
  and\ \bibinfo {author} {\bibfnamefont {U.}~\bibnamefont {Peskin}},\
  }\href@noop {} {\bibfield  {journal} {\bibinfo  {journal} {Nano Lett.}\
  }\textbf {\bibinfo {volume} {14}},\ \bibinfo {pages} {6244} (\bibinfo {year}
  {2014})}\BibitemShut {NoStop}%
\bibitem [{\citenamefont {Erpenbeck}\ \emph
  {et~al.}(2018{\natexlab{a}})\citenamefont {Erpenbeck}, \citenamefont
  {Schinabeck}, \citenamefont {Peskin},\ and\ \citenamefont
  {Thoss}}]{Erpenbeck_dissociation_2018}%
  \BibitemOpen
  \bibfield  {author} {\bibinfo {author} {\bibfnamefont {A.}~\bibnamefont
  {Erpenbeck}}, \bibinfo {author} {\bibfnamefont {C.}~\bibnamefont
  {Schinabeck}}, \bibinfo {author} {\bibfnamefont {U.}~\bibnamefont {Peskin}},
  \ and\ \bibinfo {author} {\bibfnamefont {M.}~\bibnamefont {Thoss}},\
  }\href@noop {} {\bibfield  {journal} {\bibinfo  {journal} {Phys. Rev. B}\
  }\textbf {\bibinfo {volume} {97}},\ \bibinfo {pages} {235452} (\bibinfo
  {year} {2018}{\natexlab{a}})}\BibitemShut {NoStop}%
\bibitem [{\citenamefont {Brisker}\ and\ \citenamefont
  {Peskin}(2008)}]{Brisker2008}%
  \BibitemOpen
  \bibfield  {author} {\bibinfo {author} {\bibfnamefont {D.}~\bibnamefont
  {Brisker}}\ and\ \bibinfo {author} {\bibfnamefont {U.}~\bibnamefont
  {Peskin}},\ }\href@noop {} {\bibfield  {journal} {\bibinfo  {journal} {J.
  Chem. Phys.}\ }\textbf {\bibinfo {volume} {129}},\ \bibinfo {pages} {244709}
  (\bibinfo {year} {2008})}\BibitemShut {NoStop}%
\bibitem [{\citenamefont {Gao}\ \emph {et~al.}(1995{\natexlab{a}})\citenamefont
  {Gao}, \citenamefont {Busch},\ and\ \citenamefont {Ho}}]{Gao1995_1}%
  \BibitemOpen
  \bibfield  {author} {\bibinfo {author} {\bibfnamefont {S.}~\bibnamefont
  {Gao}}, \bibinfo {author} {\bibfnamefont {D.}~\bibnamefont {Busch}}, \ and\
  \bibinfo {author} {\bibfnamefont {W.}~\bibnamefont {Ho}},\ }\href@noop {}
  {\bibfield  {journal} {\bibinfo  {journal} {Surf. Sci.}\ }\textbf {\bibinfo
  {volume} {344}},\ \bibinfo {pages} {L1252} (\bibinfo {year}
  {1995}{\natexlab{a}})}\BibitemShut {NoStop}%
\bibitem [{\citenamefont {Gao}\ \emph {et~al.}(1995{\natexlab{b}})\citenamefont
  {Gao}, \citenamefont {Lundqvist},\ and\ \citenamefont {Ho}}]{Gao1995_2}%
  \BibitemOpen
  \bibfield  {author} {\bibinfo {author} {\bibfnamefont {S.}~\bibnamefont
  {Gao}}, \bibinfo {author} {\bibfnamefont {B.}~\bibnamefont {Lundqvist}}, \
  and\ \bibinfo {author} {\bibfnamefont {W.}~\bibnamefont {Ho}},\ }\href@noop
  {} {\bibfield  {journal} {\bibinfo  {journal} {Surf. Sci.}\ }\textbf
  {\bibinfo {volume} {341}},\ \bibinfo {pages} {L1031 } (\bibinfo {year}
  {1995}{\natexlab{b}})}\BibitemShut {NoStop}%
\bibitem [{\citenamefont {Avouris}\ \emph {et~al.}(1996)\citenamefont
  {Avouris}, \citenamefont {Walkup}, \citenamefont {Rossi}, \citenamefont
  {Shen}, \citenamefont {Abeln}, \citenamefont {Tucker},\ and\ \citenamefont
  {Lyding}}]{Avouris1996}%
  \BibitemOpen
  \bibfield  {author} {\bibinfo {author} {\bibfnamefont {P.}~\bibnamefont
  {Avouris}}, \bibinfo {author} {\bibfnamefont {R.}~\bibnamefont {Walkup}},
  \bibinfo {author} {\bibfnamefont {A.}~\bibnamefont {Rossi}}, \bibinfo
  {author} {\bibfnamefont {T.-C.}\ \bibnamefont {Shen}}, \bibinfo {author}
  {\bibfnamefont {G.}~\bibnamefont {Abeln}}, \bibinfo {author} {\bibfnamefont
  {J.}~\bibnamefont {Tucker}}, \ and\ \bibinfo {author} {\bibfnamefont
  {J.}~\bibnamefont {Lyding}},\ }\href@noop {} {\bibfield  {journal} {\bibinfo
  {journal} {Chem. Phys. Lett.}\ }\textbf {\bibinfo {volume} {257}},\ \bibinfo
  {pages} {148 } (\bibinfo {year} {1996})}\BibitemShut {NoStop}%
\bibitem [{\citenamefont {Gao}(1997)}]{Gao1997}%
  \BibitemOpen
  \bibfield  {author} {\bibinfo {author} {\bibfnamefont {S.}~\bibnamefont
  {Gao}},\ }\href@noop {} {\bibfield  {journal} {\bibinfo  {journal} {Phys.
  Rev. B}\ }\textbf {\bibinfo {volume} {55}},\ \bibinfo {pages} {1876}
  (\bibinfo {year} {1997})}\BibitemShut {NoStop}%
\bibitem [{\citenamefont {Boendgen}\ and\ \citenamefont
  {Saalfrank}(1998)}]{Boendgen1998}%
  \BibitemOpen
  \bibfield  {author} {\bibinfo {author} {\bibfnamefont {G.}~\bibnamefont
  {Boendgen}}\ and\ \bibinfo {author} {\bibfnamefont {P.}~\bibnamefont
  {Saalfrank}},\ }\href@noop {} {\bibfield  {journal} {\bibinfo  {journal} {J.
  Phys. Chem. B}\ }\textbf {\bibinfo {volume} {102}},\ \bibinfo {pages} {8029}
  (\bibinfo {year} {1998})}\BibitemShut {NoStop}%
\bibitem [{\citenamefont {Seideman}(2003)}]{Seideman2003}%
  \BibitemOpen
  \bibfield  {author} {\bibinfo {author} {\bibfnamefont {T.}~\bibnamefont
  {Seideman}},\ }\href@noop {} {\bibfield  {journal} {\bibinfo  {journal} {J.
  Phys. Condens. Matter}\ }\textbf {\bibinfo {volume} {15}},\ \bibinfo {pages}
  {R521} (\bibinfo {year} {2003})}\BibitemShut {NoStop}%
\bibitem [{\citenamefont {Saalfrank}(2006)}]{Saalfrank2006}%
  \BibitemOpen
  \bibfield  {author} {\bibinfo {author} {\bibfnamefont {P.}~\bibnamefont
  {Saalfrank}},\ }\href@noop {} {\bibfield  {journal} {\bibinfo  {journal}
  {Chem. Rev.}\ }\textbf {\bibinfo {volume} {106}},\ \bibinfo {pages} {4116}
  (\bibinfo {year} {2006})}\BibitemShut {NoStop}%
\bibitem [{\citenamefont {Menzel}(2012)}]{Menzel2012}%
  \BibitemOpen
  \bibfield  {author} {\bibinfo {author} {\bibfnamefont {D.}~\bibnamefont
  {Menzel}},\ }\href@noop {} {\bibfield  {journal} {\bibinfo  {journal} {J.
  Chem. Phys.}\ }\textbf {\bibinfo {volume} {137}},\ \bibinfo {pages} {091702}
  (\bibinfo {year} {2012})}\BibitemShut {NoStop}%
\bibitem [{\citenamefont {Martel}\ \emph {et~al.}(1996)\citenamefont {Martel},
  \citenamefont {Avouris},\ and\ \citenamefont {Lyo}}]{Martel1996}%
  \BibitemOpen
  \bibfield  {author} {\bibinfo {author} {\bibfnamefont {R.}~\bibnamefont
  {Martel}}, \bibinfo {author} {\bibfnamefont {P.}~\bibnamefont {Avouris}}, \
  and\ \bibinfo {author} {\bibfnamefont {I.-W.}\ \bibnamefont {Lyo}},\
  }\href@noop {} {\bibfield  {journal} {\bibinfo  {journal} {Science}\ }\textbf
  {\bibinfo {volume} {272}},\ \bibinfo {pages} {385} (\bibinfo {year}
  {1996})}\BibitemShut {NoStop}%
\bibitem [{\citenamefont {Stipe}\ \emph {et~al.}(1997)\citenamefont {Stipe},
  \citenamefont {Rezaei}, \citenamefont {Ho}, \citenamefont {Gao},
  \citenamefont {Persson},\ and\ \citenamefont {Lundqvist}}]{Stipe1997}%
  \BibitemOpen
  \bibfield  {author} {\bibinfo {author} {\bibfnamefont {B.~C.}\ \bibnamefont
  {Stipe}}, \bibinfo {author} {\bibfnamefont {M.~A.}\ \bibnamefont {Rezaei}},
  \bibinfo {author} {\bibfnamefont {W.}~\bibnamefont {Ho}}, \bibinfo {author}
  {\bibfnamefont {S.}~\bibnamefont {Gao}}, \bibinfo {author} {\bibfnamefont
  {M.}~\bibnamefont {Persson}}, \ and\ \bibinfo {author} {\bibfnamefont
  {B.~I.}\ \bibnamefont {Lundqvist}},\ }\href@noop {} {\bibfield  {journal}
  {\bibinfo  {journal} {Phys. Rev. Lett.}\ }\textbf {\bibinfo {volume} {78}},\
  \bibinfo {pages} {4410} (\bibinfo {year} {1997})}\BibitemShut {NoStop}%
\bibitem [{\citenamefont {Lauhon}\ and\ \citenamefont
  {Ho}(2000{\natexlab{a}})}]{Lauho2000}%
  \BibitemOpen
  \bibfield  {author} {\bibinfo {author} {\bibfnamefont {L.~J.}\ \bibnamefont
  {Lauhon}}\ and\ \bibinfo {author} {\bibfnamefont {W.}~\bibnamefont {Ho}},\
  }\href@noop {} {\bibfield  {journal} {\bibinfo  {journal} {Phys. Rev. Lett.}\
  }\textbf {\bibinfo {volume} {84}},\ \bibinfo {pages} {1527} (\bibinfo {year}
  {2000}{\natexlab{a}})}\BibitemShut {NoStop}%
\bibitem [{\citenamefont {Hla}\ \emph {et~al.}(2003)\citenamefont {Hla},
  \citenamefont {Meyer},\ and\ \citenamefont {Rieder}}]{Hla2003}%
  \BibitemOpen
  \bibfield  {author} {\bibinfo {author} {\bibfnamefont {S.-W.}\ \bibnamefont
  {Hla}}, \bibinfo {author} {\bibfnamefont {G.}~\bibnamefont {Meyer}}, \ and\
  \bibinfo {author} {\bibfnamefont {K.-H.}\ \bibnamefont {Rieder}},\
  }\href@noop {} {\bibfield  {journal} {\bibinfo  {journal} {Chem. Phys.
  Lett.}\ }\textbf {\bibinfo {volume} {370}},\ \bibinfo {pages} {431 }
  (\bibinfo {year} {2003})}\BibitemShut {NoStop}%
\bibitem [{\citenamefont {Roy}\ \emph {et~al.}(2013)\citenamefont {Roy},
  \citenamefont {Mujica},\ and\ \citenamefont {Ratner}}]{Roy2013}%
  \BibitemOpen
  \bibfield  {author} {\bibinfo {author} {\bibfnamefont {S.}~\bibnamefont
  {Roy}}, \bibinfo {author} {\bibfnamefont {V.}~\bibnamefont {Mujica}}, \ and\
  \bibinfo {author} {\bibfnamefont {M.~A.}\ \bibnamefont {Ratner}},\
  }\href@noop {} {\bibfield  {journal} {\bibinfo  {journal} {J. Chem. Phys.}\
  }\textbf {\bibinfo {volume} {139}},\ \bibinfo {pages} {074702} (\bibinfo
  {year} {2013})}\BibitemShut {NoStop}%
\bibitem [{\citenamefont {Lee}\ and\ \citenamefont {Ho}(1999)}]{Lee1999}%
  \BibitemOpen
  \bibfield  {author} {\bibinfo {author} {\bibfnamefont {H.~J.}\ \bibnamefont
  {Lee}}\ and\ \bibinfo {author} {\bibfnamefont {W.}~\bibnamefont {Ho}},\
  }\href@noop {} {\bibfield  {journal} {\bibinfo  {journal} {Science}\ }\textbf
  {\bibinfo {volume} {286}},\ \bibinfo {pages} {1719} (\bibinfo {year}
  {1999})}\BibitemShut {NoStop}%
\bibitem [{\citenamefont {Brandbyge}\ \emph {et~al.}(1995)\citenamefont
  {Brandbyge}, \citenamefont {Hedeg\aa{}rd}, \citenamefont {Heinz},
  \citenamefont {Misewich},\ and\ \citenamefont {Newns}}]{Brandbyge1995}%
  \BibitemOpen
  \bibfield  {author} {\bibinfo {author} {\bibfnamefont {M.}~\bibnamefont
  {Brandbyge}}, \bibinfo {author} {\bibfnamefont {P.}~\bibnamefont
  {Hedeg\aa{}rd}}, \bibinfo {author} {\bibfnamefont {T.~F.}\ \bibnamefont
  {Heinz}}, \bibinfo {author} {\bibfnamefont {J.~A.}\ \bibnamefont {Misewich}},
  \ and\ \bibinfo {author} {\bibfnamefont {D.~M.}\ \bibnamefont {Newns}},\
  }\href@noop {} {\bibfield  {journal} {\bibinfo  {journal} {Phys. Rev. B}\
  }\textbf {\bibinfo {volume} {52}},\ \bibinfo {pages} {6042} (\bibinfo {year}
  {1995})}\BibitemShut {NoStop}%
\bibitem [{\citenamefont {Saalfrank}\ and\ \citenamefont
  {Kosloff}(1996)}]{Saalfrank1996}%
  \BibitemOpen
  \bibfield  {author} {\bibinfo {author} {\bibfnamefont {P.}~\bibnamefont
  {Saalfrank}}\ and\ \bibinfo {author} {\bibfnamefont {R.}~\bibnamefont
  {Kosloff}},\ }\href@noop {} {\bibfield  {journal} {\bibinfo  {journal} {J.
  Chem. Phys.}\ }\textbf {\bibinfo {volume} {105}},\ \bibinfo {pages} {2441}
  (\bibinfo {year} {1996})}\BibitemShut {NoStop}%
\bibitem [{\citenamefont {Vondrak}\ and\ \citenamefont
  {Zhu}(1999)}]{Vondrak1999}%
  \BibitemOpen
  \bibfield  {author} {\bibinfo {author} {\bibfnamefont {T.}~\bibnamefont
  {Vondrak}}\ and\ \bibinfo {author} {\bibfnamefont {X.-Y.}\ \bibnamefont
  {Zhu}},\ }\href@noop {} {\bibfield  {journal} {\bibinfo  {journal} {Phys.
  Rev. Lett.}\ }\textbf {\bibinfo {volume} {82}},\ \bibinfo {pages} {1967}
  (\bibinfo {year} {1999})}\BibitemShut {NoStop}%
\bibitem [{\citenamefont {Erpenbeck}\ and\ \citenamefont
  {Thoss}(2019)}]{Erpenbeck_2019_HQME}%
  \BibitemOpen
  \bibfield  {author} {\bibinfo {author} {\bibfnamefont {A.}~\bibnamefont
  {Erpenbeck}}\ and\ \bibinfo {author} {\bibfnamefont {M.}~\bibnamefont
  {Thoss}},\ }\href@noop {} {\bibfield  {journal} {\bibinfo  {journal} {J.
  Chem. Phys.}\ }\textbf {\bibinfo {volume} {151}},\ \bibinfo {pages} {191101}
  (\bibinfo {year} {2019})}\BibitemShut {NoStop}%
\bibitem [{\citenamefont {Tanimura}\ and\ \citenamefont
  {Kubo}(1989)}]{Tanimura1989}%
  \BibitemOpen
  \bibfield  {author} {\bibinfo {author} {\bibfnamefont {Y.}~\bibnamefont
  {Tanimura}}\ and\ \bibinfo {author} {\bibfnamefont {R.}~\bibnamefont
  {Kubo}},\ }\href@noop {} {\bibfield  {journal} {\bibinfo  {journal} {J. Phys.
  Soc. Jpn.}\ }\textbf {\bibinfo {volume} {58}},\ \bibinfo {pages} {101}
  (\bibinfo {year} {1989})}\BibitemShut {NoStop}%
\bibitem [{\citenamefont {Tanimura}(2006)}]{Tanimura2006}%
  \BibitemOpen
  \bibfield  {author} {\bibinfo {author} {\bibfnamefont {Y.}~\bibnamefont
  {Tanimura}},\ }\href@noop {} {\bibfield  {journal} {\bibinfo  {journal} {J.
  Phys. Soc. Jpn.}\ }\textbf {\bibinfo {volume} {75}},\ \bibinfo {pages}
  {082001} (\bibinfo {year} {2006})}\BibitemShut {NoStop}%
\bibitem [{\citenamefont {Welack}\ \emph {et~al.}(2006)\citenamefont {Welack},
  \citenamefont {Schreiber},\ and\ \citenamefont
  {Kleinekath\"ofer}}]{Welack2006}%
  \BibitemOpen
  \bibfield  {author} {\bibinfo {author} {\bibfnamefont {S.}~\bibnamefont
  {Welack}}, \bibinfo {author} {\bibfnamefont {M.}~\bibnamefont {Schreiber}}, \
  and\ \bibinfo {author} {\bibfnamefont {U.}~\bibnamefont {Kleinekath\"ofer}},\
  }\href@noop {} {\bibfield  {journal} {\bibinfo  {journal} {J. Chem. Phys.}\
  }\textbf {\bibinfo {volume} {124}},\ \bibinfo {pages} {044712} (\bibinfo
  {year} {2006})}\BibitemShut {NoStop}%
\bibitem [{\citenamefont {Popescu}\ and\ \citenamefont
  {Kleinekath\"ofer}(2013)}]{Popescu2013}%
  \BibitemOpen
  \bibfield  {author} {\bibinfo {author} {\bibfnamefont {B.}~\bibnamefont
  {Popescu}}\ and\ \bibinfo {author} {\bibfnamefont {U.}~\bibnamefont
  {Kleinekath\"ofer}},\ }\href@noop {} {\bibfield  {journal} {\bibinfo
  {journal} {Phys. Status Solidi B}\ }\textbf {\bibinfo {volume} {250}},\
  \bibinfo {pages} {2288} (\bibinfo {year} {2013})}\BibitemShut {NoStop}%
\bibitem [{\citenamefont {H\"artle}\ \emph
  {et~al.}(2013{\natexlab{b}})\citenamefont {H\"artle}, \citenamefont {Cohen},
  \citenamefont {Reichman},\ and\ \citenamefont {Millis}}]{Haertle2013a}%
  \BibitemOpen
  \bibfield  {author} {\bibinfo {author} {\bibfnamefont {R.}~\bibnamefont
  {H\"artle}}, \bibinfo {author} {\bibfnamefont {G.}~\bibnamefont {Cohen}},
  \bibinfo {author} {\bibfnamefont {D.~R.}\ \bibnamefont {Reichman}}, \ and\
  \bibinfo {author} {\bibfnamefont {A.~J.}\ \bibnamefont {Millis}},\
  }\href@noop {} {\bibfield  {journal} {\bibinfo  {journal} {Phys. Rev. B}\
  }\textbf {\bibinfo {volume} {88}},\ \bibinfo {pages} {235426} (\bibinfo
  {year} {2013}{\natexlab{b}})}\BibitemShut {NoStop}%
\bibitem [{\citenamefont {H\"artle}\ and\ \citenamefont
  {Millis}(2014)}]{Haertle2014}%
  \BibitemOpen
  \bibfield  {author} {\bibinfo {author} {\bibfnamefont {R.}~\bibnamefont
  {H\"artle}}\ and\ \bibinfo {author} {\bibfnamefont {A.~J.}\ \bibnamefont
  {Millis}},\ }\href@noop {} {\bibfield  {journal} {\bibinfo  {journal} {Phys.
  Rev. B}\ }\textbf {\bibinfo {volume} {90}},\ \bibinfo {pages} {245426}
  (\bibinfo {year} {2014})}\BibitemShut {NoStop}%
\bibitem [{\citenamefont {H\"artle}\ \emph {et~al.}(2015)\citenamefont
  {H\"artle}, \citenamefont {Cohen}, \citenamefont {Reichman},\ and\
  \citenamefont {Millis}}]{Haertle2015}%
  \BibitemOpen
  \bibfield  {author} {\bibinfo {author} {\bibfnamefont {R.}~\bibnamefont
  {H\"artle}}, \bibinfo {author} {\bibfnamefont {G.}~\bibnamefont {Cohen}},
  \bibinfo {author} {\bibfnamefont {D.~R.}\ \bibnamefont {Reichman}}, \ and\
  \bibinfo {author} {\bibfnamefont {A.~J.}\ \bibnamefont {Millis}},\
  }\href@noop {} {\bibfield  {journal} {\bibinfo  {journal} {Phys. Rev. B}\
  }\textbf {\bibinfo {volume} {92}},\ \bibinfo {pages} {085430} (\bibinfo
  {year} {2015})}\BibitemShut {NoStop}%
\bibitem [{\citenamefont {Wilkins}\ and\ \citenamefont
  {Dattani}(2015)}]{Wilkins2015}%
  \BibitemOpen
  \bibfield  {author} {\bibinfo {author} {\bibfnamefont {D.~M.}\ \bibnamefont
  {Wilkins}}\ and\ \bibinfo {author} {\bibfnamefont {N.~S.}\ \bibnamefont
  {Dattani}},\ }\href@noop {} {\bibfield  {journal} {\bibinfo  {journal} {J.
  Chem. Theory Comput.,}\ }\textbf {\bibinfo {volume} {11}},\ \bibinfo {pages}
  {3411} (\bibinfo {year} {2015})}\BibitemShut {NoStop}%
\bibitem [{\citenamefont {Wenderoth}\ \emph {et~al.}(2016)\citenamefont
  {Wenderoth}, \citenamefont {B\"atge},\ and\ \citenamefont
  {H\"artle}}]{Wenderoth2016}%
  \BibitemOpen
  \bibfield  {author} {\bibinfo {author} {\bibfnamefont {S.}~\bibnamefont
  {Wenderoth}}, \bibinfo {author} {\bibfnamefont {J.}~\bibnamefont {B\"atge}},
  \ and\ \bibinfo {author} {\bibfnamefont {R.}~\bibnamefont {H\"artle}},\
  }\href@noop {} {\bibfield  {journal} {\bibinfo  {journal} {Phys. Rev. B}\
  }\textbf {\bibinfo {volume} {94}},\ \bibinfo {pages} {121303} (\bibinfo
  {year} {2016})}\BibitemShut {NoStop}%
\bibitem [{\citenamefont {Jin}\ \emph {et~al.}(2008)\citenamefont {Jin},
  \citenamefont {Zheng},\ and\ \citenamefont {Yan}}]{Jin2008}%
  \BibitemOpen
  \bibfield  {author} {\bibinfo {author} {\bibfnamefont {J.}~\bibnamefont
  {Jin}}, \bibinfo {author} {\bibfnamefont {X.}~\bibnamefont {Zheng}}, \ and\
  \bibinfo {author} {\bibfnamefont {Y.}~\bibnamefont {Yan}},\ }\href@noop {}
  {\bibfield  {journal} {\bibinfo  {journal} {J. Chem. Phys.}\ }\textbf
  {\bibinfo {volume} {128}},\ \bibinfo {pages} {234703} (\bibinfo {year}
  {2008})}\BibitemShut {NoStop}%
\bibitem [{\citenamefont {Li}\ \emph {et~al.}(2012)\citenamefont {Li},
  \citenamefont {Tong}, \citenamefont {Zheng}, \citenamefont {Hou},
  \citenamefont {Wei}, \citenamefont {Hu},\ and\ \citenamefont {Yan}}]{Li2012}%
  \BibitemOpen
  \bibfield  {author} {\bibinfo {author} {\bibfnamefont {Z.}~\bibnamefont
  {Li}}, \bibinfo {author} {\bibfnamefont {N.}~\bibnamefont {Tong}}, \bibinfo
  {author} {\bibfnamefont {X.}~\bibnamefont {Zheng}}, \bibinfo {author}
  {\bibfnamefont {D.}~\bibnamefont {Hou}}, \bibinfo {author} {\bibfnamefont
  {J.}~\bibnamefont {Wei}}, \bibinfo {author} {\bibfnamefont {J.}~\bibnamefont
  {Hu}}, \ and\ \bibinfo {author} {\bibfnamefont {Y.}~\bibnamefont {Yan}},\
  }\href@noop {} {\bibfield  {journal} {\bibinfo  {journal} {Phys. Rev. Lett.}\
  }\textbf {\bibinfo {volume} {109}},\ \bibinfo {pages} {266403} (\bibinfo
  {year} {2012})}\BibitemShut {NoStop}%
\bibitem [{\citenamefont {Zheng}\ \emph {et~al.}(2013)\citenamefont {Zheng},
  \citenamefont {Yan},\ and\ \citenamefont {Di~Ventra}}]{Zheng2013}%
  \BibitemOpen
  \bibfield  {author} {\bibinfo {author} {\bibfnamefont {X.}~\bibnamefont
  {Zheng}}, \bibinfo {author} {\bibfnamefont {Y.}~\bibnamefont {Yan}}, \ and\
  \bibinfo {author} {\bibfnamefont {M.}~\bibnamefont {Di~Ventra}},\ }\href@noop
  {} {\bibfield  {journal} {\bibinfo  {journal} {Phys. Rev. Lett.}\ }\textbf
  {\bibinfo {volume} {111}},\ \bibinfo {pages} {086601} (\bibinfo {year}
  {2013})}\BibitemShut {NoStop}%
\bibitem [{\citenamefont {Cheng}\ \emph {et~al.}(2015)\citenamefont {Cheng},
  \citenamefont {Wei},\ and\ \citenamefont {Yan}}]{Cheng2015}%
  \BibitemOpen
  \bibfield  {author} {\bibinfo {author} {\bibfnamefont {Y.}~\bibnamefont
  {Cheng}}, \bibinfo {author} {\bibfnamefont {J.}~\bibnamefont {Wei}}, \ and\
  \bibinfo {author} {\bibfnamefont {Y.}~\bibnamefont {Yan}},\ }\href@noop {}
  {\bibfield  {journal} {\bibinfo  {journal} {Europhys. Lett.}\ }\textbf
  {\bibinfo {volume} {112}},\ \bibinfo {pages} {57001} (\bibinfo {year}
  {2015})}\BibitemShut {NoStop}%
\bibitem [{\citenamefont {Ye}\ \emph {et~al.}(2016)\citenamefont {Ye},
  \citenamefont {Wang}, \citenamefont {Hou}, \citenamefont {Xu}, \citenamefont
  {Zheng},\ and\ \citenamefont {Yan}}]{Ye2016}%
  \BibitemOpen
  \bibfield  {author} {\bibinfo {author} {\bibfnamefont {L.}~\bibnamefont
  {Ye}}, \bibinfo {author} {\bibfnamefont {X.}~\bibnamefont {Wang}}, \bibinfo
  {author} {\bibfnamefont {D.}~\bibnamefont {Hou}}, \bibinfo {author}
  {\bibfnamefont {R.}~\bibnamefont {Xu}}, \bibinfo {author} {\bibfnamefont
  {X.}~\bibnamefont {Zheng}}, \ and\ \bibinfo {author} {\bibfnamefont
  {Y.}~\bibnamefont {Yan}},\ }\href@noop {} {\bibfield  {journal} {\bibinfo
  {journal} {WIREs Comput. Mol. Sci.}\ }\textbf {\bibinfo {volume} {6}},\
  \bibinfo {pages} {608} (\bibinfo {year} {2016})}\BibitemShut {NoStop}%
\bibitem [{\citenamefont {Cheng}\ \emph {et~al.}(2017)\citenamefont {Cheng},
  \citenamefont {Wang}, \citenamefont {Wei}, \citenamefont {Zhu},\ and\
  \citenamefont {Yan}}]{Cheng2017}%
  \BibitemOpen
  \bibfield  {author} {\bibinfo {author} {\bibfnamefont {Y.}~\bibnamefont
  {Cheng}}, \bibinfo {author} {\bibfnamefont {Y.}~\bibnamefont {Wang}},
  \bibinfo {author} {\bibfnamefont {J.}~\bibnamefont {Wei}}, \bibinfo {author}
  {\bibfnamefont {Z.}~\bibnamefont {Zhu}}, \ and\ \bibinfo {author}
  {\bibfnamefont {Y.}~\bibnamefont {Yan}},\ }\href@noop {} {\bibfield
  {journal} {\bibinfo  {journal} {Phys. Rev. B}\ }\textbf {\bibinfo {volume}
  {95}},\ \bibinfo {pages} {155417} (\bibinfo {year} {2017})}\BibitemShut
  {NoStop}%
\bibitem [{\citenamefont {Hou}\ \emph {et~al.}(2017)\citenamefont {Hou},
  \citenamefont {Wang}, \citenamefont {Wei}, \citenamefont {Zhu},\ and\
  \citenamefont {Yan}}]{Hou2017}%
  \BibitemOpen
  \bibfield  {author} {\bibinfo {author} {\bibfnamefont {W.}~\bibnamefont
  {Hou}}, \bibinfo {author} {\bibfnamefont {Y.}~\bibnamefont {Wang}}, \bibinfo
  {author} {\bibfnamefont {J.}~\bibnamefont {Wei}}, \bibinfo {author}
  {\bibfnamefont {Z.}~\bibnamefont {Zhu}}, \ and\ \bibinfo {author}
  {\bibfnamefont {Y.}~\bibnamefont {Yan}},\ }\href@noop {} {\bibfield
  {journal} {\bibinfo  {journal} {Sci. Rep.}\ }\textbf {\bibinfo {volume}
  {7}},\ \bibinfo {pages} {2486} (\bibinfo {year} {2017})}\BibitemShut
  {NoStop}%
\bibitem [{\citenamefont {Erpenbeck}\ \emph {et~al.}(2019)\citenamefont
  {Erpenbeck}, \citenamefont {G{\"o}tzend{\"o}rfer}, \citenamefont
  {Schinabeck},\ and\ \citenamefont {Thoss}}]{Erpenbeck_time-dep_mol-lead}%
  \BibitemOpen
  \bibfield  {author} {\bibinfo {author} {\bibfnamefont {A.}~\bibnamefont
  {Erpenbeck}}, \bibinfo {author} {\bibfnamefont {L.}~\bibnamefont
  {G{\"o}tzend{\"o}rfer}}, \bibinfo {author} {\bibfnamefont {C.}~\bibnamefont
  {Schinabeck}}, \ and\ \bibinfo {author} {\bibfnamefont {M.}~\bibnamefont
  {Thoss}},\ }\href@noop {} {\bibfield  {journal} {\bibinfo  {journal} {Eur.
  Phys. J. Spec. Top.}\ }\textbf {\bibinfo {volume} {227}},\ \bibinfo {pages}
  {1981} (\bibinfo {year} {2019})}\BibitemShut {NoStop}%
\bibitem [{\citenamefont {Tannor}(2007)}]{Tannor}%
  \BibitemOpen
  \bibfield  {author} {\bibinfo {author} {\bibfnamefont {D.~J.}\ \bibnamefont
  {Tannor}},\ }\href@noop {} {\emph {\bibinfo {title} {Introduction to Quantum
  Mechanics - a time-dependent perspective}}}\ (\bibinfo  {publisher}
  {University Science Books},\ \bibinfo {year} {2007})\BibitemShut {NoStop}%
\bibitem [{\citenamefont {Colbert}\ and\ \citenamefont
  {Miller}(1992)}]{Colbert1992}%
  \BibitemOpen
  \bibfield  {author} {\bibinfo {author} {\bibfnamefont {D.~T.}\ \bibnamefont
  {Colbert}}\ and\ \bibinfo {author} {\bibfnamefont {W.~H.}\ \bibnamefont
  {Miller}},\ }\href@noop {} {\bibfield  {journal} {\bibinfo  {journal} {J.
  Chem. Phys.}\ }\textbf {\bibinfo {volume} {96}},\ \bibinfo {pages} {1982}
  (\bibinfo {year} {1992})}\BibitemShut {NoStop}%
\bibitem [{\citenamefont {Zheng}\ \emph {et~al.}(2012)\citenamefont {Zheng},
  \citenamefont {Xu}, \citenamefont {Xu}, \citenamefont {Jin}, \citenamefont
  {Hu},\ and\ \citenamefont {Yan}}]{Zheng2012}%
  \BibitemOpen
  \bibfield  {author} {\bibinfo {author} {\bibfnamefont {X.}~\bibnamefont
  {Zheng}}, \bibinfo {author} {\bibfnamefont {R.}~\bibnamefont {Xu}}, \bibinfo
  {author} {\bibfnamefont {J.}~\bibnamefont {Xu}}, \bibinfo {author}
  {\bibfnamefont {J.}~\bibnamefont {Jin}}, \bibinfo {author} {\bibfnamefont
  {J.}~\bibnamefont {Hu}}, \ and\ \bibinfo {author} {\bibfnamefont
  {Y.}~\bibnamefont {Yan}},\ }\href@noop {} {\bibfield  {journal} {\bibinfo
  {journal} {Prog. Chem.}\ }\textbf {\bibinfo {volume} {24}},\ \bibinfo {pages}
  {1129} (\bibinfo {year} {2012})}\BibitemShut {NoStop}%
\bibitem [{\citenamefont {Tanimura}(2020)}]{Tanimura2020}%
  \BibitemOpen
  \bibfield  {author} {\bibinfo {author} {\bibfnamefont {Y.}~\bibnamefont
  {Tanimura}},\ }\href@noop {} {\bibfield  {journal} {\bibinfo  {journal} {J.
  Chem. Phys.}\ }\textbf {\bibinfo {volume} {153}},\ \bibinfo {pages} {020901}
  (\bibinfo {year} {2020})}\BibitemShut {NoStop}%
\bibitem [{\citenamefont {Mahan}(1993)}]{Mahan}%
  \BibitemOpen
  \bibfield  {author} {\bibinfo {author} {\bibfnamefont {G.~D.}\ \bibnamefont
  {Mahan}},\ }\href@noop {} {\emph {\bibinfo {title} {Many-Particle Physics}}}\
  (\bibinfo  {publisher} {Plenum Press},\ \bibinfo {address} {New York and
  London},\ \bibinfo {year} {1993})\BibitemShut {NoStop}%
\bibitem [{\citenamefont {Hu}\ \emph {et~al.}(2010)\citenamefont {Hu},
  \citenamefont {Xu},\ and\ \citenamefont {Yan}}]{Hu2010}%
  \BibitemOpen
  \bibfield  {author} {\bibinfo {author} {\bibfnamefont {J.}~\bibnamefont
  {Hu}}, \bibinfo {author} {\bibfnamefont {R.-X.}\ \bibnamefont {Xu}}, \ and\
  \bibinfo {author} {\bibfnamefont {Y.}~\bibnamefont {Yan}},\ }\href@noop {}
  {\bibfield  {journal} {\bibinfo  {journal} {J. Chem. Phys.}\ }\textbf
  {\bibinfo {volume} {133}},\ \bibinfo {pages} {101106} (\bibinfo {year}
  {2010})}\BibitemShut {NoStop}%
\bibitem [{\citenamefont {Hu}\ \emph {et~al.}(2011)\citenamefont {Hu},
  \citenamefont {Luo}, \citenamefont {Jiang}, \citenamefont {Xu},\ and\
  \citenamefont {Yan}}]{Hu2011}%
  \BibitemOpen
  \bibfield  {author} {\bibinfo {author} {\bibfnamefont {J.}~\bibnamefont
  {Hu}}, \bibinfo {author} {\bibfnamefont {M.}~\bibnamefont {Luo}}, \bibinfo
  {author} {\bibfnamefont {F.}~\bibnamefont {Jiang}}, \bibinfo {author}
  {\bibfnamefont {R.-X.}\ \bibnamefont {Xu}}, \ and\ \bibinfo {author}
  {\bibfnamefont {Y.}~\bibnamefont {Yan}},\ }\href@noop {} {\bibfield
  {journal} {\bibinfo  {journal} {J. Chem. Phys.}\ }\textbf {\bibinfo {volume}
  {134}},\ \bibinfo {pages} {244106} (\bibinfo {year} {2011})}\BibitemShut
  {NoStop}%
\bibitem [{\citenamefont {Tian}\ and\ \citenamefont {Chen}(2012)}]{Tian2012}%
  \BibitemOpen
  \bibfield  {author} {\bibinfo {author} {\bibfnamefont {H.}~\bibnamefont
  {Tian}}\ and\ \bibinfo {author} {\bibfnamefont {G.}~\bibnamefont {Chen}},\
  }\href@noop {} {\bibfield  {journal} {\bibinfo  {journal} {J. Chem. Phys.}\
  }\textbf {\bibinfo {volume} {137}},\ \bibinfo {pages} {204114} (\bibinfo
  {year} {2012})}\BibitemShut {NoStop}%
\bibitem [{\citenamefont {Popescu}\ \emph {et~al.}(2015)\citenamefont
  {Popescu}, \citenamefont {Rahman},\ and\ \citenamefont
  {Kleinekath\"ofer}}]{Popescu2015}%
  \BibitemOpen
  \bibfield  {author} {\bibinfo {author} {\bibfnamefont {B.}~\bibnamefont
  {Popescu}}, \bibinfo {author} {\bibfnamefont {H.}~\bibnamefont {Rahman}}, \
  and\ \bibinfo {author} {\bibfnamefont {U.}~\bibnamefont {Kleinekath\"ofer}},\
  }\href@noop {} {\bibfield  {journal} {\bibinfo  {journal} {J. Chem. Phys.}\
  }\textbf {\bibinfo {volume} {142}},\ \bibinfo {pages} {154103} (\bibinfo
  {year} {2015})}\BibitemShut {NoStop}%
\bibitem [{\citenamefont {Popescu}\ \emph {et~al.}(2016)\citenamefont
  {Popescu}, \citenamefont {Rahman},\ and\ \citenamefont
  {Kleinekath\"ofer}}]{Popescu2016}%
  \BibitemOpen
  \bibfield  {author} {\bibinfo {author} {\bibfnamefont {B.}~\bibnamefont
  {Popescu}}, \bibinfo {author} {\bibfnamefont {H.}~\bibnamefont {Rahman}}, \
  and\ \bibinfo {author} {\bibfnamefont {U.}~\bibnamefont {Kleinekath\"ofer}},\
  }\href@noop {} {\bibfield  {journal} {\bibinfo  {journal} {J. Phys. Chem. A}\
  }\textbf {\bibinfo {volume} {120}},\ \bibinfo {pages} {3270} (\bibinfo {year}
  {2016})}\BibitemShut {NoStop}%
\bibitem [{\citenamefont {Tang}\ \emph {et~al.}(2015)\citenamefont {Tang},
  \citenamefont {Ouyang}, \citenamefont {Gong}, \citenamefont {Wang},\ and\
  \citenamefont {Wu}}]{Tang2015}%
  \BibitemOpen
  \bibfield  {author} {\bibinfo {author} {\bibfnamefont {Z.}~\bibnamefont
  {Tang}}, \bibinfo {author} {\bibfnamefont {X.}~\bibnamefont {Ouyang}},
  \bibinfo {author} {\bibfnamefont {Z.}~\bibnamefont {Gong}}, \bibinfo {author}
  {\bibfnamefont {H.}~\bibnamefont {Wang}}, \ and\ \bibinfo {author}
  {\bibfnamefont {J.}~\bibnamefont {Wu}},\ }\href@noop {} {\bibfield  {journal}
  {\bibinfo  {journal} {J. Chem. Phys.}\ }\textbf {\bibinfo {volume} {143}},\
  \bibinfo {pages} {224112} (\bibinfo {year} {2015})}\BibitemShut {NoStop}%
\bibitem [{\citenamefont {Ye}\ \emph {et~al.}(2017)\citenamefont {Ye},
  \citenamefont {Zhang}, \citenamefont {Wang}, \citenamefont {Zheng},\ and\
  \citenamefont {Yan}}]{Ye2017}%
  \BibitemOpen
  \bibfield  {author} {\bibinfo {author} {\bibfnamefont {L.}~\bibnamefont
  {Ye}}, \bibinfo {author} {\bibfnamefont {H.-D.}\ \bibnamefont {Zhang}},
  \bibinfo {author} {\bibfnamefont {Y.}~\bibnamefont {Wang}}, \bibinfo {author}
  {\bibfnamefont {X.}~\bibnamefont {Zheng}}, \ and\ \bibinfo {author}
  {\bibfnamefont {Y.}~\bibnamefont {Yan}},\ }\href@noop {} {\bibfield
  {journal} {\bibinfo  {journal} {J. Chem. Phys.}\ }\textbf {\bibinfo {volume}
  {147}},\ \bibinfo {pages} {074111} (\bibinfo {year} {2017})}\BibitemShut
  {NoStop}%
\bibitem [{\citenamefont {Erpenbeck}\ \emph
  {et~al.}(2018{\natexlab{b}})\citenamefont {Erpenbeck}, \citenamefont
  {Hertlein}, \citenamefont {Schinabeck},\ and\ \citenamefont
  {Thoss}}]{Erpenbeck_RSHQME}%
  \BibitemOpen
  \bibfield  {author} {\bibinfo {author} {\bibfnamefont {A.}~\bibnamefont
  {Erpenbeck}}, \bibinfo {author} {\bibfnamefont {C.}~\bibnamefont {Hertlein}},
  \bibinfo {author} {\bibfnamefont {C.}~\bibnamefont {Schinabeck}}, \ and\
  \bibinfo {author} {\bibfnamefont {M.}~\bibnamefont {Thoss}},\ }\href@noop {}
  {\bibfield  {journal} {\bibinfo  {journal} {J. Chem. Phys.}\ }\textbf
  {\bibinfo {volume} {149}},\ \bibinfo {pages} {064106} (\bibinfo {year}
  {2018}{\natexlab{b}})}\BibitemShut {NoStop}%
\bibitem [{\citenamefont {Croy}\ and\ \citenamefont
  {Saalmann}(2009)}]{Croy2009}%
  \BibitemOpen
  \bibfield  {author} {\bibinfo {author} {\bibfnamefont {A.}~\bibnamefont
  {Croy}}\ and\ \bibinfo {author} {\bibfnamefont {U.}~\bibnamefont
  {Saalmann}},\ }\href@noop {} {\bibfield  {journal} {\bibinfo  {journal}
  {Phys. Rev. B}\ }\textbf {\bibinfo {volume} {80}},\ \bibinfo {pages} {245311}
  (\bibinfo {year} {2009})}\BibitemShut {NoStop}%
\bibitem [{\citenamefont {Zheng}\ \emph {et~al.}(2010)\citenamefont {Zheng},
  \citenamefont {Chen}, \citenamefont {Mo}, \citenamefont {Koo}, \citenamefont
  {Tian}, \citenamefont {Yam},\ and\ \citenamefont {Yan}}]{Zheng2010}%
  \BibitemOpen
  \bibfield  {author} {\bibinfo {author} {\bibfnamefont {X.}~\bibnamefont
  {Zheng}}, \bibinfo {author} {\bibfnamefont {G.}~\bibnamefont {Chen}},
  \bibinfo {author} {\bibfnamefont {Y.}~\bibnamefont {Mo}}, \bibinfo {author}
  {\bibfnamefont {S.}~\bibnamefont {Koo}}, \bibinfo {author} {\bibfnamefont
  {H.}~\bibnamefont {Tian}}, \bibinfo {author} {\bibfnamefont {C.}~\bibnamefont
  {Yam}}, \ and\ \bibinfo {author} {\bibfnamefont {Y.}~\bibnamefont {Yan}},\
  }\href@noop {} {\bibfield  {journal} {\bibinfo  {journal} {J. Chem. Phys.}\
  }\textbf {\bibinfo {volume} {133}},\ \bibinfo {pages} {114101} (\bibinfo
  {year} {2010})}\BibitemShut {NoStop}%
\bibitem [{\citenamefont {Zhang}\ \emph {et~al.}(2013)\citenamefont {Zhang},
  \citenamefont {Chen},\ and\ \citenamefont {Chen}}]{Zhang2013}%
  \BibitemOpen
  \bibfield  {author} {\bibinfo {author} {\bibfnamefont {Y.}~\bibnamefont
  {Zhang}}, \bibinfo {author} {\bibfnamefont {S.}~\bibnamefont {Chen}}, \ and\
  \bibinfo {author} {\bibfnamefont {G.}~\bibnamefont {Chen}},\ }\href@noop {}
  {\bibfield  {journal} {\bibinfo  {journal} {Phys. Rev. B}\ }\textbf {\bibinfo
  {volume} {87}},\ \bibinfo {pages} {085110} (\bibinfo {year}
  {2013})}\BibitemShut {NoStop}%
\bibitem [{\citenamefont {Kwok}\ \emph {et~al.}(2014)\citenamefont {Kwok},
  \citenamefont {Zhang},\ and\ \citenamefont {Chen}}]{Kwok2014}%
  \BibitemOpen
  \bibfield  {author} {\bibinfo {author} {\bibfnamefont {Y.}~\bibnamefont
  {Kwok}}, \bibinfo {author} {\bibfnamefont {Y.}~\bibnamefont {Zhang}}, \ and\
  \bibinfo {author} {\bibfnamefont {G.}~\bibnamefont {Chen}},\ }\href@noop {}
  {\bibfield  {journal} {\bibinfo  {journal} {Front. Phys.}\ }\textbf {\bibinfo
  {volume} {9}},\ \bibinfo {pages} {698} (\bibinfo {year} {2014})}\BibitemShut
  {NoStop}%
\bibitem [{\citenamefont {Tanimura}\ and\ \citenamefont
  {Wolynes}(1991)}]{Tanimura1991}%
  \BibitemOpen
  \bibfield  {author} {\bibinfo {author} {\bibfnamefont {Y.}~\bibnamefont
  {Tanimura}}\ and\ \bibinfo {author} {\bibfnamefont {P.~G.}\ \bibnamefont
  {Wolynes}},\ }\href@noop {} {\bibfield  {journal} {\bibinfo  {journal} {Phys.
  Rev. A}\ }\textbf {\bibinfo {volume} {43}},\ \bibinfo {pages} {4131}
  (\bibinfo {year} {1991})}\BibitemShut {NoStop}%
\bibitem [{\citenamefont {Xu}\ \emph {et~al.}(2005)\citenamefont {Xu},
  \citenamefont {Cui}, \citenamefont {Li}, \citenamefont {Mo},\ and\
  \citenamefont {Yan}}]{Xu2005}%
  \BibitemOpen
  \bibfield  {author} {\bibinfo {author} {\bibfnamefont {R.-X.}\ \bibnamefont
  {Xu}}, \bibinfo {author} {\bibfnamefont {P.}~\bibnamefont {Cui}}, \bibinfo
  {author} {\bibfnamefont {X.-Q.}\ \bibnamefont {Li}}, \bibinfo {author}
  {\bibfnamefont {Y.}~\bibnamefont {Mo}}, \ and\ \bibinfo {author}
  {\bibfnamefont {Y.}~\bibnamefont {Yan}},\ }\href@noop {} {\bibfield
  {journal} {\bibinfo  {journal} {J. Chem. Phys.}\ }\textbf {\bibinfo {volume}
  {122}},\ \bibinfo {eid} {041103} (\bibinfo {year} {2005})}\BibitemShut
  {NoStop}%
\bibitem [{\citenamefont {Shi}\ \emph {et~al.}(2009)\citenamefont {Shi},
  \citenamefont {Chen}, \citenamefont {Nan}, \citenamefont {Xu},\ and\
  \citenamefont {Yan}}]{Shi2009}%
  \BibitemOpen
  \bibfield  {author} {\bibinfo {author} {\bibfnamefont {Q.}~\bibnamefont
  {Shi}}, \bibinfo {author} {\bibfnamefont {L.}~\bibnamefont {Chen}}, \bibinfo
  {author} {\bibfnamefont {G.}~\bibnamefont {Nan}}, \bibinfo {author}
  {\bibfnamefont {R.-X.}\ \bibnamefont {Xu}}, \ and\ \bibinfo {author}
  {\bibfnamefont {Y.}~\bibnamefont {Yan}},\ }\href@noop {} {\bibfield
  {journal} {\bibinfo  {journal} {J. Chem. Phys.}\ }\textbf {\bibinfo {volume}
  {130}},\ \bibinfo {pages} {084105} (\bibinfo {year} {2009})}\BibitemShut
  {NoStop}%
\bibitem [{\citenamefont {Selst{\o}}\ and\ \citenamefont
  {Kvaal}(2010)}]{Selsto2010}%
  \BibitemOpen
  \bibfield  {author} {\bibinfo {author} {\bibfnamefont {S.}~\bibnamefont
  {Selst{\o}}}\ and\ \bibinfo {author} {\bibfnamefont {S.}~\bibnamefont
  {Kvaal}},\ }\href@noop {} {\bibfield  {journal} {\bibinfo  {journal} {J.
  Phys. B: At., Mol. Opt. Phys.}\ }\textbf {\bibinfo {volume} {43}},\ \bibinfo
  {pages} {065004} (\bibinfo {year} {2010})}\BibitemShut {NoStop}%
\bibitem [{\citenamefont {Kvaal}(2011)}]{Kvaal2011}%
  \BibitemOpen
  \bibfield  {author} {\bibinfo {author} {\bibfnamefont {S.}~\bibnamefont
  {Kvaal}},\ }\href@noop {} {\bibfield  {journal} {\bibinfo  {journal} {Phys.
  Rev. A}\ }\textbf {\bibinfo {volume} {84}},\ \bibinfo {pages} {022512}
  (\bibinfo {year} {2011})}\BibitemShut {NoStop}%
\bibitem [{\citenamefont {Prucker}\ \emph {et~al.}(2018)\citenamefont
  {Prucker}, \citenamefont {Bockstedte}, \citenamefont {Thoss},\ and\
  \citenamefont {Coto}}]{Prucker2018}%
  \BibitemOpen
  \bibfield  {author} {\bibinfo {author} {\bibfnamefont {V.}~\bibnamefont
  {Prucker}}, \bibinfo {author} {\bibfnamefont {M.}~\bibnamefont {Bockstedte}},
  \bibinfo {author} {\bibfnamefont {M.}~\bibnamefont {Thoss}}, \ and\ \bibinfo
  {author} {\bibfnamefont {P.~B.}\ \bibnamefont {Coto}},\ }\href@noop {}
  {\bibfield  {journal} {\bibinfo  {journal} {J. Chem. Phys.}\ }\textbf
  {\bibinfo {volume} {148}},\ \bibinfo {pages} {124705} (\bibinfo {year}
  {2018})}\BibitemShut {NoStop}%
\bibitem [{\citenamefont {Fabrikant}(1991)}]{Fabrikant1991}%
  \BibitemOpen
  \bibfield  {author} {\bibinfo {author} {\bibfnamefont {I.~I.}\ \bibnamefont
  {Fabrikant}},\ }\href@noop {} {\bibfield  {journal} {\bibinfo  {journal} {J.
  Phys. B}\ }\textbf {\bibinfo {volume} {24}},\ \bibinfo {pages} {2213}
  (\bibinfo {year} {1991})}\BibitemShut {NoStop}%
\bibitem [{\citenamefont {Fabrikant}(1994)}]{Fabrikant1994}%
  \BibitemOpen
  \bibfield  {author} {\bibinfo {author} {\bibfnamefont {I.~I.}\ \bibnamefont
  {Fabrikant}},\ }\href@noop {} {\bibfield  {journal} {\bibinfo  {journal} {J.
  Phys. B}\ }\textbf {\bibinfo {volume} {27}},\ \bibinfo {pages} {4325}
  (\bibinfo {year} {1994})}\BibitemShut {NoStop}%
\bibitem [{\citenamefont {Gertitschke}\ and\ \citenamefont
  {Domcke}(1993)}]{Gertitschke1993}%
  \BibitemOpen
  \bibfield  {author} {\bibinfo {author} {\bibfnamefont {P.~L.}\ \bibnamefont
  {Gertitschke}}\ and\ \bibinfo {author} {\bibfnamefont {W.}~\bibnamefont
  {Domcke}},\ }\href@noop {} {\bibfield  {journal} {\bibinfo  {journal} {Phys.
  Rev. A}\ }\textbf {\bibinfo {volume} {47}},\ \bibinfo {pages} {1031}
  (\bibinfo {year} {1993})}\BibitemShut {NoStop}%
\bibitem [{\citenamefont {Hahndorf}\ \emph {et~al.}(1994)\citenamefont
  {Hahndorf}, \citenamefont {Illenberger}, \citenamefont {Lehr},\ and\
  \citenamefont {Manz}}]{Hahndorf1994}%
  \BibitemOpen
  \bibfield  {author} {\bibinfo {author} {\bibfnamefont {I.}~\bibnamefont
  {Hahndorf}}, \bibinfo {author} {\bibfnamefont {E.}~\bibnamefont
  {Illenberger}}, \bibinfo {author} {\bibfnamefont {L.}~\bibnamefont {Lehr}}, \
  and\ \bibinfo {author} {\bibfnamefont {J.}~\bibnamefont {Manz}},\ }\href@noop
  {} {\bibfield  {journal} {\bibinfo  {journal} {Chem. Phys. Lett.}\ }\textbf
  {\bibinfo {volume} {231}},\ \bibinfo {pages} {460 } (\bibinfo {year}
  {1994})}\BibitemShut {NoStop}%
\bibitem [{\citenamefont {Wilde}\ \emph {et~al.}(1999)\citenamefont {Wilde},
  \citenamefont {Gallup},\ and\ \citenamefont {Fabrikant}}]{Wilde1999}%
  \BibitemOpen
  \bibfield  {author} {\bibinfo {author} {\bibfnamefont {R.~S.}\ \bibnamefont
  {Wilde}}, \bibinfo {author} {\bibfnamefont {G.~A.}\ \bibnamefont {Gallup}}, \
  and\ \bibinfo {author} {\bibfnamefont {I.~I.}\ \bibnamefont {Fabrikant}},\
  }\href@noop {} {\bibfield  {journal} {\bibinfo  {journal} {J. Phys. B}\
  }\textbf {\bibinfo {volume} {32}},\ \bibinfo {pages} {663} (\bibinfo {year}
  {1999})}\BibitemShut {NoStop}%
\bibitem [{\citenamefont {Dou}\ \emph {et~al.}(2016)\citenamefont {Dou},
  \citenamefont {Nitzan},\ and\ \citenamefont {Subotnik}}]{Dou2016}%
  \BibitemOpen
  \bibfield  {author} {\bibinfo {author} {\bibfnamefont {W.}~\bibnamefont
  {Dou}}, \bibinfo {author} {\bibfnamefont {A.}~\bibnamefont {Nitzan}}, \ and\
  \bibinfo {author} {\bibfnamefont {J.~E.}\ \bibnamefont {Subotnik}},\
  }\href@noop {} {\bibfield  {journal} {\bibinfo  {journal} {J. Chem. Phys.}\
  }\textbf {\bibinfo {volume} {144}},\ \bibinfo {pages} {074109} (\bibinfo
  {year} {2016})}\BibitemShut {NoStop}%
\bibitem [{\citenamefont {Todorov}\ \emph {et~al.}(2001)\citenamefont
  {Todorov}, \citenamefont {Hoekstra},\ and\ \citenamefont
  {Sutton}}]{Todorov2001}%
  \BibitemOpen
  \bibfield  {author} {\bibinfo {author} {\bibfnamefont {T.~N.}\ \bibnamefont
  {Todorov}}, \bibinfo {author} {\bibfnamefont {J.}~\bibnamefont {Hoekstra}}, \
  and\ \bibinfo {author} {\bibfnamefont {A.~P.}\ \bibnamefont {Sutton}},\
  }\href@noop {} {\bibfield  {journal} {\bibinfo  {journal} {Phys. Rev. Lett.}\
  }\textbf {\bibinfo {volume} {86}},\ \bibinfo {pages} {3606} (\bibinfo {year}
  {2001})}\BibitemShut {NoStop}%
\bibitem [{\citenamefont {Kopf}\ and\ \citenamefont
  {Saalfrank}(2004)}]{Kopf2004}%
  \BibitemOpen
  \bibfield  {author} {\bibinfo {author} {\bibfnamefont {A.}~\bibnamefont
  {Kopf}}\ and\ \bibinfo {author} {\bibfnamefont {P.}~\bibnamefont
  {Saalfrank}},\ }\href@noop {} {\bibfield  {journal} {\bibinfo  {journal}
  {Chem. Phys. Lett.}\ }\textbf {\bibinfo {volume} {386}},\ \bibinfo {pages}
  {17 } (\bibinfo {year} {2004})}\BibitemShut {NoStop}%
\bibitem [{\citenamefont {Pshenichnyuk}\ \emph {et~al.}(2013)\citenamefont
  {Pshenichnyuk}, \citenamefont {Coto}, \citenamefont {Leitherer},\ and\
  \citenamefont {Thoss}}]{Pshenichnyuk2013}%
  \BibitemOpen
  \bibfield  {author} {\bibinfo {author} {\bibfnamefont {I.~A.}\ \bibnamefont
  {Pshenichnyuk}}, \bibinfo {author} {\bibfnamefont {P.~B.}\ \bibnamefont
  {Coto}}, \bibinfo {author} {\bibfnamefont {S.}~\bibnamefont {Leitherer}}, \
  and\ \bibinfo {author} {\bibfnamefont {M.}~\bibnamefont {Thoss}},\
  }\href@noop {} {\bibfield  {journal} {\bibinfo  {journal} {J. Phys. Chem.
  Lett.}\ }\textbf {\bibinfo {volume} {4}},\ \bibinfo {pages} {809} (\bibinfo
  {year} {2013})}\BibitemShut {NoStop}%
\bibitem [{\citenamefont {Leitherer}\ \emph {et~al.}(2017)\citenamefont
  {Leitherer}, \citenamefont {J\"ager}, \citenamefont {Krause}, \citenamefont
  {Halik}, \citenamefont {Clark},\ and\ \citenamefont {Thoss}}]{Leitherer2017}%
  \BibitemOpen
  \bibfield  {author} {\bibinfo {author} {\bibfnamefont {S.}~\bibnamefont
  {Leitherer}}, \bibinfo {author} {\bibfnamefont {C.~M.}\ \bibnamefont
  {J\"ager}}, \bibinfo {author} {\bibfnamefont {A.}~\bibnamefont {Krause}},
  \bibinfo {author} {\bibfnamefont {M.}~\bibnamefont {Halik}}, \bibinfo
  {author} {\bibfnamefont {T.}~\bibnamefont {Clark}}, \ and\ \bibinfo {author}
  {\bibfnamefont {M.}~\bibnamefont {Thoss}},\ }\href@noop {} {\bibfield
  {journal} {\bibinfo  {journal} {Phys. Rev. Materials}\ }\textbf {\bibinfo
  {volume} {1}},\ \bibinfo {pages} {064601} (\bibinfo {year}
  {2017})}\BibitemShut {NoStop}%
\bibitem [{\citenamefont {Dujardin}\ \emph {et~al.}(1992)\citenamefont
  {Dujardin}, \citenamefont {Walkup},\ and\ \citenamefont
  {Avouris}}]{Dujardin1992}%
  \BibitemOpen
  \bibfield  {author} {\bibinfo {author} {\bibfnamefont {G.}~\bibnamefont
  {Dujardin}}, \bibinfo {author} {\bibfnamefont {R.~E.}\ \bibnamefont
  {Walkup}}, \ and\ \bibinfo {author} {\bibfnamefont {P.}~\bibnamefont
  {Avouris}},\ }\href@noop {} {\bibfield  {journal} {\bibinfo  {journal}
  {Science}\ }\textbf {\bibinfo {volume} {255}},\ \bibinfo {pages} {1232}
  (\bibinfo {year} {1992})}\BibitemShut {NoStop}%
\bibitem [{\citenamefont {Lauhon}\ and\ \citenamefont
  {Ho}(2000{\natexlab{b}})}]{Lauhon2000}%
  \BibitemOpen
  \bibfield  {author} {\bibinfo {author} {\bibfnamefont {L.~J.}\ \bibnamefont
  {Lauhon}}\ and\ \bibinfo {author} {\bibfnamefont {W.}~\bibnamefont {Ho}},\
  }\href@noop {} {\bibfield  {journal} {\bibinfo  {journal} {J. Phys. Chem. A}\
  }\textbf {\bibinfo {volume} {104}},\ \bibinfo {pages} {2463} (\bibinfo {year}
  {2000}{\natexlab{b}})}\BibitemShut {NoStop}%
\bibitem [{\citenamefont {Lauhon}\ and\ \citenamefont
  {Ho}(2000{\natexlab{c}})}]{Lauhon2000_2}%
  \BibitemOpen
  \bibfield  {author} {\bibinfo {author} {\bibfnamefont {L.}~\bibnamefont
  {Lauhon}}\ and\ \bibinfo {author} {\bibfnamefont {W.}~\bibnamefont {Ho}},\
  }\href@noop {} {\bibfield  {journal} {\bibinfo  {journal} {Surf. Sci.}\
  }\textbf {\bibinfo {volume} {451}},\ \bibinfo {pages} {219 } (\bibinfo {year}
  {2000}{\natexlab{c}})}\BibitemShut {NoStop}%
\bibitem [{\citenamefont {Preston}\ \emph {et~al.}(2020)\citenamefont
  {Preston}, \citenamefont {Kershaw},\ and\ \citenamefont
  {Kosov}}]{Preston2020}%
  \BibitemOpen
  \bibfield  {author} {\bibinfo {author} {\bibfnamefont {R.~J.}\ \bibnamefont
  {Preston}}, \bibinfo {author} {\bibfnamefont {V.~F.}\ \bibnamefont
  {Kershaw}}, \ and\ \bibinfo {author} {\bibfnamefont {D.~S.}\ \bibnamefont
  {Kosov}},\ }\href@noop {} {\bibfield  {journal} {\bibinfo  {journal} {Phys.
  Rev. B}\ }\textbf {\bibinfo {volume} {101}},\ \bibinfo {pages} {155415}
  (\bibinfo {year} {2020})}\BibitemShut {NoStop}%
\bibitem [{\citenamefont {Horsfield}\ \emph
  {et~al.}(2004{\natexlab{a}})\citenamefont {Horsfield}, \citenamefont
  {Bowler}, \citenamefont {Fisher}, \citenamefont {Todorov},\ and\
  \citenamefont {Sanchez}}]{Horsfield2004_3}%
  \BibitemOpen
  \bibfield  {author} {\bibinfo {author} {\bibfnamefont {A.~P.}\ \bibnamefont
  {Horsfield}}, \bibinfo {author} {\bibfnamefont {D.~R.}\ \bibnamefont
  {Bowler}}, \bibinfo {author} {\bibfnamefont {A.~J.}\ \bibnamefont {Fisher}},
  \bibinfo {author} {\bibfnamefont {T.~N.}\ \bibnamefont {Todorov}}, \ and\
  \bibinfo {author} {\bibfnamefont {C.~G.}\ \bibnamefont {Sanchez}},\
  }\href@noop {} {\bibfield  {journal} {\bibinfo  {journal} {J. Phys. Condens.
  Matter}\ }\textbf {\bibinfo {volume} {16}},\ \bibinfo {pages} {8251}
  (\bibinfo {year} {2004}{\natexlab{a}})}\BibitemShut {NoStop}%
\bibitem [{\citenamefont {Verdozzi}\ \emph {et~al.}(2006)\citenamefont
  {Verdozzi}, \citenamefont {Stefanucci},\ and\ \citenamefont
  {Almbladh}}]{Verdozzi2006}%
  \BibitemOpen
  \bibfield  {author} {\bibinfo {author} {\bibfnamefont {C.}~\bibnamefont
  {Verdozzi}}, \bibinfo {author} {\bibfnamefont {G.}~\bibnamefont
  {Stefanucci}}, \ and\ \bibinfo {author} {\bibfnamefont {C.-O.}\ \bibnamefont
  {Almbladh}},\ }\href@noop {} {\bibfield  {journal} {\bibinfo  {journal}
  {Phys. Rev. Lett.}\ }\textbf {\bibinfo {volume} {97}},\ \bibinfo {pages}
  {046603} (\bibinfo {year} {2006})}\BibitemShut {NoStop}%
\bibitem [{\citenamefont {Dundas}\ \emph {et~al.}(2009)\citenamefont {Dundas},
  \citenamefont {McEniry},\ and\ \citenamefont {Todorov}}]{Dundas2009}%
  \BibitemOpen
  \bibfield  {author} {\bibinfo {author} {\bibfnamefont {D.}~\bibnamefont
  {Dundas}}, \bibinfo {author} {\bibfnamefont {E.~J.}\ \bibnamefont {McEniry}},
  \ and\ \bibinfo {author} {\bibfnamefont {T.~N.}\ \bibnamefont {Todorov}},\
  }\href@noop {} {\bibfield  {journal} {\bibinfo  {journal} {Nat. Nano}\
  }\textbf {\bibinfo {volume} {4}},\ \bibinfo {pages} {99} (\bibinfo {year}
  {2009})}\BibitemShut {NoStop}%
\bibitem [{\citenamefont {Todorov}\ \emph {et~al.}(2014)\citenamefont
  {Todorov}, \citenamefont {Dundas}, \citenamefont {L\"u}, \citenamefont
  {Brandbyge},\ and\ \citenamefont {Hedeg\aa{}rd}}]{Todorov2014}%
  \BibitemOpen
  \bibfield  {author} {\bibinfo {author} {\bibfnamefont {T.}~\bibnamefont
  {Todorov}}, \bibinfo {author} {\bibfnamefont {D.}~\bibnamefont {Dundas}},
  \bibinfo {author} {\bibfnamefont {J.-T.}\ \bibnamefont {L\"u}}, \bibinfo
  {author} {\bibfnamefont {M.}~\bibnamefont {Brandbyge}}, \ and\ \bibinfo
  {author} {\bibfnamefont {P.}~\bibnamefont {Hedeg\aa{}rd}},\ }\href@noop {}
  {\bibfield  {journal} {\bibinfo  {journal} {Eur. Phys. J. Spec. Top.}\
  }\textbf {\bibinfo {volume} {35}},\ \bibinfo {pages} {065004} (\bibinfo
  {year} {2014})}\BibitemShut {NoStop}%
\bibitem [{\citenamefont {Cunningham}\ \emph {et~al.}(2015)\citenamefont
  {Cunningham}, \citenamefont {Todorov},\ and\ \citenamefont
  {Dundas}}]{Cunningham2015}%
  \BibitemOpen
  \bibfield  {author} {\bibinfo {author} {\bibfnamefont {B.}~\bibnamefont
  {Cunningham}}, \bibinfo {author} {\bibfnamefont {T.~N.}\ \bibnamefont
  {Todorov}}, \ and\ \bibinfo {author} {\bibfnamefont {D.}~\bibnamefont
  {Dundas}},\ }\href@noop {} {\bibfield  {journal} {\bibinfo  {journal}
  {Beilstein J. Nanotechnol.}\ }\textbf {\bibinfo {volume} {6}},\ \bibinfo
  {pages} {2140} (\bibinfo {year} {2015})}\BibitemShut {NoStop}%
\bibitem [{\citenamefont {Leitherer}\ \emph {et~al.}(2019)\citenamefont
  {Leitherer}, \citenamefont {Papior},\ and\ \citenamefont
  {Brandbyge}}]{Leitherer2019}%
  \BibitemOpen
  \bibfield  {author} {\bibinfo {author} {\bibfnamefont {S.}~\bibnamefont
  {Leitherer}}, \bibinfo {author} {\bibfnamefont {N.}~\bibnamefont {Papior}}, \
  and\ \bibinfo {author} {\bibfnamefont {M.}~\bibnamefont {Brandbyge}},\
  }\href@noop {} {\bibfield  {journal} {\bibinfo  {journal} {Phys. Rev. B}\
  }\textbf {\bibinfo {volume} {100}},\ \bibinfo {pages} {035415} (\bibinfo
  {year} {2019})}\BibitemShut {NoStop}%
\bibitem [{\citenamefont {Horsfield}\ \emph
  {et~al.}(2004{\natexlab{b}})\citenamefont {Horsfield}, \citenamefont
  {Bowler}, \citenamefont {Fisher}, \citenamefont {Todorov},\ and\
  \citenamefont {Montgomery}}]{Horsfield2004_2}%
  \BibitemOpen
  \bibfield  {author} {\bibinfo {author} {\bibfnamefont {A.~P.}\ \bibnamefont
  {Horsfield}}, \bibinfo {author} {\bibfnamefont {D.~R.}\ \bibnamefont
  {Bowler}}, \bibinfo {author} {\bibfnamefont {A.~J.}\ \bibnamefont {Fisher}},
  \bibinfo {author} {\bibfnamefont {T.~N.}\ \bibnamefont {Todorov}}, \ and\
  \bibinfo {author} {\bibfnamefont {M.~J.}\ \bibnamefont {Montgomery}},\
  }\href@noop {} {\bibfield  {journal} {\bibinfo  {journal} {J. Phys. Condens.
  Matter}\ }\textbf {\bibinfo {volume} {16}},\ \bibinfo {pages} {3609}
  (\bibinfo {year} {2004}{\natexlab{b}})}\BibitemShut {NoStop}%
\bibitem [{\citenamefont {Shen}\ \emph {et~al.}(1995)\citenamefont {Shen},
  \citenamefont {Wang}, \citenamefont {Abeln}, \citenamefont {Tucker},
  \citenamefont {Lyding}, \citenamefont {Avouris},\ and\ \citenamefont
  {Walkup}}]{Shen1995}%
  \BibitemOpen
  \bibfield  {author} {\bibinfo {author} {\bibfnamefont {T.~C.}\ \bibnamefont
  {Shen}}, \bibinfo {author} {\bibfnamefont {C.}~\bibnamefont {Wang}}, \bibinfo
  {author} {\bibfnamefont {G.~C.}\ \bibnamefont {Abeln}}, \bibinfo {author}
  {\bibfnamefont {J.~R.}\ \bibnamefont {Tucker}}, \bibinfo {author}
  {\bibfnamefont {J.~W.}\ \bibnamefont {Lyding}}, \bibinfo {author}
  {\bibfnamefont {P.}~\bibnamefont {Avouris}}, \ and\ \bibinfo {author}
  {\bibfnamefont {R.~E.}\ \bibnamefont {Walkup}},\ }\href@noop {} {\bibfield
  {journal} {\bibinfo  {journal} {Science}\ }\textbf {\bibinfo {volume}
  {268}},\ \bibinfo {pages} {1590} (\bibinfo {year} {1995})}\BibitemShut
  {NoStop}%
\bibitem [{\citenamefont {Schmidt}\ \emph {et~al.}(2008)\citenamefont
  {Schmidt}, \citenamefont {Werner}, \citenamefont {M\"uhlbacher},\ and\
  \citenamefont {Komnik}}]{Schmidt2008}%
  \BibitemOpen
  \bibfield  {author} {\bibinfo {author} {\bibfnamefont {T.~L.}\ \bibnamefont
  {Schmidt}}, \bibinfo {author} {\bibfnamefont {P.}~\bibnamefont {Werner}},
  \bibinfo {author} {\bibfnamefont {L.}~\bibnamefont {M\"uhlbacher}}, \ and\
  \bibinfo {author} {\bibfnamefont {A.}~\bibnamefont {Komnik}},\ }\href@noop {}
  {\bibfield  {journal} {\bibinfo  {journal} {Phys. Rev. B}\ }\textbf {\bibinfo
  {volume} {78}},\ \bibinfo {pages} {235110} (\bibinfo {year}
  {2008})}\BibitemShut {NoStop}%
\bibitem [{\citenamefont {Jorn}\ and\ \citenamefont
  {Seideman}(2009)}]{Jorn2009}%
  \BibitemOpen
  \bibfield  {author} {\bibinfo {author} {\bibfnamefont {R.}~\bibnamefont
  {Jorn}}\ and\ \bibinfo {author} {\bibfnamefont {T.}~\bibnamefont
  {Seideman}},\ }\href@noop {} {\bibfield  {journal} {\bibinfo  {journal} {J.
  Chem. Phys.}\ }\textbf {\bibinfo {volume} {131}},\ \bibinfo {pages} {244114}
  (\bibinfo {year} {2009})}\BibitemShut {NoStop}%
\bibitem [{\citenamefont {Albrecht}\ \emph {et~al.}(2013)\citenamefont
  {Albrecht}, \citenamefont {Soller}, \citenamefont {M\"uhlbacher},\ and\
  \citenamefont {Komnik}}]{Albrecht2013}%
  \BibitemOpen
  \bibfield  {author} {\bibinfo {author} {\bibfnamefont {K.}~\bibnamefont
  {Albrecht}}, \bibinfo {author} {\bibfnamefont {H.}~\bibnamefont {Soller}},
  \bibinfo {author} {\bibfnamefont {L.}~\bibnamefont {M\"uhlbacher}}, \ and\
  \bibinfo {author} {\bibfnamefont {A.}~\bibnamefont {Komnik}},\ }\href@noop {}
  {\bibfield  {journal} {\bibinfo  {journal} {Physica E}\ }\textbf {\bibinfo
  {volume} {54}},\ \bibinfo {pages} {15 } (\bibinfo {year} {2013})}\BibitemShut
  {NoStop}%
\bibitem [{\citenamefont {Gergs}\ \emph {et~al.}(2015)\citenamefont {Gergs},
  \citenamefont {H\"orig}, \citenamefont {Wegewijs},\ and\ \citenamefont
  {Schuricht}}]{Gergs2015}%
  \BibitemOpen
  \bibfield  {author} {\bibinfo {author} {\bibfnamefont {N.~M.}\ \bibnamefont
  {Gergs}}, \bibinfo {author} {\bibfnamefont {C.~B.~M.}\ \bibnamefont
  {H\"orig}}, \bibinfo {author} {\bibfnamefont {M.~R.}\ \bibnamefont
  {Wegewijs}}, \ and\ \bibinfo {author} {\bibfnamefont {D.}~\bibnamefont
  {Schuricht}},\ }\href@noop {} {\bibfield  {journal} {\bibinfo  {journal}
  {Phys. Rev. B}\ }\textbf {\bibinfo {volume} {91}},\ \bibinfo {pages} {201107}
  (\bibinfo {year} {2015})}\BibitemShut {NoStop}%
\bibitem [{\citenamefont {Gaudenzi}\ \emph {et~al.}(2017)\citenamefont
  {Gaudenzi}, \citenamefont {Misiorny}, \citenamefont {Burzurí}, \citenamefont
  {Wegewijs},\ and\ \citenamefont {van~der Zant}}]{Gaudenzi2017}%
  \BibitemOpen
  \bibfield  {author} {\bibinfo {author} {\bibfnamefont {R.}~\bibnamefont
  {Gaudenzi}}, \bibinfo {author} {\bibfnamefont {M.}~\bibnamefont {Misiorny}},
  \bibinfo {author} {\bibfnamefont {E.}~\bibnamefont {Burzurí}}, \bibinfo
  {author} {\bibfnamefont {M.~R.}\ \bibnamefont {Wegewijs}}, \ and\ \bibinfo
  {author} {\bibfnamefont {H.~S.~J.}\ \bibnamefont {van~der Zant}},\
  }\href@noop {} {\bibfield  {journal} {\bibinfo  {journal} {J. Chem. Phys.}\
  }\textbf {\bibinfo {volume} {146}},\ \bibinfo {pages} {092330} (\bibinfo
  {year} {2017})}\BibitemShut {NoStop}%
\end{thebibliography}%

\end{document}